\documentclass[final,3p,times]{elsarticle}

\usepackage{amssymb}
\usepackage[]{amsmath}
\usepackage{graphics}
\usepackage{graphicx}
\usepackage{amssymb}
\usepackage[]{amsmath}
\usepackage{amsthm}
\usepackage{setspace}
\usepackage{epsfig}
\usepackage{subfigure}

 \usepackage{amsthm}
 \usepackage{amsmath}
\newtheorem{proposition}{Proposition}[section]
\newtheorem{theorem}{Theorem}[section]

\allowdisplaybreaks

\biboptions{numbers,sort&compress}
\usepackage[dvipdfm,colorlinks,linkcolor=red,anchorcolor=blue,citecolor=green]{hyperref}
\usepackage{multirow}
\usepackage{amsmath}
\usepackage{anysize}
\marginsize{2cm}{2cm}{0cm}{1.5cm}
\selectfont

\begin{document}

\begin{frontmatter}



\title{Modulation instability, conservation laws and localized waves for the generalized coupled Fokas-Lenells equation}

\author{Yunfei Yue$^{a}$ }
\author{Yong Chen$^{a,b,c}$ \corref{cor1}}
\ead{ychen@sei.ecnu.edu.cn}

\cortext[cor1]{Corresponding author. }

\address[label1]{School of Mathematical Sciences, Shanghai Key Laboratory of PMMP, East China Normal University, Shanghai, 200062, China}
\address[label2]{College of mathematics and systems science, shandong university of science and technology, Qingdao,
266590, China}
\address[label3]{Department of Physics, Zhejiang Normal University,
    Jinhua, 321004,  China}


\begin{abstract}
This paper focuses on the modulation instability, conservation laws and localized wave solutions of the generalized coupled Fokas-Lenells equation. Based on the theory of linear stability analysis, distribution pattern of modulation instability gain $G$ in the $(K,k)$ frequency plane is depicted, and the constraints for the existence of rogue waves are derived. Subsequently, we construct the infinitely many conservation laws for the generalized coupled Fokas-Lenells equation from the Riccati-type formulas of the Lax pair. In addition, the compact determinant expressions of the $N$-order localized wave solutions are given via generalized Darboux transformation, including higher-order rogue waves and interaction solutions among rogue waves with bright-dark solitons or breathers. These solutions are parameter controllable: $(m_i,n_i)$ and $(\alpha,\beta)$ control the structure and ridge deflection of solution respectively, while the value of $|d|$ controls the strength of interaction to realize energy exchange. Especially, when $d=0$, the interaction solutions degenerate into the corresponding order of rogue waves.

%
%
\end{abstract}

\end{frontmatter}


\section{Introduction}

There are extremely complex dynamical processes in nature all the time, and many physical phenomena are nonlinear in our daily life. Research on modulation instability (MI) and dynamical behavior of the localized wave solutions for the nonlinear systems is a fascinating subject in the field of contemporary nonlinear science, which has attracted widespread attention from many experts and scholars.
MI \cite{j-zakharovve-pd-2009} is the instability of a monochromatic wave or constant background for disturbances. In experiments, MI is usually used to generate typical nonlinear waves such as soliton, rogue wave and breather. Theoretically, MI is divided into two stages. The first stage can be described by linear theory, and the second stage is called nonlinear stage. Linear MI analysis can only give the instability criterion of small perturbation and the change rate of linear instability. The linear phase disturbance grows exponentially with time and quickly reaches a scale comparable to the background wave, thus entering the nonlinear phase. As early as the 1960s, MI, known as Benjamin-Feir instability \cite{j-benjamintb-jfm-1967}, was discovered in deep water wave theory. As is known to all, MI is a natural phenomenon that can be discovered in many fields, such as nonlinear optics \cite{j-bespalovvi-jetpl-1966}, fluid mechanics \cite{j-bonnefoyf-prf-2020}, Bose-Einstein condensation \cite{j-nguyenjhv-science-2017}, plasma \cite{j-taniutit-prl-1968} and so on.


Localized wave is one of the main research hotspots in contemporary nonlinear mathematical physics. Soliton, breather and rogue wave are localized waves with obvious dynamical and physical characteristics in nonlinear systems.
Soliton \cite{j-ablowitzmj-sam-1974,j-yangb-nd-2018} has locality and particle properties, and all integrable equations have soliton solutions that reflect the universal nonlinear phenomena in nature. Breather and rogue wave are two typical localized structures with obvious instability on the plane wave backgrounds.
Breather \cite{j-wangl-pre-2016,j-yueyf-aml-2019} not only has a periodic structure in a particular direction, but also can be used to explain the phenomenon of rogue wave.
rogue wave \cite{j-akhmedievn-pre-2009,j-chenjc-pre-2019,
j-huangll-cnsns-2019,j-yueyf-cnsns-2020,j-mug-pdnp-2020} is dual localized structures in both time evolution and spatial distribution directions, which comes and disappears without a trace \cite{j-akhmedievn-pla-2009}.
The mechanism of rogue wave can be regarded as the high amplitude wave, which generated by the collision of soliton and breather \cite{j-akhmedievn-pra-2009}.

Among various nonlinear systems, one of the most prominent integrable examples, which is widely used to model the propagation of nonlinear wave in monomode optical fibers, is the Fokas-Lenells (FL) equation
\begin{equation}\label{gcfl-fl}
    iu_t-\nu u_{xt}+\gamma u_{xx}+\rho|u|^2(u+ivu_x)=0,~~\rho=\pm1, ~~x\in\mathbb{R}.
\end{equation}
It was proposed by Fokas \cite{j-fokasas-pd-1995} and Lenells \cite{j-lenellsj-sam-2009}, is a completely integrable equation. In \cite{j-fokasas-pd-1995}, Eq. \eqref{gcfl-fl} was derived by means of the bi-Hamiltonian method. In \cite{j-lenellsj-sam-2009}, the physical derivation of Eq. \eqref{gcfl-fl} was presented. Eq. \eqref{gcfl-fl} is an integrable generalization of the nonlinear Schr\"{o}dinger (NLS) equation and relates to the first negative flow of the derivative NLS equation hierarchy, which is similar with the relationship between the Camassa-Holm equation and the first negative flow of KdV hierarchy \cite{j-mckeanhp-cpam-2003}. When $\nu=0$, Eq. \eqref{gcfl-fl} can be reduced to the NLS equation \cite{j-agrawalgp-2007}, which is used to describe the nonlinear deep water model, and it occupies a particularly prominent position in nonlinear physics.
In \cite{j-lenellsj-n-2009}, Lenells and Fokas not only derived the conservation laws and Lax pair of Eq. \eqref{gcfl-fl} via its bi-Hamiltonian structure, but also solved the initial value problem via the inverse scattering transform. Subsequently, there are many results on the localized wave solutions of the FL equation, such as soliton, breather and rogue wave have been constructed by the Darboux transformation (DT) method \cite{j-chensh-pla-2014,j-hejs-jpsj-2012}, the dressing method \cite{j-lenellsj-jns-2010}, the Hirota direct method \cite{j-matsunoy-jpamt-2012} and the complex envelope function method \cite{j-trikih-wrc-2017}. In \cite{j-xuj-jde-2015}, long-time asymptotic behavior for the cauchy problem of the FL equation has been discussed by the Deift-Zhou method. General rogue waves and some new rogue wave patterns for the generalized derivative NLS equations \cite{j-yangb-jns-2020} have been constructed by the bilinear KP reduction method.

Similar to the case of the NLS equation, it is necessary to consider the two-component or multi-component generalization of the FL equation for describing the effects of polarization or anisotropy. The coupled FL equation \cite{j-guobl-jmp-2012} can be written as
\begin{equation}\label{cfl-eq}
    \begin{split}
     p_{xt}+p+i(|p|^2+\frac{1}{2}\sigma|q|^2)p_x+\frac{1}{2}i\sigma pq^*q_x=0,\\
     q_{xt}+q+i(\sigma|q|^2+\frac{1}{2}|p|^2)q_x+\frac{1}{2}i\sigma qp^*p_x=0,
     \end{split}
\end{equation}
here $\sigma=\pm1$. The coupled FL equation shares the same spatial part of the spectral problem with the coupled derivative NLS equation \cite{j-morrishc-ps-1979}, which is the first nontrivial negative flow of the vector Kaup-Newell hierarchy and relevants in the theory of polarized Alfv\'{e}n waves and the propagation of the ultra-short pulse. Many effective methods, such as Riemann-Hilbert method \cite{j-hubb-arxive-2017}, DT \cite{j-zhangy-narwa-2017}, non-recursive DT \cite{j-yeyl-prsa-2019}, generalized DT \cite{j-wangmm-nd-2019}, etc., have been developed to study on the coupled FL equation. The coupled FL equation is one of the integrable systems as shown in \cite{j-zhangmx-jnmp-2015} and of course admit other integrable properties including multi-Hamiltonian structure, infinitely many conservation laws, the general soliton solutions \cite{j-linglm-narwa-2018} and optical soliton \cite{j-biswasa-o-2018}. In addition, the baseband MI, rogue wave solutions and state transitions for a deformed FL equation \cite{j-wangx-nd-2019} and semi-rational solutions for the coupled derivative NLS equation \cite{j-xut-nd-2020} have been studied by the generalized DT method. Fundamental Peregrine solitons with unprecedentedly ultrahigh peak amplitude have been discovered for the vector derivative NLS equation \cite{j-chensh-prl-2020}.

In this paper, we will use the following generalized coupled Fokas-Lenells (gc-FL) equation \cite{j-yangjw-nd-2018}
\begin{equation}\label{gcfl-eq}
    \begin{split}
     iu_{xt}-i\alpha u_{xx}+2\gamma u_x-2\beta(2|u|^2+|v|^2)u_x-2\beta uv^*v_x+4i\beta u=0,\\
     iv_{xt}-i\alpha v_{xx}+2\gamma v_x-2\beta(2|v|^2+|u|^2)v_x-2\beta vu^*u_x+4i\beta v=0,
     \end{split}
\end{equation}
as the reference system, where $\alpha,\beta$ and $\gamma$ are real parameters. It associates with two conserved quantities
\begin{equation}\label{I-1-2}
\begin{split}
I_1[u,v]&=\int_{-\infty}^{\infty}(u_xu_x^*+v_xv_x^*)dx,\\
I_2[u,v]&=\int_{-\infty}^{\infty}\left(-(u_xu_x^*+v_xv_x^*)^2+2i(u_xu_{xx}^*+v_xv_{xx}^*)\right)dx,
\end{split}
\end{equation}
When $\alpha=0$,~~$\beta=\frac{1}{4}$, and $\gamma=0$, Eq. \eqref{gcfl-eq} degenerates to the coupled FL equation \eqref{cfl-eq}. It should be pointed out that the coupled FL equation studied in \cite{j-zhangy-narwa-2017} is equivalent to the coupled FL equation \eqref{cfl-eq} by a gauge transformation. In \cite{j-zhangy-narwa-2017}, taking $\alpha=2,~~\gamma=2$, and $\beta=\frac{1}{2}$, soliton, first-order breather and rogue wave solutions of Eq. \eqref{gcfl-eq} have been derived. Owing to the practical significance of the FL equation, further study the dynamical characteristics of the higher-order localized wave solutions for the gc-FL equation \eqref{gcfl-eq}, especially the interaction among different localized wave solutions, is therefore of great importance.

The motivation of the present paper is to investigate modulation instability of continuous waves, conservation laws and localized wave solutions for the gc-FL equation \eqref{gcfl-eq}. Based on the theory of linear stability analysis, we depict the distribution pattern of the MI gain $G$ on the frequency plane $(K,k)$ and derive the constraint condition for the existence of rogue waves. According to Riccati-type formulas of the spectral problem, infinitely many conservation laws and recursion relations for the conserved density and conserved flux of the gc-FL equation \eqref{gcfl-eq} are derived, respectively. Furthermore, the concrete formula of the higher-order localized wave solution is derived by generalized DT method, and the effects of arbitrary parameters on the structure of localized wave solutions are discussed in detail.


The remainder of our paper is organized as follows. In Section 2, we discusses MI distribution features of the gc-FL equation according to the theory of modulation instability analysis. In Section 3, we construct infinitely many conservation laws and the generalized DT of the gc-FL equation, and give a concrete formula of the $N$-order localized wave solution. Interactional localized waves are presented and their dynamical behavior are analyzed in Section 4. In the final Section, some conclusions are given.

\section{Modulation instability of continuous waves}

The dispersion term and the nonlinear term play different roles in the nonlinear systems, and both of them affect the instability of the solution in the system. In general, the nonlinear partial differential equation satisfying the fiber communication model, for the stable solution, under the interaction of the nonlinear term and the dispersive term, there will be instability, which is called modulation instability. In nonlinear dispersive media, modulation instability is a nonlinear process, in which a continuous plane wave generates amplitude and frequency self-modulation through a nonlinear dispersive medium, resulting in exponential growth of small perturbations superimposed on a plane wave \cite{j-agrawalgp-2007}. The study of modulation instability regions is crucial in many fields and is the basis for interpreting or regulating various models or phenomena in different fields.

In this section, we focus on the modulation instability of continuous waves in the gc-FL equation. The plane wave solution of equation (\ref{gcfl-eq}) has the following form
\begin{equation}\label{gcfl-mi-1}
    \begin{split}
     u_{cw}=c_1e^{i\theta}=c_1e^{i(kx+\omega t)},\\
     v_{cw}=c_2e^{i\theta}=c_2e^{i(kx+\omega t)},\\
     \end{split}
\end{equation}
where $c_i(i=1,2)$, $k$ and $\omega$ represent the amplitude, frequency and wave number of background, respectively. Substituting Eq. (\ref{gcfl-mi-1}) into Eq. (\ref{gcfl-eq}), it yields the dispersion relation
\begin{equation*}
\omega=-\frac{4c_1^2\beta k+4c_2^2\beta k-\alpha k^2-2\gamma k-4\beta}{k}.
\end{equation*}
According to the modulation instability theory, adding small perturbations ($p,q\ll1$) to the plane-wave solution, then a perturbation solution can be given as
\begin{equation}\label{gcfl-mi-2}
    \begin{split}
    u_{pert}=\left(c_1+\epsilon p(x,t)\right)e^{i(kx+\omega t)},\\
    v_{pert}=\left(c_2+\epsilon q(x,t)\right)e^{i(kx+\omega t)},\\
    \end{split}
\end{equation}
with
\begin{equation*}
    \begin{split}
    p(z,t)=m_1e^{i(Kx+\Omega t)}+n_1e^{-i(Kx+\Omega t)},\\
    q(z,t)=m_2e^{i(Kx+\Omega t)}+n_2e^{-i(Kx+\Omega t)},\\
    \end{split}
\end{equation*}
where $m_i$ and $n_i(i=1,2)$ are small real parameters, and $K$ represents the perturbed frequency. Substituting the perturbation solution (\ref{gcfl-mi-2}) into system (\ref{gcfl-eq}) generates a system of linear homogeneous equations for $m_i$ and $n_i$, that is
\begin{equation}
   \begin{pmatrix}
     D_{1,1}& D_{1,2} & D_{1,3} & D_{1,4} \\
    D_{2,1}& D_{2,2} & D_{2,3} & D_{2,4}  \\
   D_{3,1}& D_{3,2} & D_{3,3} & D_{3,4}  \\
   D_{4,1}& D_{4,2} & D_{4,3} & D_{4,4}
    \end{pmatrix}
    \begin{pmatrix}
    m_1 \\
    m_2 \\
    n_1 \\
    n_2
    \end{pmatrix}=0.
\end{equation}
with
\begin{equation*}
\begin{split}
  D_{1,1}&=(-4\beta c_1^2+K\alpha-\Omega)k^2+Kk(2\beta c_2^2+K\alpha-\Omega)-4K\beta,\\
  D_{2,2}&=(-4\beta c_2^2+K\alpha-\Omega)k^2+Kk(2\beta c_1^2+K\alpha-\Omega)-4K\beta,\\
  D_{3,3}&=(-4\beta c_1^2-K\alpha+\Omega)k^2+Kk(-2\beta c_2^2+K\alpha-\Omega)+4K\beta,\\
  D_{4,4}&=(-4\beta c_2^2-K\alpha+\Omega)k^2+Kk(-2\beta c_1^2+K\alpha-\Omega)+4K\beta,\\
  D_{1,3}&=D_{3,1}=-4k^2c_1^2\beta,~~D_{2,4}=D_{4,2}=-4k^2c_2^2\beta,\\
  D_{1,2}&=D_{2,1}=-2\beta c_1c_2k(K+2k),\\
  D_{3,4}&=D_{4,3}=2\beta c_1c_2k(K-2k),\\
  D_{1,4}&=D_{4,1}=D_{2,3}=D_{3,2}=-4c_1c_2\beta k^2.\\
\end{split}
\end{equation*}

Based on the existence condition for solutions of linear homogeneous equations, and $m_i$, $n_i(i=1,2)$ are all nonzero real parameters, then the determinant of coefficient matrix $D$ for $m_i$ and $n_i$ is equal to 0, namely
\begin{equation}
    det(D)=\begin{vmatrix}
     D_{1,1}& D_{1,2} & D_{1,3} & D_{1,4} \\
    D_{2,1}& D_{2,2} & D_{2,3} & D_{2,4}  \\
   D_{3,1}& D_{3,2} & D_{3,3} & D_{3,4}  \\
   D_{4,1}& D_{4,2} & D_{4,3} & D_{4,4}
    \end{vmatrix}=0,
\end{equation}
here $D=(d_{ij})_{4\times4}$ with $d_{ij}$ being the polynomials about $c_i, \alpha, \beta, \gamma, k, K, \Omega$, which gives rise to a dispersion relation equation. By solving this equation, MI gain can be obtained
\begin{equation}\label{gcfl-mi-3}
\begin{split}
  G &= |Im(\Omega)|=\Big|Im\Big(\frac{K|\beta|\sqrt{(C^2k^4-2Ck^3+K^2)}}{(K^2-k^2)k}\Big)\Big|,\\
  C &= c_1^2+c_2^2.\\
\end{split}
\end{equation}

Below we analyze the above expressions of Eq. (\ref{gcfl-mi-3}) and discuss the MI distribution characteristics of the gc-FL equation (\ref{gcfl-eq}). Obviously, MI region occurs if and only if the perturbed frequency $K$ satisfies the following inequality
\begin{equation*}
    |K|<|k|\sqrt{2Ck-C^2k^2}.
\end{equation*}

\begin{figure*}[!htbp]
\centering
\subfigure[]{\includegraphics[height=2.5in,width=2.5in]{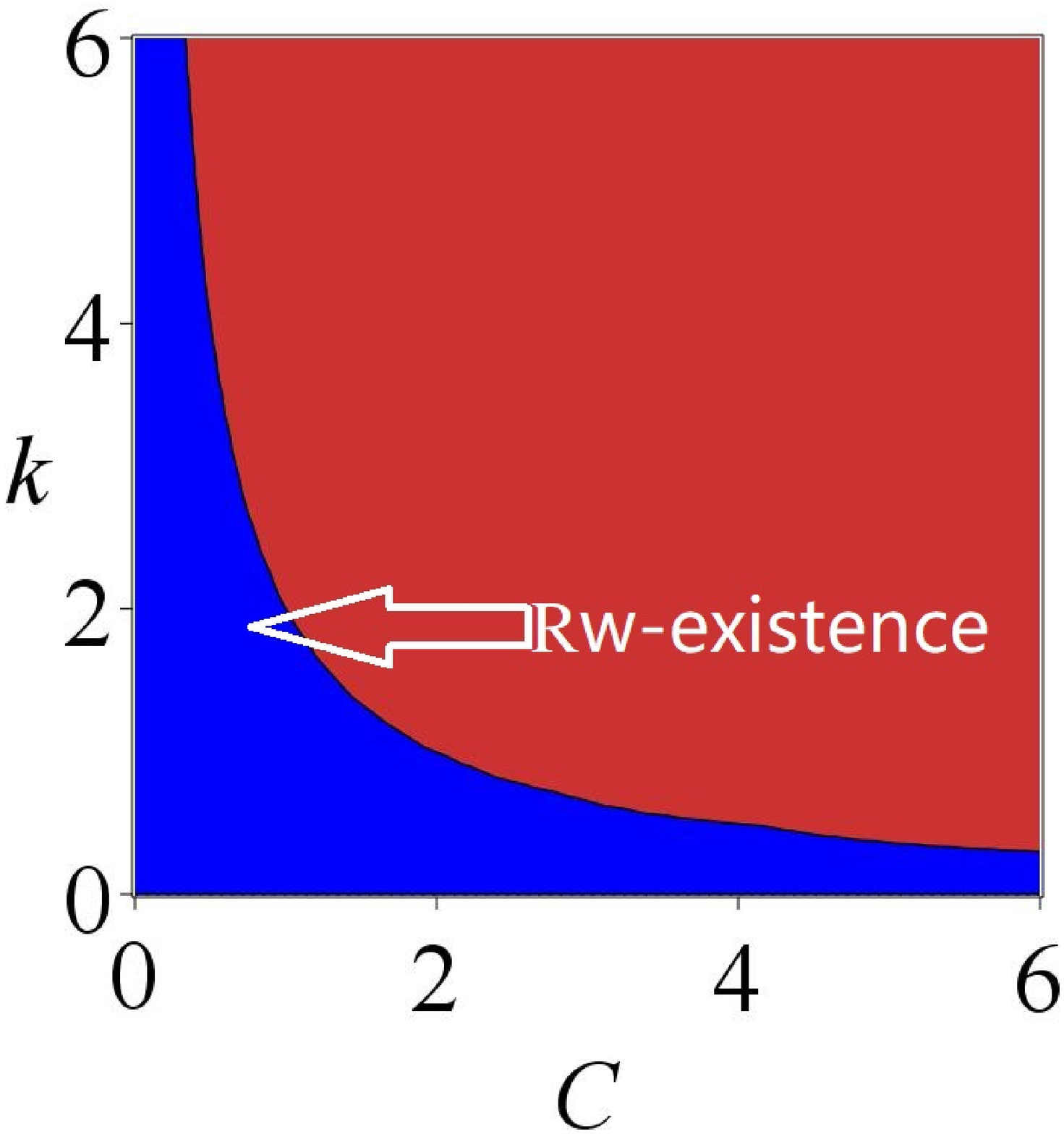}}\hspace{0.5cm}
\subfigure[]{\includegraphics[height=2.5in,width=2.5in]{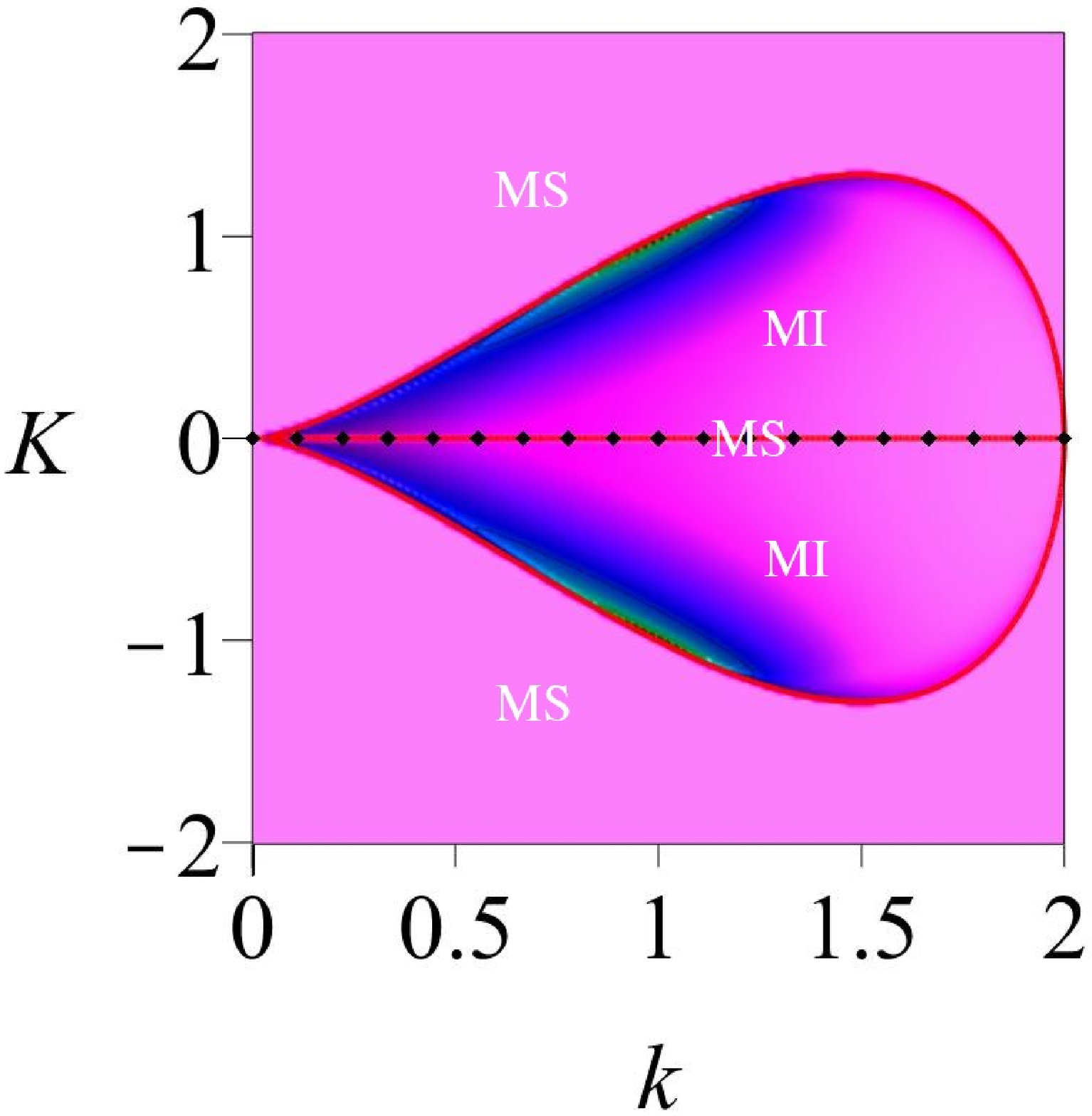}}
\caption{Left panel: condition for the existence of rogue wave at the plane $(k,C)$; Right panel: modulation instability  distribution of gain $G$ on the background frequency $k$ and perturbation frequency $K$ plane, choosing free parameters $C=1$ and $\beta=1$.}
\label{Fig-gcfl-1}
\end{figure*}

Moreover, based on the MI theory, when perturbed frequency $K=0$, we can obtain inequations
\begin{equation*}
\begin{cases}
    2Ck-C^2k^2>0,\\
    C\neq0,k\neq0,
\end{cases}
\end{equation*}
namely
\begin{equation*}
    0<k<\frac{2}{C},
\end{equation*}
this parameter condition can give rise to rogue wave solution, which is described as the left panel of Fig. \ref{Fig-gcfl-1}. Obviously, the gain $G$ is only related to four parameters $\beta,C,K$ and $k$, and has nothing to do with parameters $\alpha$ and $\gamma$.
When setting $\beta=0$, Eq. (\ref{gcfl-mi-3}) is reduced to $G=0$, which is modulation stability. Since $\beta\neq0$ is mentioned in the previous section, then setting $C=1$ and $\beta=1$, as shown in Fig. \ref{Fig-gcfl-1}, all the areas inside the red line are MI region except $K=0$. The red dashed line in MI region is a stability region, namely, MI gain rate $G=0$. In addition, the red line is also stability. With the increase of amplitude $C$, the area of MI region also decreases gradually. According to the gain expression, it can be seen that it is an even function with respect to $K$, so its MI distribution pattern is symmetrical with respect to $K=0$ as right panel in Fig.\ref{Fig-gcfl-1}.

\section{Infinitely many conservation laws and Darboux transformation for the gc-FL equation}

In this section, our main aim is to construct infinitely many conservation laws and $N$-fold Darboux transformation of the gc-FL equation \eqref{gcfl-eq} with the following $3\times3$ matrix spectral problem:
\begin{equation}\label{gcfl-dt-1}
\begin{split}
\Psi_x &= U(\lambda;x,t)\Psi,\\
\Psi_t &= V(\lambda;x,t)\Psi,\\
\end{split}
\end{equation}
where
\begin{equation}\label{gcfl-dt-1.1}
\begin{split}
U(\lambda;x,t)&=i\lambda^{-2}J+\lambda^{-1}Q_x,  \\
V(\lambda;x,t)&=\alpha i\lambda^{-2}J+\alpha\lambda^{-1}Q_x+2\beta iJQ^2-\gamma iJ-2i\beta\lambda JQ+i\beta\lambda^{2}J, \\
\end{split}
\end{equation}
and
\begin{equation*}
  J=
  \begin{pmatrix}
    -1 & 0 &0 \\
    0 & 1 & 0 \\
    0 & 0 & 1
    \end{pmatrix},~~~~
    Q=\begin{pmatrix}
    0 & u & v \\
    u^\ast & 0 & 0 \\
    v^\ast & 0 & 0
    \end{pmatrix}.\\
\end{equation*}
Here, $\Psi=(\phi(x,t),\psi(x,t),\chi(x,t))^T$ is the vector eigenfunction, $u$ and $v$ are potential function, $i$ is the imaginary unit, $\lambda$ is the spectral parameter, $\alpha$, $\beta$, $\gamma$ are real parameters. The upper corner symbols $T$ and $\ast$ denote transposition and conjugation, respectively. According to compatibility condition, Eq. \eqref{gcfl-eq} can be easily deduced from the zero-curvature equation: $U_t-V_x+[U,V]=0$.

\subsection{Infinitely many conservation laws}

For the gc-FL equation \eqref{gcfl-eq}, $u=u(t,x)$ and $v=v(t,x)$ are the solutions. If there exists a pair of continuously differentiable functions $\omega(t,x,u,v)$ and $J(t,x,u,v)$, such that
\begin{equation}\label{w-J}
\partial_t\omega(t,x,u,v)=\partial_xJ(t,x,u,v),
\end{equation}
the relationship is said to be a conservation law of Eq. \eqref{gcfl-eq}, $\omega(t,x,u,v)$ is called conserved density and $J(t,x,u,v)$ is called conserved flux. If the conservaed density and flux approach zero sufficiently fast as $|x|$ goes to infinity, then by integrating the conservation law \eqref{w-J} with respect to the independent variable $x$ in the whole real domain, we can deduce that
\begin{equation}\label{w-J-dt}
I[u,v]=\frac{d}{dt}\int_{-\infty}^{\infty}\omega(t,x,u,v)dx=0,
\end{equation}
which is a constant of motion.

So far, many classical equations, such as KdV equation \cite{j-miurarm-jmp-1968}, mKdV equation \cite{j-konnok-ptp-1974} and Sine-Gordon equation \cite{j-scottac-pi-1973}, have been proved to possess infinitely many conservation laws.
In 1975, Wadati, Sanuki and Konno \cite{j-wadatim-ptp-1975} proved that the AKNS equation also has infinitely many conservation laws from the linear problems corresponding to this family of equations. In 2018, Ling, Feng and Zhu \cite{j-linglm-narwa-2018} established infinitely many conservation laws for the vector positive and negative orders Kaup-Newell hierarchy, and pointed out that the first nontrivial negative flow was corresponding to the coupled Fokas-Lenells equation.

Based on the Lax pair \eqref{gcfl-dt-1}, now we will investigate the infinitely many conservation laws of the gc-FL equation \eqref{gcfl-eq}. First, we introduce two quantities $\Gamma_1=\Gamma_1(t,x)=\psi/\phi$ and $\Gamma_2=\Gamma_2(t,x)=\chi/\phi$. Then substituting them into the spectral problems \eqref{gcfl-dt-1}, two Riccati-type equations can be generated
\begin{equation}\label{gcfl-Riccati-type}
\begin{split}
\Gamma_{1x}&=2i\lambda^{-2}\Gamma_1+\lambda^{-1}(u_x^*-u_x\Gamma_1^2-v_x\Gamma_1\Gamma_2),\\
\Gamma_{2x}&=2i\lambda^{-2}\Gamma_2+\lambda^{-1}(v_x^*-v_x\Gamma_2^2-u_x\Gamma_1\Gamma_2).
\end{split}
\end{equation}

Substituting the following suitable expansions
\begin{equation}\label{gamma-1-2-expansion}
\begin{split}
u_x\Gamma_1&=\sum_{n=1}^{\infty}(\frac{i}{2}\lambda)^{2n-1}\Gamma_1^{(n)},\\
v_x\Gamma_2&=\sum_{n=1}^{\infty}(\frac{i}{2}\lambda)^{2n-1}\Gamma_2^{(n)},
\end{split}
\end{equation}
into the Riccati-type equations \eqref{gcfl-Riccati-type}, we can obtain the recursion formulas for $\Gamma_1^{(n)}$ and $\Gamma_2^{(n)}$, $(n=1,2,3,\cdots)$, which means all the coefficients of the orders $O(\lambda^{-1}),O(\lambda^{1}),O(\lambda^{3}),\cdots,O(\lambda^{2k-1}), (k=0,1,2,\cdots)$ will be computed, respectively. It then follows
\begin{eqnarray}
&&\Gamma_1^{(1)}=u_xu_x^*,\label{gamma-1--1}\\
&&\Gamma_1^{(2)}=-\Gamma_1^{(1)}\Gamma_2^{(1)}-\Gamma_1^{(1)2}
+2i(\Gamma_{1x}^{(1)}-\frac{u_{xx}}{u_x}\Gamma_1^{(1)})=-u_xu_x^*v_xv_x^*-(u_xu_x^*)^2+2iu_xu_{xx}^*,\label{gamma-1-1}\\
&&\Gamma_1^{(3)}=-\sum_{j=1}^{2}\Gamma_1^{(j)}(\Gamma_1^{(3-j)}+\Gamma_2^{(3-j)})
+2i(\Gamma_{1x}^{(2)}-\frac{u_{xx}}{u_x}\Gamma_1^{(2)})\nonumber\\
&&~~~~~~=2u_xu_x^*(u_xu_x^*+v_xv_x^*)^2-4u_xu_{xxx}^*
-2iu_x[2u_{xx}^*(2u_xu_x^*+v_xv_x^*)+u_x^*(u_{xx}u_x^*+2v_xv_{xx}^*+v_{xx}v_x^*)],\label{gamma-1-3}\\
&&\vdots \nonumber \\
&&\Gamma_1^{(k+1)}=-\sum_{j=1}^{k}\Gamma_1^{(j)}(\Gamma_1^{(k+1-j)}+\Gamma_2^{(k+1-j)})
+2i(\Gamma_{1x}^{(k)}-\frac{u_{xx}}{u_x}\Gamma_1^{(k)})~ (k=3,4,5,\cdots),\label{gamma-1-k+1}
\end{eqnarray}
and
\begin{eqnarray}
&&\Gamma_2^{(1)}=v_xv_x^*,\label{gamma-2--1}\\
&&\Gamma_2^{(2)}=-\Gamma_1^{(1)}\Gamma_2^{(1)}-\Gamma_2^{(1)2}
+2i(\Gamma_{2x}^{(1)}-\frac{v_{xx}}{v_x}\Gamma_2^{(1)})
=-u_xu_x^*v_xv_x^*-(v_xv_x^*)^2+2iv_xv_{xx}^*,\label{gamma-2-1}\\
&&\Gamma_2^{(3)}=-\sum_{j=1}^{2}\Gamma_2^{(j)}(\Gamma_1^{(3-j)}+\Gamma_2^{(3-j)})
+2i(\Gamma_{2x}^{(2)}-\frac{u_{xx}}{u_x}\Gamma_2^{(2)})\nonumber\\
&&~~~~~~=2v_xv_x^*(u_xu_x^*+v_xv_x^*)^2-4v_xv_{xxx}^*
-2iv_x[2v_{xx}^*(u_xu_x^*+2v_xv_x^*)+v_x^*(2u_xu_{xx}^*+u_{xx}u_x^*+v_{xx}v_x^*)],\label{gamma-2-3}\\
&&\vdots \nonumber \\
&&\Gamma_2^{(k+1)}=-\sum_{j=1}^{k}\Gamma_2^{(j)}(\Gamma_1^{(k+1-j)}+\Gamma_2^{(k+1-j)})
+2i(\Gamma_{2x}^{(k)}-\frac{v_{xx}}{v_x}\Gamma_2^{(k)})~ (k=3,4,5,\cdots).\label{gamma-2-k+1}
\end{eqnarray}

From the first equation of the $3\times3$ matrix spectral problem \eqref{gcfl-dt-1}, we have
\begin{equation}\label{ln-phi-x-t}
\begin{split}
(\ln\phi)_x&=-i\lambda^{-2}+\lambda^{-1}(u_x\Gamma_1+v_x\Gamma_2),\\
(\ln\phi)_t&=-i\alpha\lambda^{-2}+i\gamma-i\beta\lambda^2-2i\beta(uu^*+vv^*)
+\alpha\lambda^{-1}(u_x\Gamma_1+v_x\Gamma_2)+2i\beta\lambda(u\Gamma_1+v\Gamma_2).
\end{split}
\end{equation}
Through the compatibility condition of \eqref{ln-phi-x-t}, it yields
\begin{equation}\label{w-J-compatibility-condition}
\lambda^{-1}(u_x\Gamma_1+v_x\Gamma_2)_t=-2i\beta(uu^*+vv^*)_x
+\alpha\lambda^{-1}(u_x\Gamma_1+v_x\Gamma_2)_x+2i\beta\lambda(u\Gamma_1+v\Gamma_2)_x.
\end{equation}
Let $\omega(t,x,u,v)=\sum_{k=1}^{\infty}\lambda^{2k-1}\omega_k$ and $J(t,x,u,v)=\sum_{k=1}^{\infty}\lambda^{2k-1}J_k$. With the relations \eqref{gamma-1--1}-\eqref{gamma-2-k+1}, it is natural to obtain the infinitely many conservation laws of the gc-FL equation \eqref{gcfl-eq} as follows
\begin{equation}\label{w-J-consevative-law}
\frac{\partial \omega_k}{\partial t}=\frac{\partial J_k}{\partial x} ~(k=1,2,3,\cdots),
\end{equation}
here $\omega_k$ and $J_k$ represent conserved density and conserved flux, respectively. So the first two conserved densities and conserved fluxes are
\begin{equation}\label{w-J-1}
\begin{split}
\omega_1&=u_xu_x^*+v_xv_x^*,\\
J_1&=-4\beta(uu^*+vv^*)+\alpha(u_xu_x^*+v_xv_x^*),
\end{split}
\end{equation}
and
\begin{equation}\label{w-J-2}
\begin{split}
\omega_2&=-(u_xu_x^*+v_xv_x^*)^2+2i(u_xu_{xx}^*+v_xv_{xx}^*),\\
J_2&=-\alpha(u_xu_x^*+v_xv_x^*)^2+2i\alpha(u_xu_{xx}^*+v_xv_{xx}^*)-8i\beta(uu_x^*+vv_x^*).
\end{split}
\end{equation}
Finally, we can obtain the recursion relations for the conserved density $\omega_k$ and conserved flux $J_k$ with $k\geq3$,
\begin{equation}\label{w-J-k}
\begin{split}
\omega_k&=\Gamma_1^{(k)}+\Gamma_2^{(k)},\\
J_k&=\alpha(\Gamma_1^{(k)}+\Gamma_2^{(k)})-8i\beta(\Gamma_1^{(k-1)}+\Gamma_2^{(k-1)}),
\end{split}
\end{equation}
here $\Gamma_{1,2}^{(k)}$ and $\Gamma_{1,2}^{(k-1)}$ are provided in the relations \eqref{gamma-1--1}-\eqref{gamma-2-k+1}.

\subsection{Darboux transformation for the gc-FL equation}

Here, we derive the $N$-fold Darboux transformation of the gc-FL equation \eqref{gcfl-eq}.
Motivated by the results mentioned in \cite{j-linglm-narwa-2018}, we can get the following symmetric relations
\begin{itemize}
  \item $U(\lambda)=JU(-\lambda)J,~~V(\lambda)=JV(-\lambda)J$,~~$U(\lambda)=-JU(\lambda^\ast)^\dagger J,~~V(\lambda)=-JV(\lambda^\ast)^\dagger J$,
  \item $\Psi(-\lambda;x,t)=J\Psi(\lambda;x,t)J$,~~$J[\Psi(\lambda;x,t)]^\dagger J=[\Psi(\lambda^\ast;x,t)]^{-1}$,
  \item $T(\lambda)=JT(-\lambda)J,~~[T(\lambda)]^{-1}=J[T(\lambda^\ast)]^\dagger J$,
\end{itemize}
here, the upper corner symbol $\dagger$ indicates the Hermitian adjoint. The above relationship equations can be deduced by direct calculation. Due to above symmetric relations between \{$U(\lambda),V(\lambda)$\} and \{$U(\lambda^*),V(\lambda^*)$\}, it  is easy to find that if $\Psi_i$ is a nonzero solution of the spectral problems \eqref{gcfl-dt-1} at spectral parameter $\lambda=\lambda_i$, then $\Psi_i^\dagger J$ is a nonzero solution of the conjugate form of system \eqref{gcfl-dt-1}
\begin{equation*}
    \Phi_x=-\Phi U, ~~\Phi_t=-\Phi V
\end{equation*}
at spectral parameter $\lambda=\lambda_i^*$, and here $U,V$ have the same forms as in system \eqref{gcfl-dt-1}.
Through Darboux matrix $T[1]$ (under above symmetric relations), we obtain a new system in the same form as Eq. \eqref{gcfl-dt-1}, which can be written as follows
\begin{equation}\label{gcfl-dt-2}
\begin{split}
\Psi[1]_x &= U[1](\lambda;x,t)\Psi[1],~~~U[1](\lambda;x,t)=i\lambda^{-2}J+\lambda^{-1}Q[1]_x,  \\
\Psi[1]_t &= V[1](\lambda;x,t)\Psi[1],~~~V[1](\lambda;x,t)=\alpha i\lambda^{-2}J+\alpha\lambda^{-1}Q[1]_x+2\beta iJQ[1]^2-\gamma iJ-2i\beta\lambda JQ[1]+i\beta\lambda^{2}J, \\
\end{split}
\end{equation}
where
\begin{equation}\label{gcfl-dt-3}
\begin{split}
U[1]&=T[1]_xT[1]^{-1}+T[1]UT[1]^{-1},  \\
V[1]&=T[1]_tT[1]^{-1}+T[1]VT[1]^{-1}. \\
\end{split}
\end{equation}

Based on the above Darboux matrix symmetric relations and the loop method, $T[1]$, $T[1]^{-1}$ can be constructed as
\begin{equation}\label{gcfl-dt-4}
\begin{split}
T[1]&=E+\frac{A_1}{\lambda-\lambda_1^\ast}-\frac{JA_1J}{\lambda+\lambda_1^\ast},  \\
T[1]^{-1}&=E+\frac{JA_1^\dagger J}{\lambda-\lambda_1}-\frac{A_1^\dagger}{\lambda+\lambda_1}, \\
\end{split}
\end{equation}
with
\begin{equation}\label{gcfl-dt-5}
\begin{split}
A_1=&|x_1><y_1|J,~~|x_1>=K_1|y_1>,\\
K_1=&diag(-\Theta_1^{\ast-1},\Theta_1^{-1},\Theta_1^{-1}),\\
\Theta_1=&\frac{2[\lambda_1^\ast(|\psi_1|^2+|\chi_1|^2)-\lambda_1|\phi_1|^2]}{\lambda_1^{\ast2}-\lambda_1^2},\\
\end{split}
\end{equation}
where $E$ is the identify matrix, $|y_1>=\Psi_1=(\phi_1,\psi_1,\chi_1)^T$ is a special solution with $u=u[0],~v=v[0]$, which satisfies system \eqref{gcfl-dt-1} at spectral parameter $\lambda=\lambda_1$, and $<y_1|=(|y_1>)^\dagger$. Substituting Eq. \eqref{gcfl-dt-4} into Eq. \eqref{gcfl-dt-3} and collecting all the coefficients, the relationship between $Q_1$ and $Q$ can be deduced as
\begin{equation}\label{gcfl-dt-6}
Q[1]=Q+A_1-JA_1J.
\end{equation}
Furthermore, the concrete expressions of the relationship between the old and the new solutions for the gc-FL equation \eqref{gcfl-eq} can be rewritten as
\begin{equation}\label{gcfl-dt-7}
\begin{split}
u[1]=&u[0]-\frac{2\phi_1\psi_1^\ast}{\Theta_1^\ast},\\
v[1]=&v[0]-\frac{2\phi_1\chi_1^\ast}{\Theta_1^\ast}.\\
\end{split}
\end{equation}
\begin{proposition}\label{prop1}
The $N$-fold DT for the gc-FL equation \eqref{gcfl-eq} has the following form
    \begin{equation}\label{pro-1}
    T_N=T[N]T[N-1]\cdots T[1]=E+\sum_{i=1}^N (\frac{C_i}{\lambda-\lambda_i^\ast}-\frac{JC_iJ}{\lambda+\lambda_i^\ast}),
    \end{equation}
and
    \begin{equation}\label{pro-2}
    T_N^{-1}=T[1]^{-1}T[2]^{-1}\cdots T[N]^{-1}=E+\sum_{i=1}^N (\frac{JD_i^\dagger J}{\lambda-\lambda_i}-\frac{D_i^\dagger}{\lambda+\lambda_i}),
    \end{equation}
where
\begin{equation*}
\begin{split}
C_i=&{\rm Res}|_{\lambda=\lambda_i^*}(T_N),\\
D_i=&{\rm Res}|_{\lambda=\lambda_i}(T_N^{-1}),\\
A_i=&|x_i><y_i|J,~~|x_i>=K_i|y_i>,\\
K_i=&diag(-\Theta_i^{\ast-1},\Theta_i^{-1},\Theta_i^{-1}),~~|y_i>=\Psi_i=(\phi_i,\psi_i,\chi_i)^T,\\
\Theta_i=&\frac{2[\lambda_i^\ast(|\psi_i|^2+|\chi_i|^2)-\lambda_i|\phi_i|^2]}{\lambda_i^{\ast2}-\lambda_i^2},~~(i=1,2,\ldots,N).\\
\end{split}
\end{equation*}
\end{proposition}
\proof{}
Taking the residue on both sides of Eq. \eqref{pro-1}, it yields
    \begin{equation}\label{proof-pro-1}
    \begin{split}
    {\rm Res}|_{\lambda=\lambda_i^*}(T_N)&=(E+\frac{A_N}{\lambda_i^*-\lambda_N^*}-\frac{JA_NJ}{\lambda_i^*+\lambda_N^*})\cdots A_i\cdots (E+\frac{A_1}{\lambda_i^*-\lambda_N^*}-\frac{JA_1J}{\lambda_i^*+\lambda_N^*})\\
    &=C_i,\\
    {\rm Res}|_{\lambda=-\lambda_i^*}(T_N)&=-(E+\frac{A_N}{-\lambda_i^*-\lambda_N^*}-\frac{JA_NJ}{-\lambda_i^*+\lambda_N^*})\cdots JA_iJ\cdots (E+\frac{A_1}{-\lambda_i^*-\lambda_N^*}-\frac{JA_1J}{-\lambda_i^*+\lambda_N^*}).\\
    \end{split}
    \end{equation}
Because of ${\rm Res}|_{\lambda=\lambda_i^*}(T_N)=-J{\rm Res}|_{\lambda=-\lambda_i^*}(T_N)J$, Eq. \eqref{pro-1} is valid. Similarly, Eq. \eqref{pro-2} can also be proved.

This completes the proof of Proposition \ref{prop1}.

According to the Darboux matrix $T[N]$, we can get the transformations between seed solutions $u[0],v[0]$ and new potential functions $u[N],v[N]$, which is described as the following Theorem \ref{the1}.

\begin{theorem}\label{the1}
 The $N$-fold Darboux transformation of the spectral problems \eqref{gcfl-dt-1} gives rise to the following compact recurrence formulas
    \begin{equation}\label{n-dt}
        \begin{split}
        u[N]=&u[0]-\frac{2
            }{|P|}\left|
                                                       \begin{array}{cc}
                                                         P & \eta_2^\dagger \\
                                                         \eta_1 & 0 \\
                                                       \end{array}
                                                     \right|,\\
        v[N]=&v[0]-\frac{2
            }{|P|}\left|
                                                       \begin{array}{cc}
                                                         P & \eta_3^\dagger \\
                                                         \eta_1 & 0 \\
                                                       \end{array}
                                                     \right|,\\
        \end{split}
    \end{equation}
where
    \begin{equation*}
        \begin{split}
        &\eta_1=(\phi_1,\phi2,...,\phi_N),\\
        &\eta_2=(\psi_1,\psi_2,...,\psi_N),\\
        &\eta_3=(\chi_1,\chi_2,...,\chi_N),\\
        &P=(p_{ij})_{N\times N},~~p_{ij}=\frac{\Psi_i^\dagger J\Psi_j}{\lambda_i^\ast-\lambda_j}+\frac{\Psi_i^\dagger\Psi_j}{\lambda_i^\ast+\lambda_j},\\
        &\Psi_i=(\phi_i^\ast,\psi_i^\ast,\chi_i^\ast),~~\Psi_j=(\phi_j^\ast,\psi_j^\ast,\chi_j^\ast)~(1\leq i,j\leq N).\\
        \end{split}
    \end{equation*}
\end{theorem}

\proof{}
According to the derivation process of one-fold Darboux transformation, there has the following relation after $N$-fold Darboux transformation, that is
\begin{equation}
T_{N,x}+T_{N}U=U[N]T_{N}.
\end{equation}
Substituting Eq. \eqref{pro-1} into above equation, which results into the following potential function relation
\begin{equation}
Q[N]=Q+\sum_{i=1}^N(C_i-JC_iJ).
\end{equation}
Then according to the relative properties of matrix rank, it is not difficult to deduce Rank($C_i$)=1. so it can be assumed that $C_i=|x_i><y_i|$ and $|y_i>=\Psi_i=(\phi_i,\psi_i,\chi_i)^T$. Since $T_NT_N^{-1}=E$ and $\Psi_i^\dagger J$ is a nonzero solution of the conjugate form of system \eqref{gcfl-dt-1} at $\lambda=\lambda_i^*$, thus
\begin{equation*}
    <y_i|T_N^{-1}\mid_{\lambda=\lambda_i^*}=0,~~\Psi_i^\dagger JT_N^{-1}\mid_{\lambda=\lambda_i^*}=0.
\end{equation*}
Comparing the above two expressions, without loss of generality, we can choose $<y_i|=\Psi_i^\dagger J$. Furthermore, according to  $T_N\mid_{\lambda=\lambda_j}\Psi_j=0$ and Eq.\eqref{pro-1}, it holds
\begin{equation}\label{n-dt-eq}
    \Psi_j+\sum_{i=1}^N\big(\frac{|x_i>\Psi_i^\dagger J\Psi_j}{\lambda_j-\lambda_i^*}-\frac{J|x_i>\Psi_i^\dagger\Psi_j}{\lambda_j+\lambda_i^*}\big)=0,~~(i=1,2,\ldots N).
\end{equation}
Solving Eq. \eqref{n-dt-eq}, it yields
\begin{equation}\label{gcfl-dt-9}
\begin{split}
{\rm First~~row:}~~(|x_{1,1}>,|x_{2,1}>,...,|x_{N,1}>)&=(|y_{1,1}>,|y_{2,1}>,...,|y_{N,1}>)P^{-1}\\
&=(\phi_1,\phi_2,...,\phi_N)P^{-1},\\
{\rm Other~~rows:}~~(|x_{1,i}>,|x_{2,i}>,...,|x_{N,i}>)&=(|y_{1,i}>,|y_{2,i}>,...,|y_{N,i}>)S^{-1}\\
&=(\phi_1,\phi_2,...,\phi_N)S^{-1}~~(i=2,3),\\
\end{split}
\end{equation}
with
\begin{equation}\label{gcfl-dt-10}
\begin{split}
&P=(p_{ij})_{N\times N},~~p_{ij}=\frac{\Psi_i^\dagger J\Psi_j}{\lambda_i^\ast-\lambda_j}+\frac{\Psi_i^\dagger\Psi_j}{\lambda_i^\ast+\lambda_j},\\
&S=(s_{ij})_{N\times N},~~s_{ij}=\frac{\Psi_i^\dagger J\Psi_j}{\lambda_i^\ast-\lambda_j}-\frac{\Psi_i^\dagger\Psi_j}{\lambda_i^\ast+\lambda_j}~(1\leq i,j\leq N).\\
\end{split}
\end{equation}
Finally, substituting Eqs. \eqref{gcfl-dt-9}-\eqref{gcfl-dt-10}, one gets \{$u[N],~~v[N]$\} and \{$u[0],~~v[0]$\} has the following relations
\begin{equation}\label{gcfl-dt-12}
\begin{split}
u[N]=&u[0]+(\sum_{i=1}^NC_i-JC_iJ)_{12}\\
    =&u[0]+2\sum_{i=1}^{N}(|x_{i,1}>\psi_i^\ast)\\
    =&u[0]-\frac{2
    }{|P|}\left|
                                               \begin{array}{cc}
                                                 P & \eta_2^\dagger \\
                                                 \eta_1 & 0 \\
                                               \end{array}
                                             \right|,\\
v[N]=&v[0]+(\sum_{i=1}^NC_i-JC_iJ)_{13}\\
    =&v[0]+2\sum_{i=1}^{N}(|x_{i,1}>\chi_i^\ast)\\
    =&v[0]-\frac{2
    }{|P|}\left|
                                               \begin{array}{cc}
                                                 P & \eta_3^\dagger \\
                                                 \eta_1 & 0 \\
                                               \end{array}
                                             \right|,\\
\end{split}
\end{equation}
with
\begin{equation*}
  \eta_1=(\phi_1,\phi_2,\ldots,\phi_N),~~\eta_2=(\psi_1,\psi_2,\ldots,\psi_N),~~\eta_3=(\chi_1,\chi_2,\ldots,\chi_N).
\end{equation*}

This completes the proof of Theorem \ref{the1}.

\section{Interaction solutions between localized waves for the gc-FL equation}

In this section, we turn our attention to the construction of higher-order localized waves and their interaction solutions for the gc-FL equation \eqref{gcfl-eq}.
On account of a seed solution under the same spectral parameter $\lambda$ can not realize the iterative process of classical DT, it is necessary to introduce a limit process to avoid this deficiency and obtain higher-order localized wave solutions of Eq. \eqref{gcfl-eq}. To this end, assuming the seed solutions as follows
\begin{equation}\label{gcfl-lws-1}
\begin{split}
u[0]=c_1\exp(i\theta),~~v[0]=c_2\exp(i\theta),~~\theta=ax+bt,\\
b=-4\beta(c_1^2+c_2^2)+(a\alpha+2\gamma)+\frac{4\beta}{a},~~(a, c_1, c_2\in\mathbb{R}).\\
\end{split}
\end{equation}
By substituting the seed solutions \eqref{gcfl-lws-1} into the Lax pair \eqref{gcfl-dt-1} of the gc-FL equation \eqref{gcfl-dt-1.1}, there appears a variable coefficient differential system. Then we can convert this variable coefficient differential equations to constant coefficient differential equations through a gauge transformation $L$=diag($e^{-\frac{2i}{3}\theta},e^{\frac{i}{3}\theta},e^{\frac{i}{3}\theta}$). Finally, the corresponding fundamental vector solution of the gc-FL equation \eqref{gcfl-eq}  is derived, that is
\begin{equation}\label{gcfl-lws-2}
  \Psi=
  \begin{pmatrix}
    \phi \\
    \psi \\
    \chi \\
    \end{pmatrix}
    =
  \begin{pmatrix}
    (C_1e^{A}-C_2e^{-A})e^{\frac{i}{2}\theta} \\
    \rho_1(C_2e^{A}-C_1e^{-A})e^{-\frac{i}{2}\theta}-dc_2e^B\\
    \rho_2(C_2e^{A}-C_1e^{-A})e^{-\frac{i}{2}\theta}+dc_1e^B\\
    \end{pmatrix},
\end{equation}
with
\begin{equation*}
\begin{split}
  C_1 &= \frac{\Big(a\lambda^2+2-\sqrt{(a\lambda^2+2)^2-4a^2\lambda^2(c_1^2+c_2^2)}\Big)^{\frac{1}{2}}}{\sqrt{(a\lambda^2+2)^2-4a^2\lambda^2(c_1^2+c_2^2)}},~~
  C_2 =\frac{\Big(a\lambda^2+2+\sqrt{(a\lambda^2+2)^2-4a^2\lambda^2(c_1^2+c_2^2)}\Big)^{\frac{1}{2}}}{\sqrt{(a\lambda^2+2)^2-4a^2\lambda^2(c_1^2+c_2^2)}},\\
  A&=\frac{i\sqrt{(a\lambda^2+2)^2-4a^2\lambda^2(c_1^2+c_2^2)}}{2a\lambda^2}\Big[ax+(a\alpha+2\beta\lambda^2)t+\sum_{i=1}^{N}\kappa_i\xi^{2i}\Big],~~
  B=\frac{i(x+(\alpha-\gamma\lambda^2+\beta\lambda^4)t)}{\lambda^2},\\
    \rho_1&=\frac{c_1}{\sqrt{c_1^2+c_2^2}},~~\rho_2=\frac{c_2}{\sqrt{c_1^2+c_2^2}},~~\kappa_i=m_i+in_i,~~(d,m_i,n_i\in\mathbb{R}),\\
\end{split}
\end{equation*}
where $\xi$ represents a small disturbance parameter. Next, we consider the double roots of the characteristic polynomial of the time partial matrix $U_0$, and then we can get a constraint condition satisfied by the spectral parameter as follows
\begin{equation*}
       (a\lambda^2+2)^2-4a^2C\lambda^2=0,
\end{equation*}
where
\begin{equation*}
  C=c_1^2+c_2^2.
\end{equation*}
By solving the above constraint condition, it follows spectral parameter $\lambda=\pm(\sqrt{C}\pm\sqrt{\frac{aC-2}{a}})$. It is not difficult to find that whether there is an imaginary part of the spectral parameter $\lambda$ depends on the frequency $a$ and the amplitude $C$, while the real part completely depends on the amplitude $C$. However, when the spectral parameter $\lambda$ contains the imaginary part, namely $0<a<\frac{2}{C}$, then there are rogue wave solutions of the gc-FL equation \eqref{gcfl-eq}, which is consistent with the constraint condition for the existence of rogue waves obtained from the MI analysis in the section 2 of this article. To obtain the rogue wave solutions, for convenience of calculation, we choose $a=1$ and $C=1$, it thus transpires that the corresponding spectral parameter $\lambda=\lambda_0=1+i$.
Applying limitation approach to construct the $N$-fold generalized Darboux transformation of the gc-FL equation, we can get the relations between the new solutions $u[N],~v[N]$ and the seed solutions $u[0],~v[0]$, whose detailed process is shown in the following Theorem \ref{the2}.
\begin{theorem}\label{the2}
    The $N$-fold generalized Darboux transformation of the spectral problems \eqref{gcfl-dt-1} gives rise to the following compact recurrence formulas
\begin{equation}\label{gcfl-lws-8}
\begin{split}
    u[N]=&u[0]-\frac{2
    }{|\hat{P}|}\left|
                                               \begin{array}{cc}
                                                 \hat{P} & \hat{\eta}_2^\dagger \\
                                                 \hat{\eta}_1 & 0 \\
                                               \end{array}
                                             \right|,\\
    v[N]=&v[0]-\frac{2
    }{|\hat{P}|}\left|
                                               \begin{array}{cc}
                                                 \hat{P} & \hat{\eta}_3^\dagger \\
                                                 \hat{\eta}_1 & 0 \\
                                               \end{array}
                                             \right|,\\
\end{split}
\end{equation}
where
\begin{equation}\label{gcfl-lws-9}
\begin{split}
   \hat{P}=&(\hat{p}_{kl})_{N\times N},~~1\leq k,l\leq N,~~\hat{\eta}_1=(\phi_1^{[0]},\phi_1^{[1]},\ldots,\phi_1^{[N-1]}),\\
   \hat{\eta}_2=&(\psi_1^{[0]},\psi_1^{[1]},\ldots,\psi_1^{[N-1]}),~~\hat{\eta}_3=(\chi_1^{[0]},\chi_1^{[1]},\ldots,\chi_1^{[N-1]}).
\end{split}
\end{equation}
\end{theorem}
\proof{}
Let the spectral parameter $\lambda=\lambda_1=1+i+\xi^2$ and put it into Eq. \eqref{gcfl-lws-1}. By expanding $\Psi_1$ at $\xi=0$, it arrives at
\begin{equation}\label{gcfl-lws-3}
  \Psi_1(\xi)=\Psi|_{\lambda=\lambda_1}=\sum_{k=0}^{N-1}\Psi_1^{[k]}\xi^{2k}+O(\xi^{2N}),
\end{equation}
with
\begin{equation}\label{gcfl-lws-4}
  \Psi_1^{[k]}=\left(
                 \begin{array}{c}
                   \phi_1^{[k]} \\
                   \psi_1^{[k]} \\
                   \chi_1^{[k]} \\
                 \end{array}
               \right)
               =\frac{1}{(2k)!}\frac{\partial^k\Psi_1}{\partial\xi^k}|_{\xi=0}.
\end{equation}
Additionally, when $\lambda_i=\lambda_j=\lambda_1$, through the same limit processing technique, $p_{ij}$ in Eq. \eqref{gcfl-dt-10} can be rewritten as
\begin{equation}\label{gcfl-lws-7}
\begin{split}
    p_{ij}=&-\frac{2[(1+\xi^2+i)(|\psi_1|^2+|\chi_1|^2)-(1+\xi^{\ast2}-i)|\phi_1|^2]}{(2+\xi^2+\xi^{\ast2})(2i+\xi^2-\xi^{\ast2})}\\
    =&\sum_{k,l=1}^{N}\hat{p}_{kl}\xi^{2(k-1)}\xi^{2(l-1)}+O(|\xi|^{4N}),\\
\end{split}
\end{equation}
with
\begin{equation*}
    \hat{p}_{kl}=\lim_{\xi,\xi^\ast\rightarrow0}\frac{1}{(2(k-1))!(2(l-1))!}\frac{\partial^{2(k+l-2)}p_{ij}}{\partial\xi^{2(k-1)}\xi^{\ast2(l-1)}}|_{(\lambda_i=\lambda_1,\lambda_j=\lambda_1)}.
\end{equation*}
So $N$-order localized wave solutions of the gc-FL equation \eqref{gcfl-eq}, namely, the formulas \eqref{gcfl-lws-8} can be deduced.

This completes the proof of Theorem \ref{the2}.

As $k$ increases, the expression of vector function $\Psi_1^{[k]}$ in Eq. \eqref{gcfl-lws-4} will be more complicated. Here we only give the explicit expressions for $k=0,1$, that is
\begin{equation}\label{gcfl-lws-5}
    \begin{split}
  \Psi_1^{[0]}=&\left(
                 \begin{array}{c}
                   \phi_1^{[0]} \\
                   \psi_1^{[0]} \\
                   \chi_1^{[0]} \\
                 \end{array}
               \right)
  =\left(
                 \begin{array}{c}
                   \frac{1}{4}[(8i\beta+2\alpha)t+2x+i-1]\sqrt{2+2i} e^{\frac{i}{2}\theta} \\
                   \frac{1}{4}c_1[(8i\beta+2\alpha)t+2x-i+1]\sqrt{2+2i}e^{-\frac{i}{2}\theta}-dc_2e^{\frac{1}{2}\theta_0} \\
                   \frac{1}{4}c_2[(8i\beta+2\alpha)t+2x-i+1]\sqrt{2+2i}e^{-\frac{i}{2}\theta}+dc_1e^{\frac{1}{2}\theta_0} \\
                 \end{array}
               \right),\\
  \Psi_1^{[1]}=&\left(
                 \begin{array}{c}
                   \phi_1^{[1]} \\
                   \psi_1^{[1]} \\
                   \chi_1^{[1]} \\
                 \end{array}
               \right)
  =\left(
                 \begin{array}{c}
                   \frac{1}{48}\Omega_1\sqrt{2+2i}e^{\frac{i}{2}\theta} \\
                   \frac{1}{48}c_1\Omega_2\sqrt{2+2i}e^{-\frac{i}{2}\theta}+\frac{1}{2}dc_2[\left( 1-i \right)  \left( \alpha+4\,\beta \right) t+ \left( 1-i \right) x]e^{\frac{1}{2}\theta_0} \\
                   \frac{1}{48}c_2\Omega_2\sqrt{2+2i}e^{-\frac{i}{2}\theta}-\frac{1}{2}dc_1[\left( 1-i \right)  \left( \alpha+4\,\beta \right) t+ \left( 1-i \right) x]e^{\frac{1}{2}\theta_0} \\
                 \end{array}
               \right),\\
    \end{split}
\end{equation}
with
\begin{equation}\label{gcfl-lws-6}
    \begin{split}
    \theta_0=&(-2i\gamma-4\beta+\alpha)t+x, \\
    \Omega_1=&(2i\alpha^3-24i\alpha^2\beta-96i\alpha\beta^2+128i\beta^3-2\alpha^3-24\beta\alpha^2+96\beta^2\alpha+128\beta^3)t^3\\
             &+(6i\alpha^2-48i\alpha\beta-96i\beta^2-6\alpha^2-48\beta\alpha+96\beta^2)t^2x+(-6i\alpha^2+96i\beta^2+48\beta\alpha)t^2\\
             &+(6i\alpha-24i\beta-6\alpha-24\beta)tx^2+(-12i\alpha+48\beta)tx+(21i\alpha+36i\beta-15\alpha+12\beta)t\\
             &+(-2+2i)x^3-6ix^2+(-15+21i)x+24in_1+24m_1+3-6i,\\
    \Omega_2=&(2i\alpha^3-24i\alpha^2\beta-96i\alpha\beta^2+128i\beta^3-2\alpha^3-24\beta\alpha^2+96\beta^2\alpha+128\beta^3)t^3\\
             &+(6i\alpha^2-48i\alpha\beta-96i\beta^2-6\alpha^2-48\beta\alpha+96\beta^2)t^2x-(-6i\alpha^2+96i\beta^2+48\beta\alpha)t^2\\
             &+(6i\alpha-24i\beta-6\alpha-24\beta)tx^2-(-12i\alpha+48\beta)tx+(21i\alpha+36i\beta-15\alpha+12\beta)t\\
             &+(-2+2i)x^3+6ix^2+(-15+21i)x+24in_1+24m_1-3+6i.\\
    \end{split}
\end{equation}


Therefore, utilizing the above formulas with different parameter $N=1,2,3$ in Theorem \ref{the2}, the first-order, second-order and third-order semirational solution of the gc-FL equation \eqref{gcfl-eq} can be obtained. As $N$ becomes larger, the expression of the solution becomes more complicated. For simplicity, here only the specific expression for the first-order  semirational solution is given
\begin{equation}\label{gcfl-lws-10}
\begin{split}
   u[1]=&c_1e^{i\theta}+\frac{2ic_1\Xi_1e^{i\theta}+2i\sqrt{1+i}dc_2\Xi_2e^{\frac{\theta_0^\ast+i\theta}{2}}}{\sqrt(2)(1+i)d^2e^{\frac{\theta_0+\theta_0^\ast}{2}}+\Xi_3},\\
   v[1]=&c_2e^{i\theta}+\frac{2ic_2\Xi_1e^{i\theta}-2i\sqrt{1+i}dc_1\Xi_2e^{\frac{\theta_0^\ast+i\theta}{2}}}{\sqrt{2}(1+i)d^2e^{\frac{\theta_0+\theta_0^\ast}{2}}+\Xi_3},\\
\end{split}
\end{equation}
with
\begin{equation*}
\begin{split}
   \theta=&x+(\alpha+2\gamma)t, ~~\theta_0=(\alpha-2i\gamma-4\beta)t+x,\\
   \Xi_1=&(2\alpha^2+32\beta^2)t^2+2x^2+4\alpha xt+2i(1+4\beta)t+2ix-1,\\
   \Xi_2=&(8i\beta+2\alpha)t+2x-1+i,\\
   \Xi_3=&2i(16\beta^2+\alpha^2)t^2+2ix^2+4i\alpha xt+2(\alpha-4\beta)t+2x+i.\\
\end{split}
\end{equation*}

Next, we will analyze the dynamical behavior characteristics of the above-mentioned first- to third-order solutions of the gc-FL equation \eqref{gcfl-eq}. It is not difficult to see that $u[1]$ and $v[1]$ in Eqs. \eqref{gcfl-lws-10} are controllable with parameters $\alpha$, $\beta$, $\gamma$, $d$ and $c_i~(i=1..2)$, the latter two parameters play a leading role. According to the selection of parameters $d$ and $c_i$, there are usually the following three structural solutions.


\textbf{Case 1. Rogue waves}

Choosing the parameters $d=0$ and $c_i\neq0~(i=1..2)$, then Eqs. \eqref{gcfl-lws-10} can be rewritten as
\begin{equation}\label{gcfl-lws-11}
\begin{split}
   u[1]=&c_1e^{i\theta}\Big(1-\frac{2i\Xi_1}{\Xi_3}\Big),\\
   v[1]=&c_2e^{i\theta}\Big(1-\frac{2i\Xi_1}{\Xi_3}\Big),\\
\end{split}
\end{equation}
which are all first-order rogue wave solution of the gc-FL equation \eqref{gcfl-eq}. From the above expression, it is obvious that these two components $u[1],v[1]$ are proportional, the ratio is $|\frac{c_1}{c_2}|$ and these maximum amplitudes have nearly three times the amplitudes of the background wave. As $N$ increases, there are two additional parameters $(m_i,n_i)$, which can control the structure of the higher-order rogue waves solution.

\begin{figure*}[!htbp]
\centering
\subfigure{\includegraphics[height=3.6in,width=3.6in]{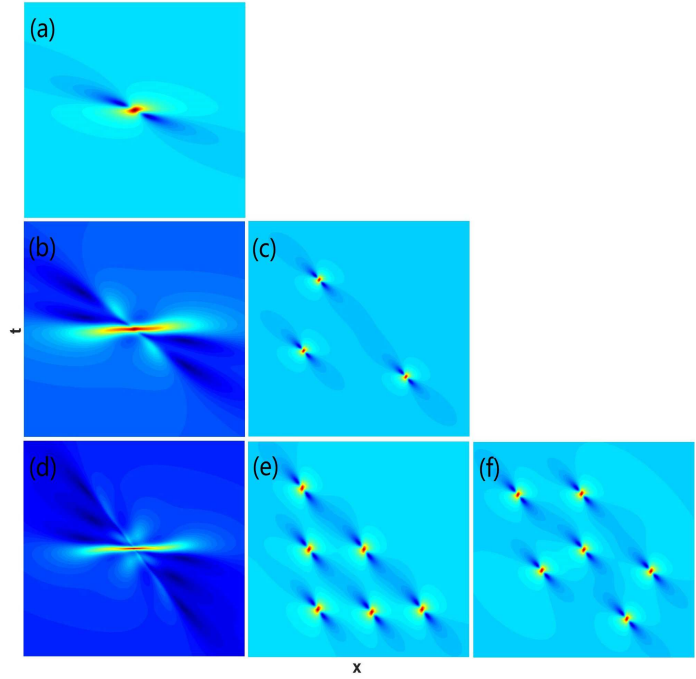}}
\caption{ The first- to third-order rogue waves of the gc-FL equation \eqref{gcfl-eq} with parameters:$(\alpha,\beta,\gamma,c_1,c_2)=(3,1,2,\frac{\sqrt{3}}{3},\frac{\sqrt{6}}{3})$; and $(m_i,n_i)=(0,0)$ in the first column, $(m_1,n_1)=(100,0)$ and $(m_1,m_2,n_1,n_2)=(100,0,100,0)$ in the second column, $(m_1,m_2,n_1,n_2)=(10,10000,0,0)$ in the third column. \label{Fig-gcfl-rw}}
\end{figure*}

For $N=2$, there are two structure patterns: a second-order fundamental rogue wave and a second-order triangular rogue wave composed of three standard first-order rogue waves (also known as the triplet structure). When $N=3$, in addition to above two structures: the third-order fundamental rogue wave and the third-order triangle rogue waves, there is also an additional pentagon rogue waves pattern. The so-called triangle and pentagon are both composed of six standard first-order rogue waves, but they are distributed differently. According to the above analysis, the number of corresponding parameters $m_i,~n_i$ increase with $N$ increases, and then there are a variety of combination forms about parameters $m_i,~n_i$, which give rise to more abundant structures of higher-order rogue wave solutions. In short, higher-order rogue wave solutions can be roughly divided into three types: the $N$-order fundamental structure, the combination of $N$-triplet of the standard first-order rogue wave and the combination structure of different order rogue waves, whose order less than $\frac{N(N+1)}{2}$. Since $u$ and $v$ components are similar in structure, only the density distribution of rogue waves for the $u$ component is given here, as presented in Fig. \ref{Fig-gcfl-rw}. From top to bottom, the first- to third-order rogue waves are depicted, respectively. And from left to right, the different structures of the same order rogue waves are represented.

\textbf{Case 2. Soliton+Rogue waves}

\begin{figure*}[!htbp]
\centering
\subfigure[]{\includegraphics[height=1.5in,width=1.9in]{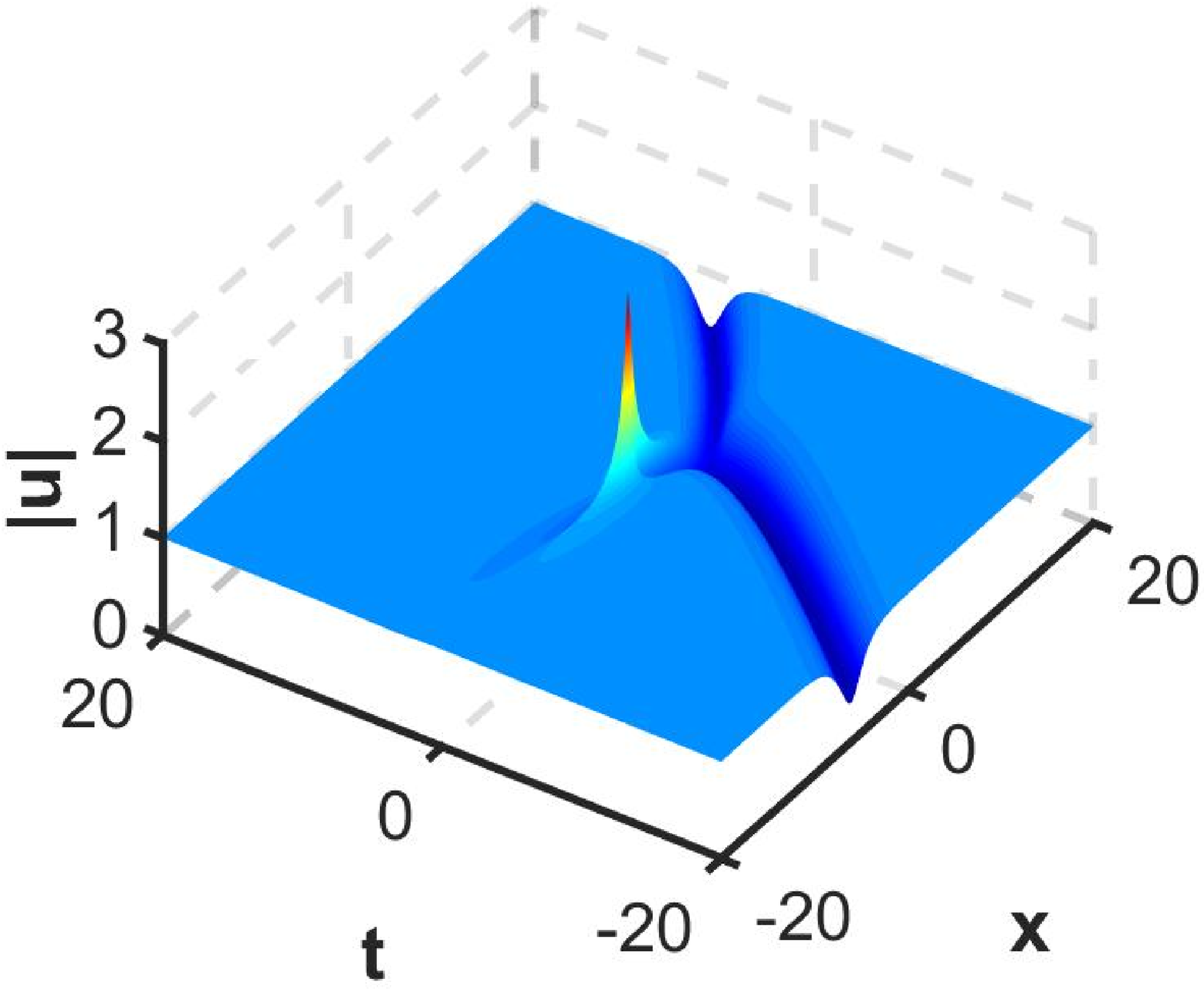}}\hspace{0.5cm}
\subfigure[]{\includegraphics[height=1.5in,width=1.6in]{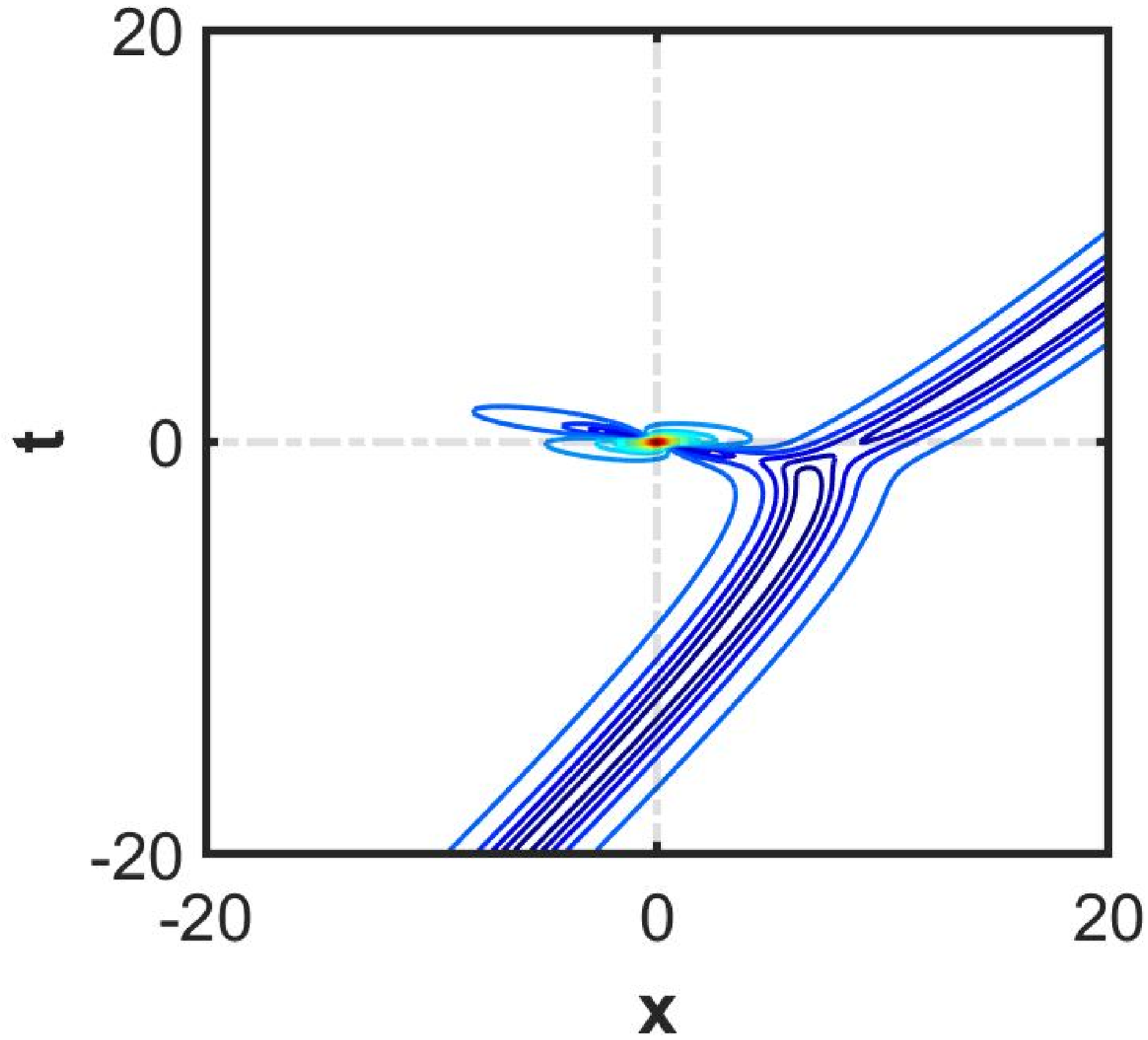}}\hspace{0.5cm}
\subfigure[]{\includegraphics[height=1.5in,width=1.6in]{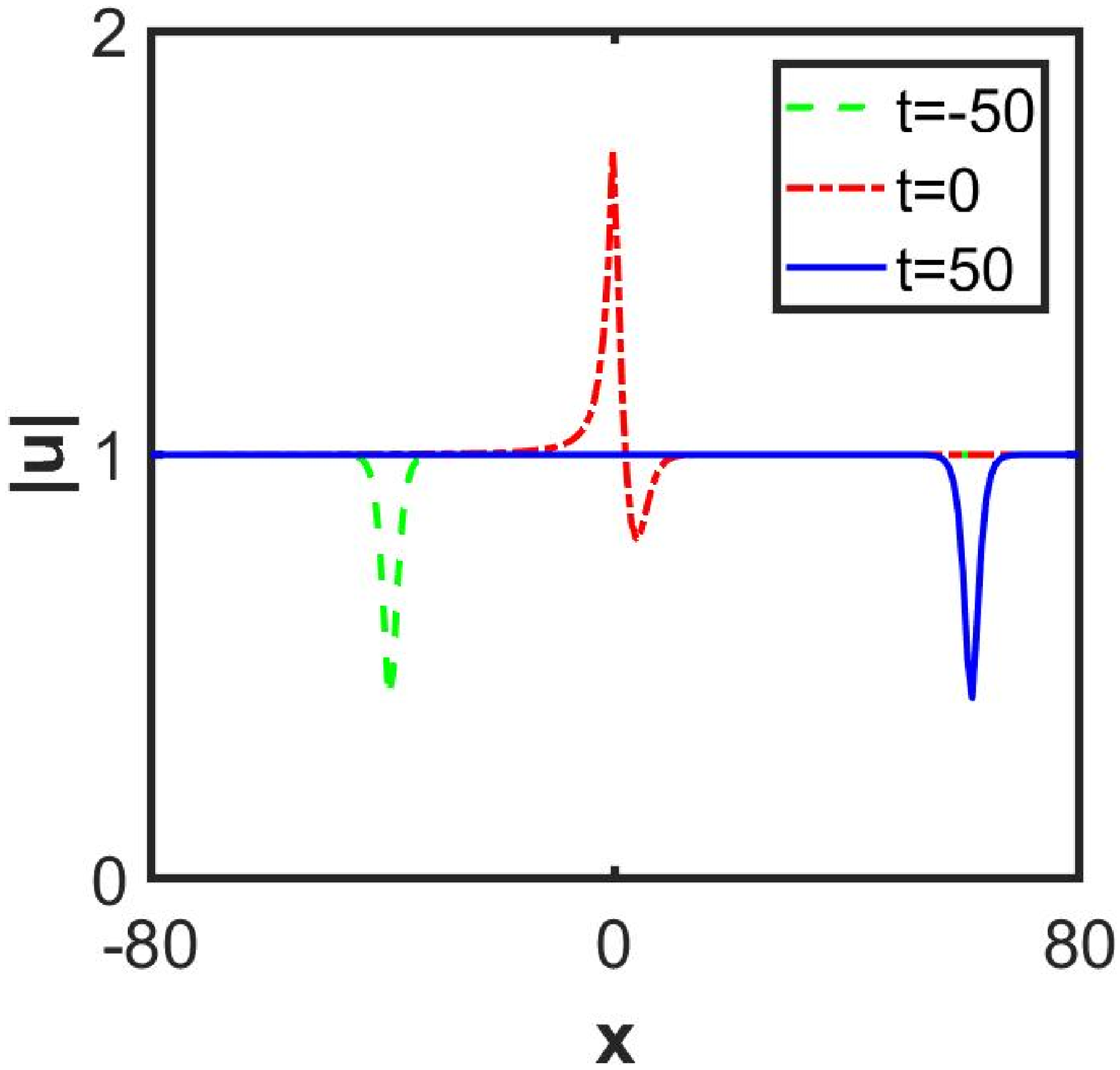}}\hspace{0.5cm}\\
\subfigure[]{\includegraphics[height=1.5in,width=1.9in]{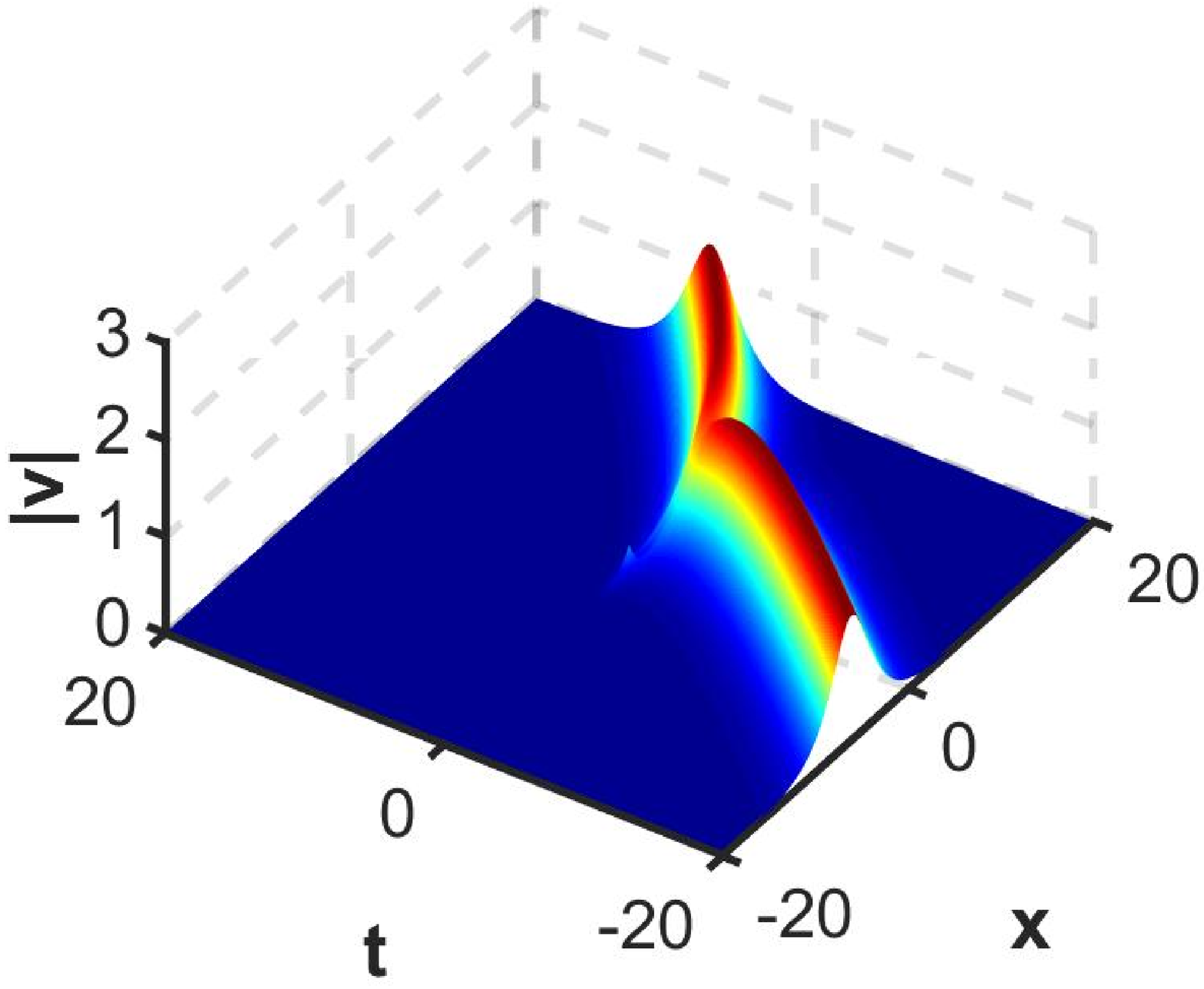}}\hspace{0.5cm}
\subfigure[]{\includegraphics[height=1.5in,width=1.6in]{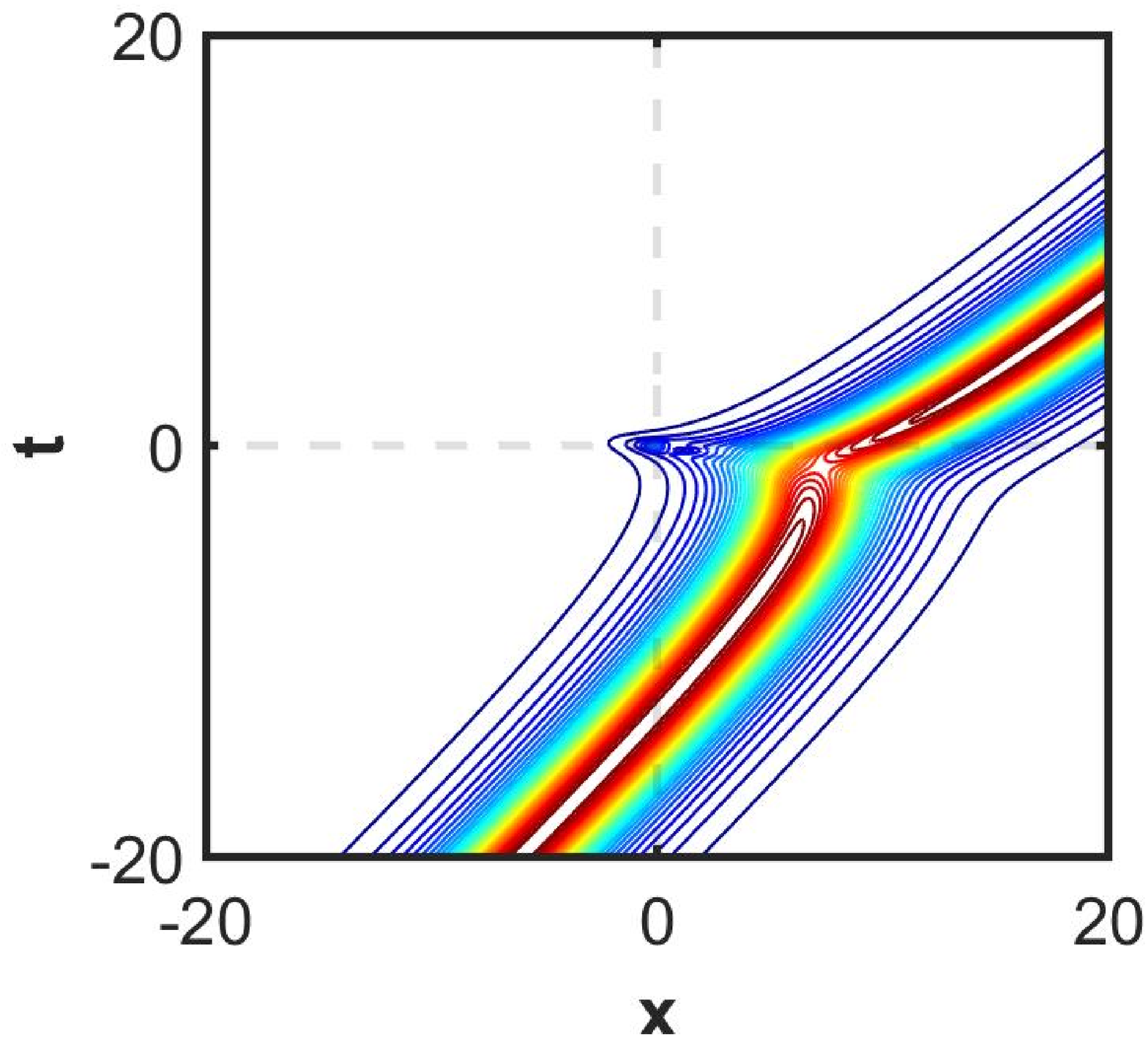}}\hspace{0.5cm}
\subfigure[]{\includegraphics[height=1.5in,width=1.6in]{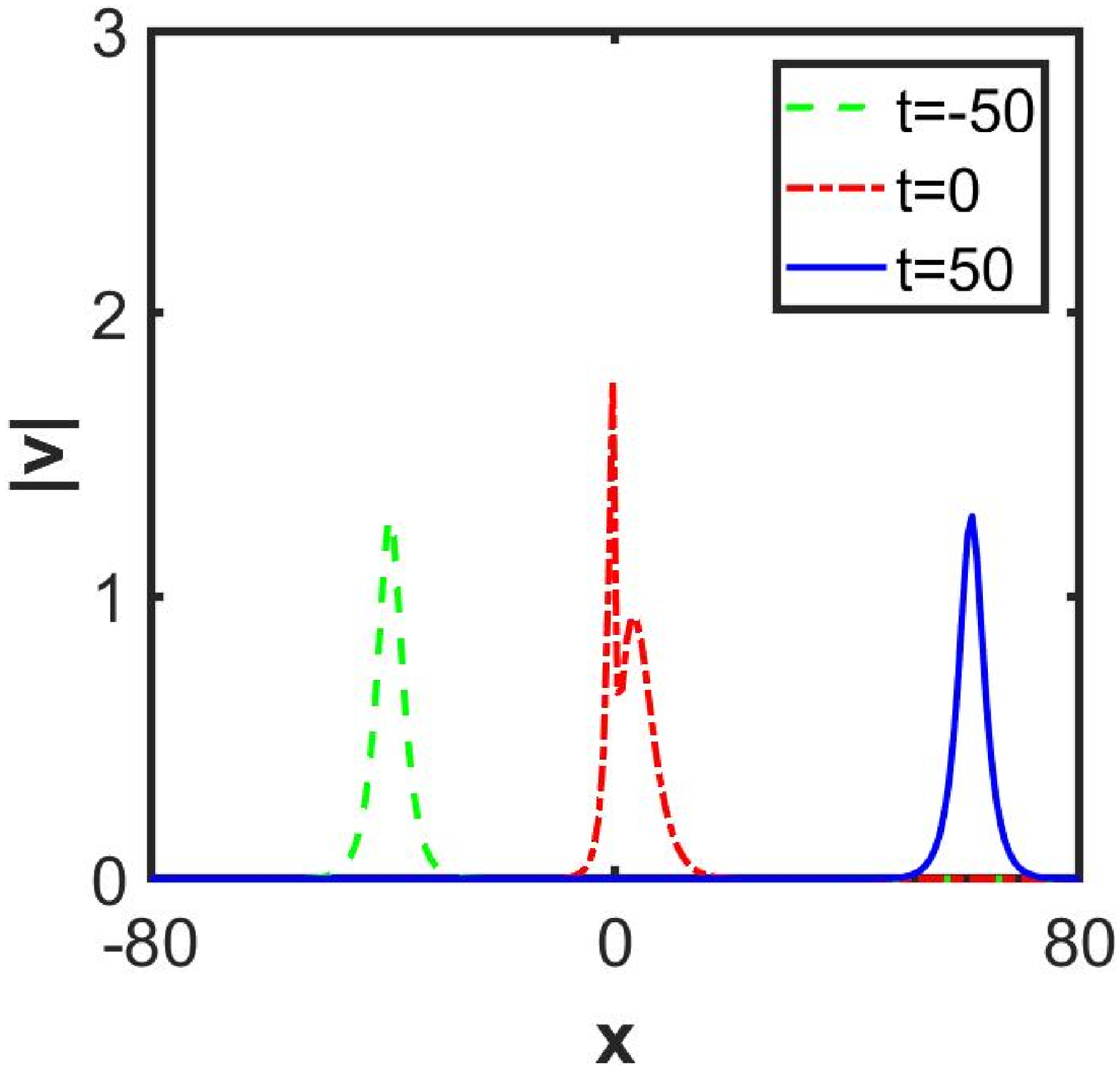}}
\caption{ The first-order interaction solution between a soliton and a rogue wave of the gc-FL equation \eqref{gcfl-eq} with parameters: $(\alpha,\beta,\gamma,c_1,c_2)=(3,1,2,1,0)$. The first and second columns show 3D separate structure and the corresponding contour map with $d=\frac{1}{10}$, respectively; the third column shows collision process over time with $d=1$. \label{Fig-gcfl-sr1}}
\end{figure*}

When parameters $(d,c_1)\neq(0,0)$ and $c_2=0$ in Eqs. \eqref{gcfl-lws-10}, the first-order semi-rational solution of the gc-FL equation can be obtained. At this time, there are two kinds of semi-rational solutions with different structures: bright soliton interacts with rogue wave and dark soliton interacts with rogue wave. The 3D diagrams and contour maps of $u$ and $v$ components are depicted in the first two columns in Fig. \ref{Fig-gcfl-sr1}. The soliton propagates below the background wave in the $u$ component, while the soliton  in the $v$ component propagates above the background wave. Obviously, the background amplitude of the $u$ component is $1$, and the background amplitude of the $v$ component is $0$, which depending entirely on their corresponding seed solutions. In addition, by changing the value of $|d|$ to adjust the strength of the interactional structure of the solution, it results in the variation of the distance between the soliton (dark or bright) and the rogue wave, that is, merge or separation.

The last column of Fig. \ref{Fig-gcfl-sr1} depicts the collision and interaction process of the dark soliton in the $u$ component and the bright soliton in the $v$ component with rogue wave at different times, respectively. Obviously, both $u$ and $v$ components propagate along the positive direction of the $x$-axis with time, and at the beginning, only bright and dark soliton solutions located on the negative half of the $x$-axis exist at $t=-50$. At $t=0$, a rogue wave suddenly appears and interacts with the soliton and casues the energy transfer between them, which resulting in an increase in the amplitude of the rogue wave and a decrease in the amplitude of the soliton. Then, at $t=50$, the rogue wave disappears, and the soliton continues to propagate along the positive direction of the $x$-axis without changing its shape, wave width and amplitude. Since the background wave height of the rogue wave in the $v$ component is $0$, the amplitude of the rogue wave relative to the bright soliton is not obvious when it is close to the bright soliton, but the amplitude of the rogue wave will increase sharply when the complete collision occurs.

\begin{figure*}[!htbp]
\centering
\subfigure[]{\includegraphics[height=1.3in,width=1.6in]{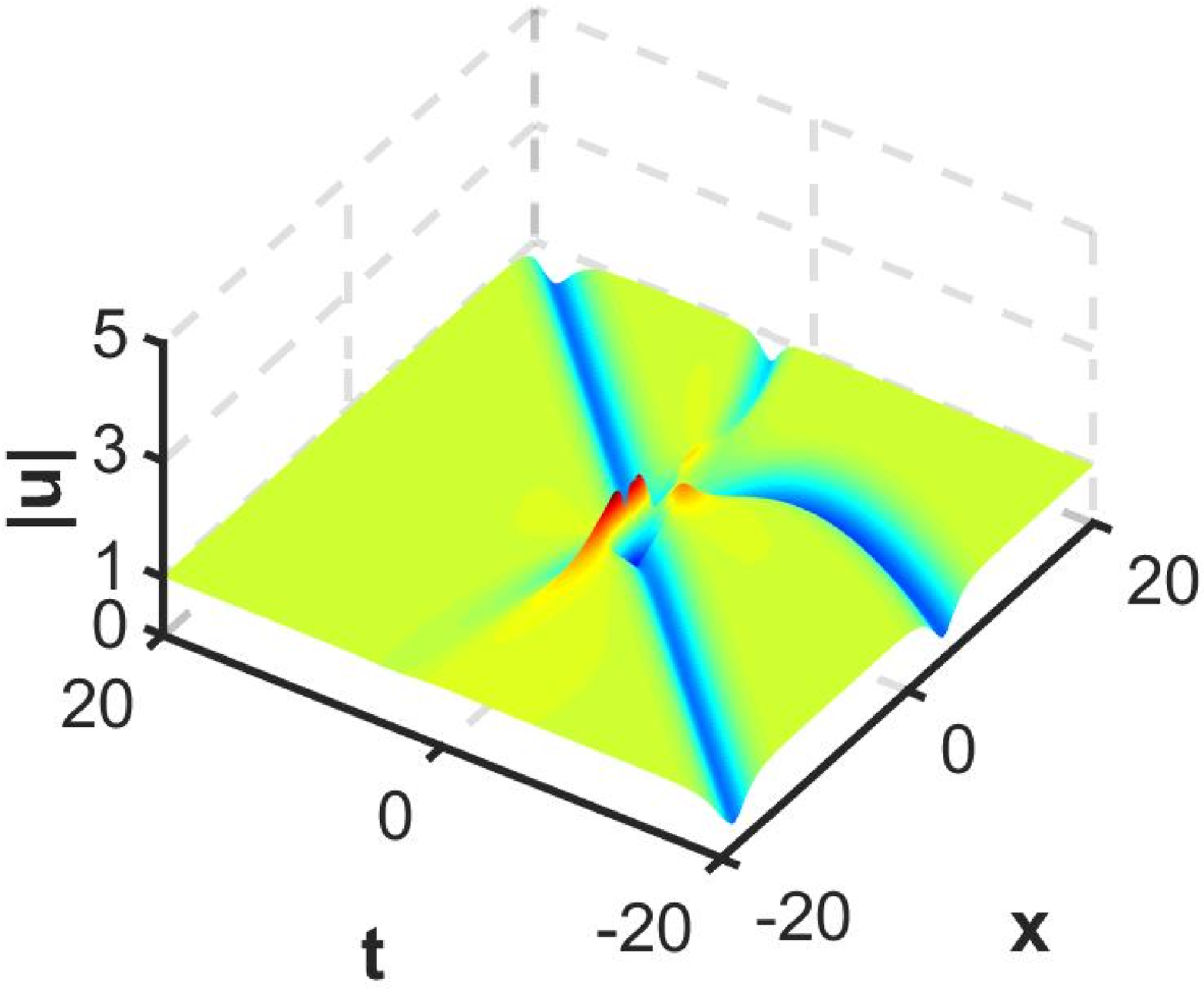}}\hspace{0.1cm}
\subfigure[]{\includegraphics[height=1.3in,width=1.6in]{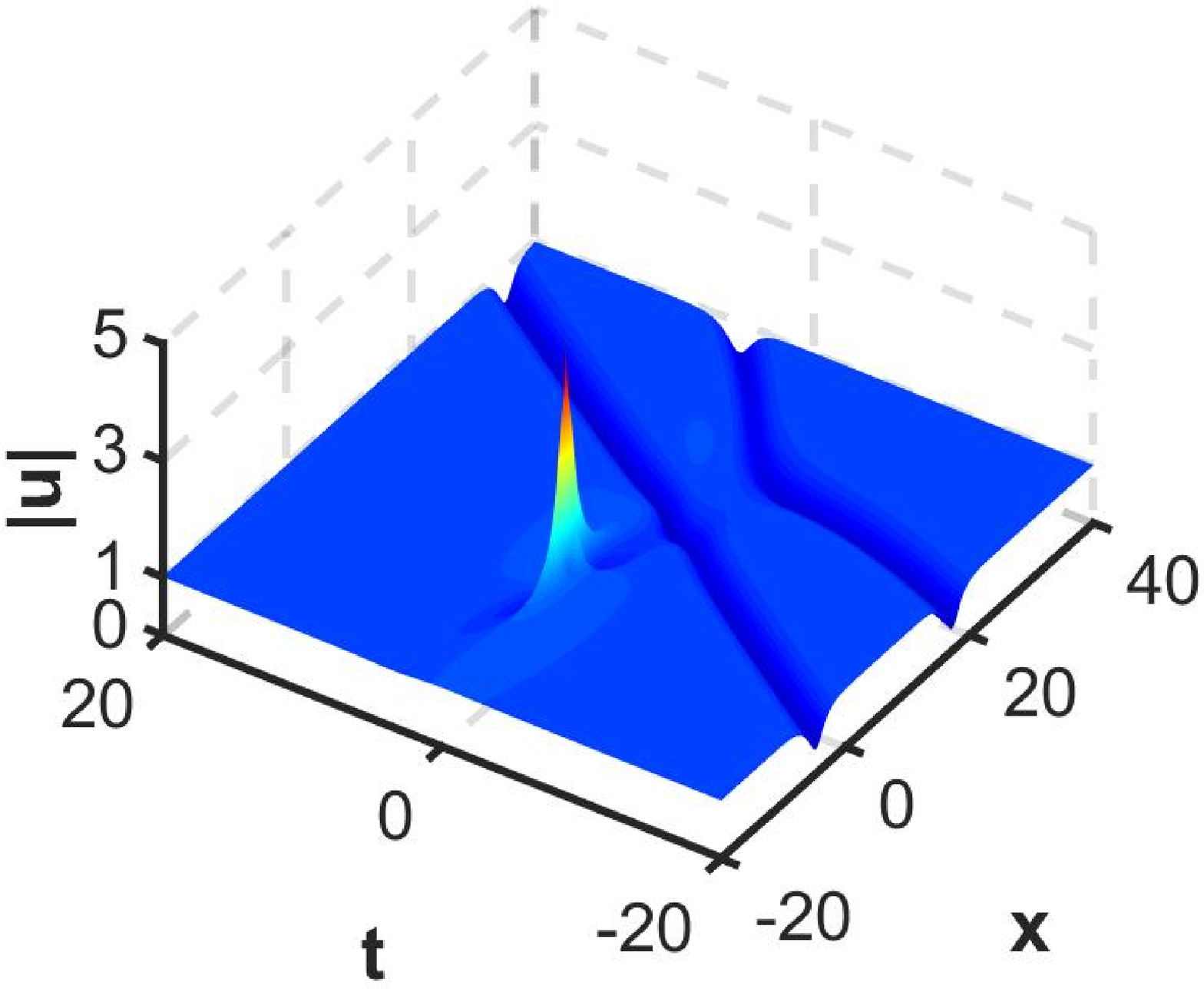}}\hspace{0.1cm}
\subfigure[]{\includegraphics[height=1.3in,width=1.6in]{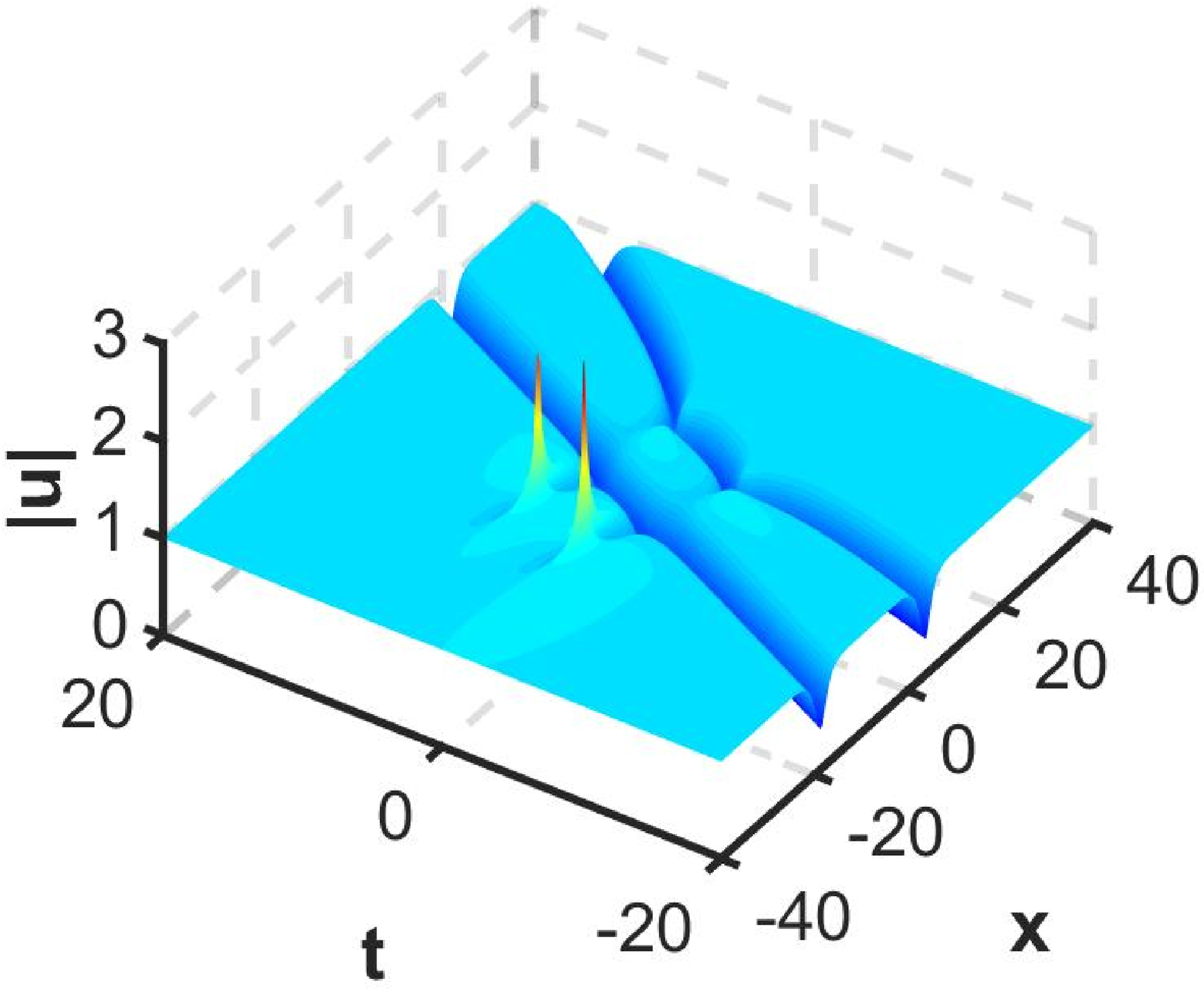}}\hspace{0.1cm}
\subfigure[]{\includegraphics[height=1.3in,width=1.6in]{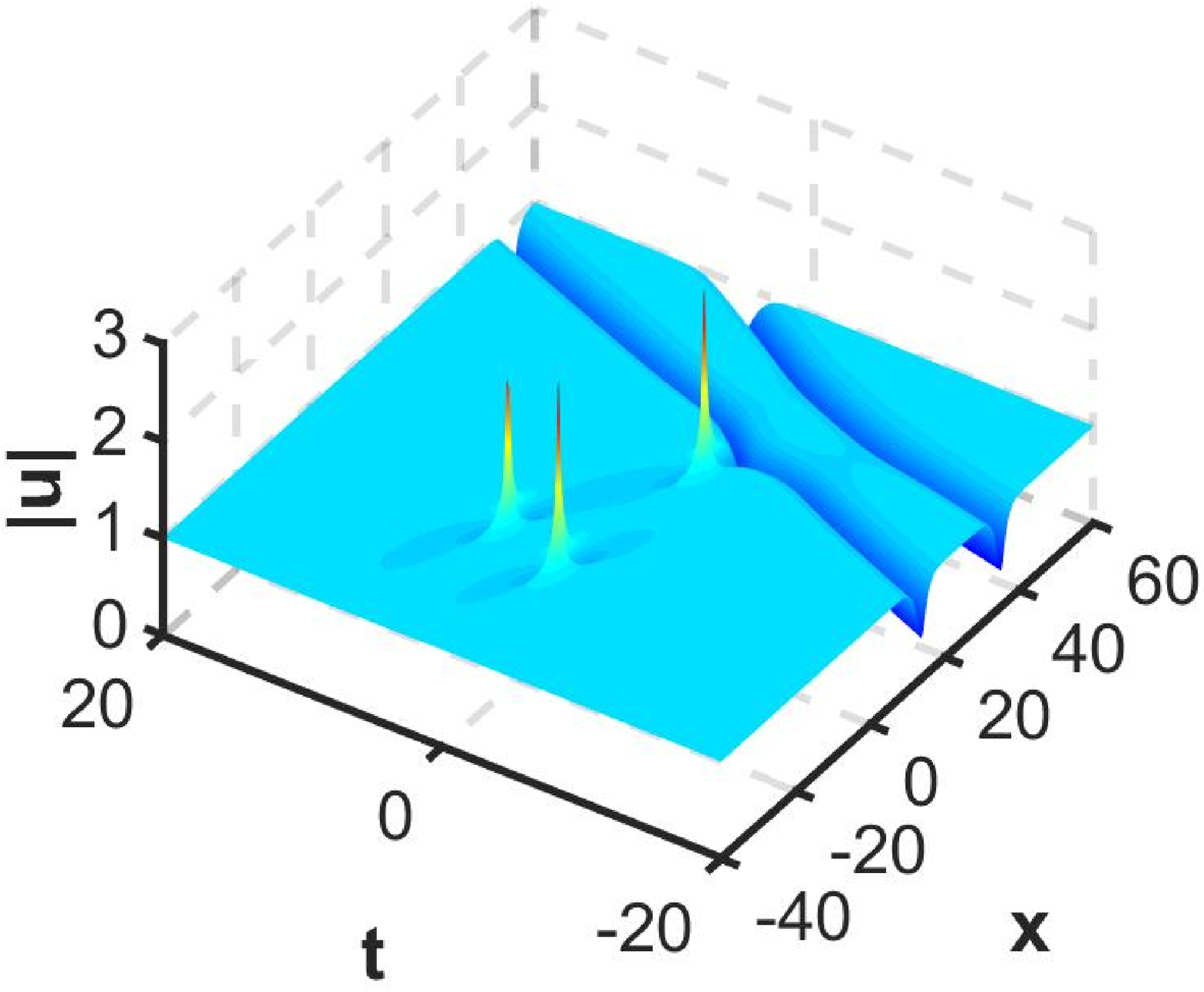}}\hspace{0.1cm}\\
\subfigure[]{\includegraphics[height=1.3in,width=1.6in]{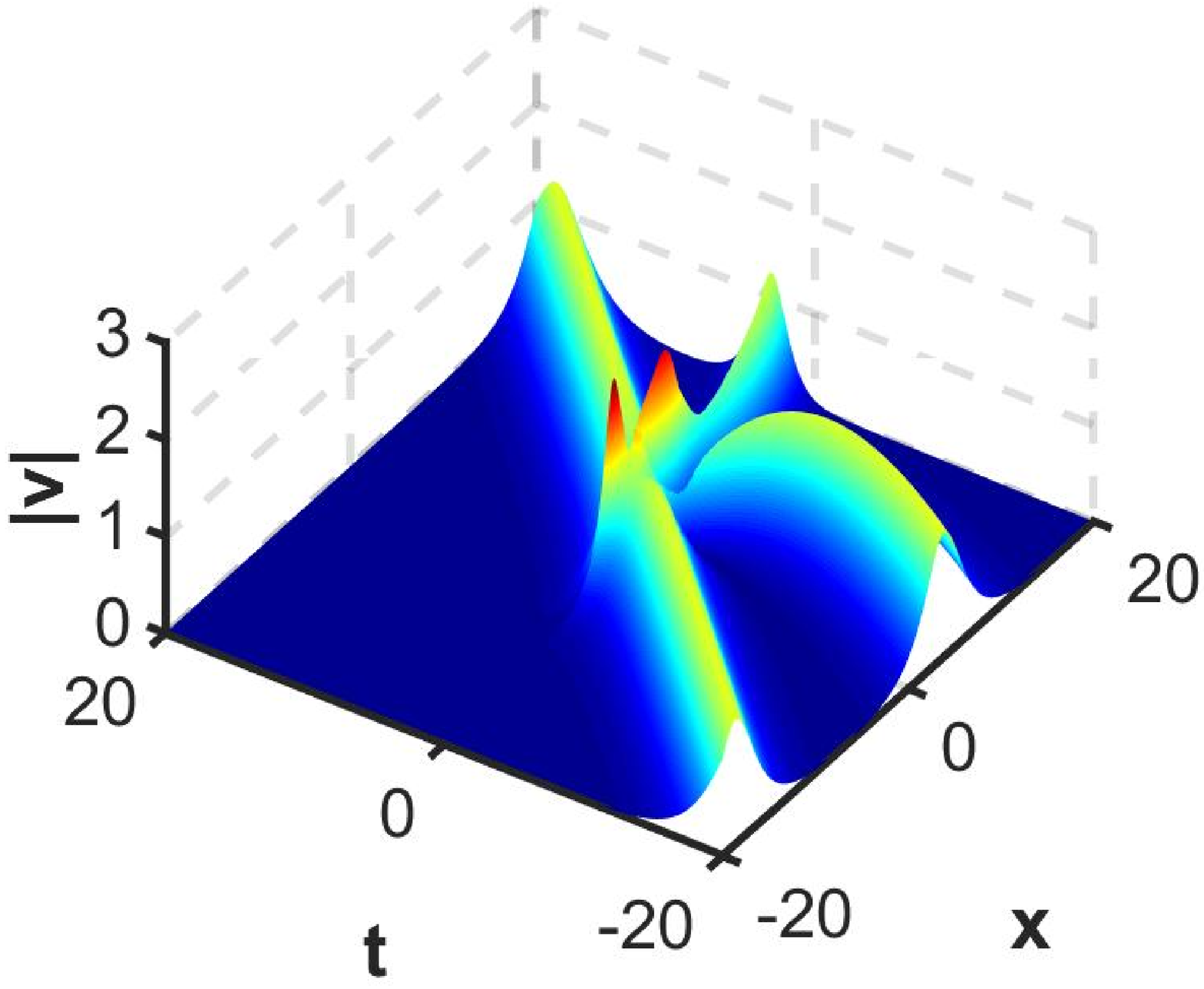}}\hspace{0.1cm}
\subfigure[]{\includegraphics[height=1.3in,width=1.6in]{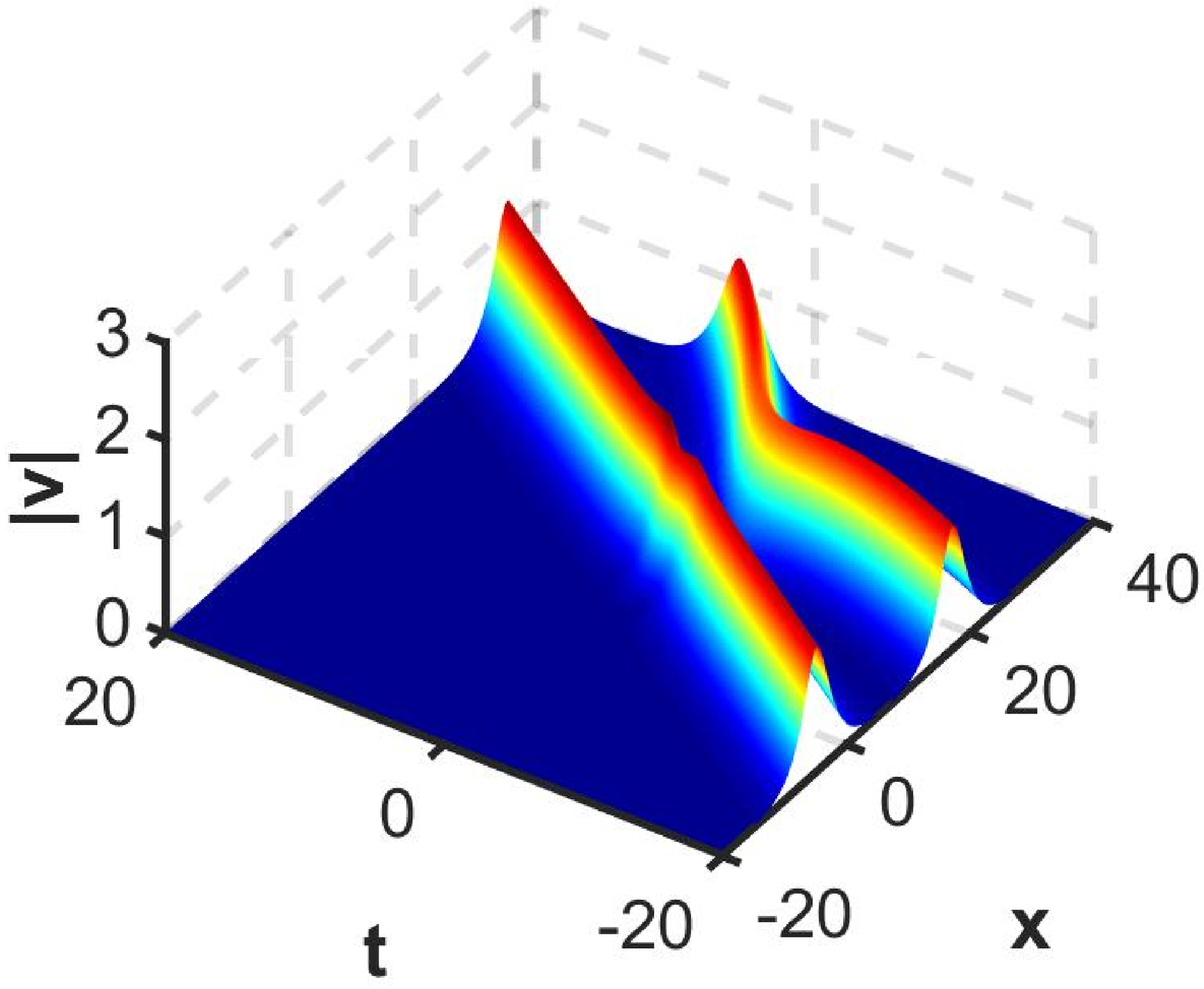}}\hspace{0.1cm}
\subfigure[]{\includegraphics[height=1.3in,width=1.6in]{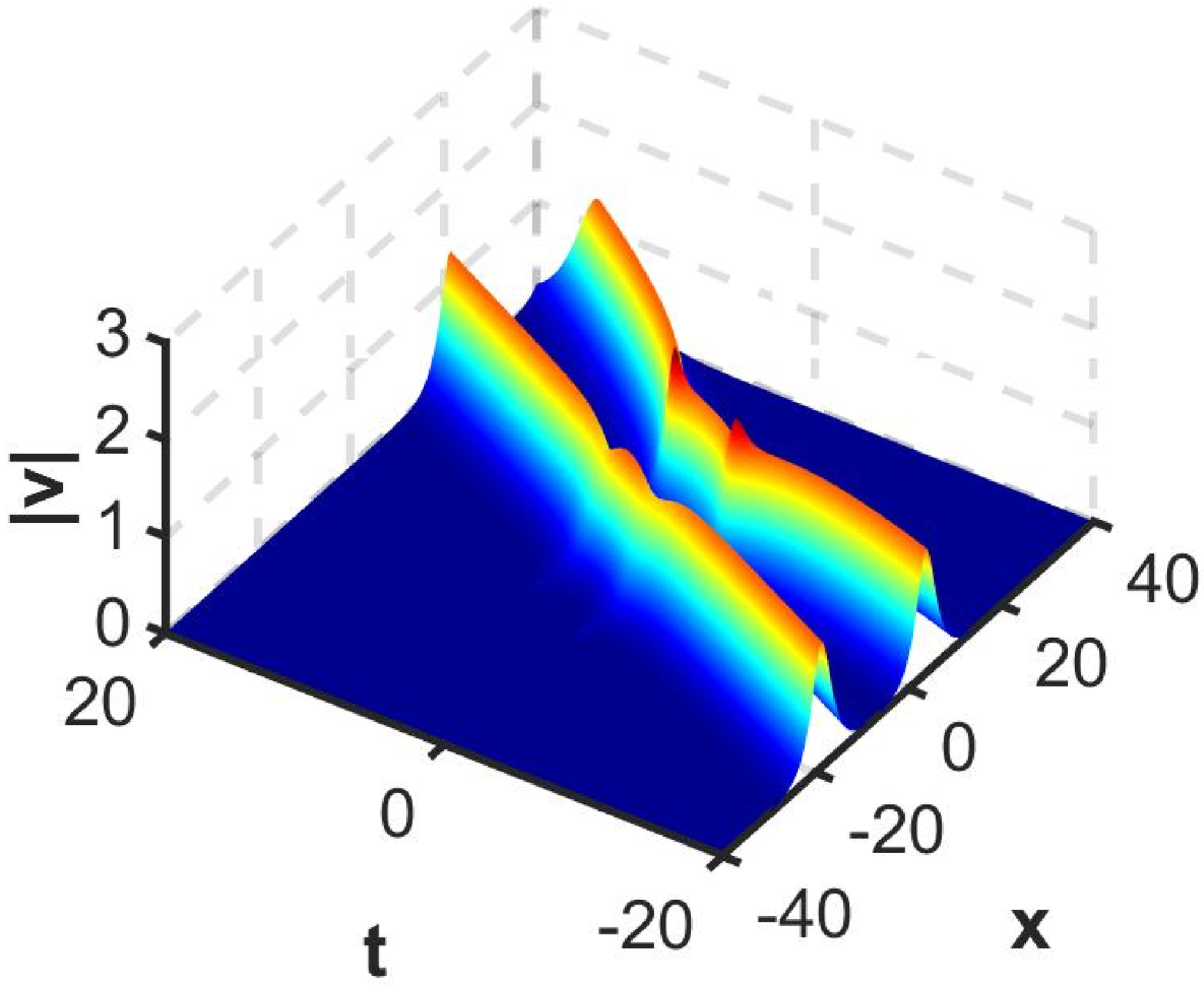}}\hspace{0.1cm}
\subfigure[]{\includegraphics[height=1.3in,width=1.6in]{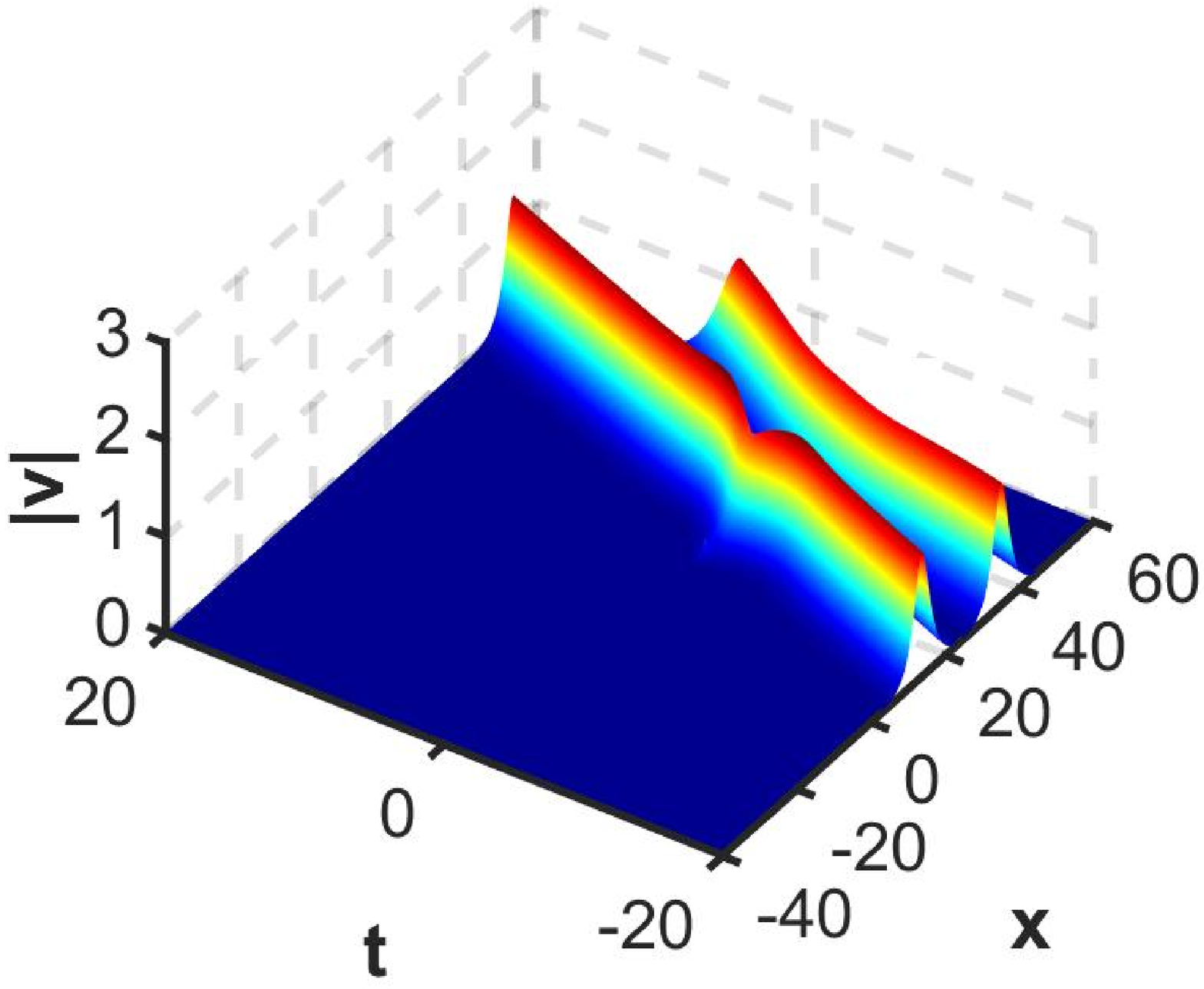}}
\caption{ The second-order interaction solution between two solitons and two-order rogue wave of the gc-FL equation with parameters:$(\alpha,\beta,\gamma,c_1,c_2)=(3,1,2,1,0)$;
From the left column to the right column the parameters $(d,m_1,n_1)$ are (1,0,0), $(\frac{1}{1000},0,0)$, (1,1000,0) and $(\frac{1}{10000000},1000,0)$, respectively. The odd columns show merge structure; the even columns show separate structure. \label{Fig-gcfl-sr2}}
\end{figure*}

According to the formula \eqref{gcfl-lws-8} for $N$-order localized wave solution of the gc-FL equation given in Theorem \ref{the2}, the second-order localized wave solution can be obtained by setting the parameter $N=2$. There are two structures with parameters $(\alpha,\beta,\gamma,c_1,c_2)=(3,1,2,1,0)$: a second-order fundamental rogue wave interacts with two solitons and  triplet rogue wave interacts with two solitons. When $d=1$, the two components $u_2$ and $v_2$ show strong interaction as odd columns (merge structure) in Fig. \ref{Fig-gcfl-sr2}; while $d=\frac{1}{1000}$ or $\frac{1}{10000000}$, the two components $u_2$ and $v_2$ show weak interaction as even columns (separate structure) in Fig. \ref{Fig-gcfl-sr2}. For the $v$ component, it is not difficult to find that the rogue wave can be clearly seen in the case of strong interaction, while the rogue wave is not obvious in the case of weak interaction. The main reason is that the energy surge of the rogue wave driven by soliton under the strong interaction. However, during the weak interaction, the soliton does not completely collide with the rogue wave, so the soliton has more energy relative to the background wave than the rogue wave relative to the background wave, which is demonstrated as the even columns in the last row of Fig. \ref{Fig-gcfl-sr2}. 

\begin{figure*}[!htbp]
\centering
\subfigure[]{\includegraphics[height=1.5in,width=1.9in]{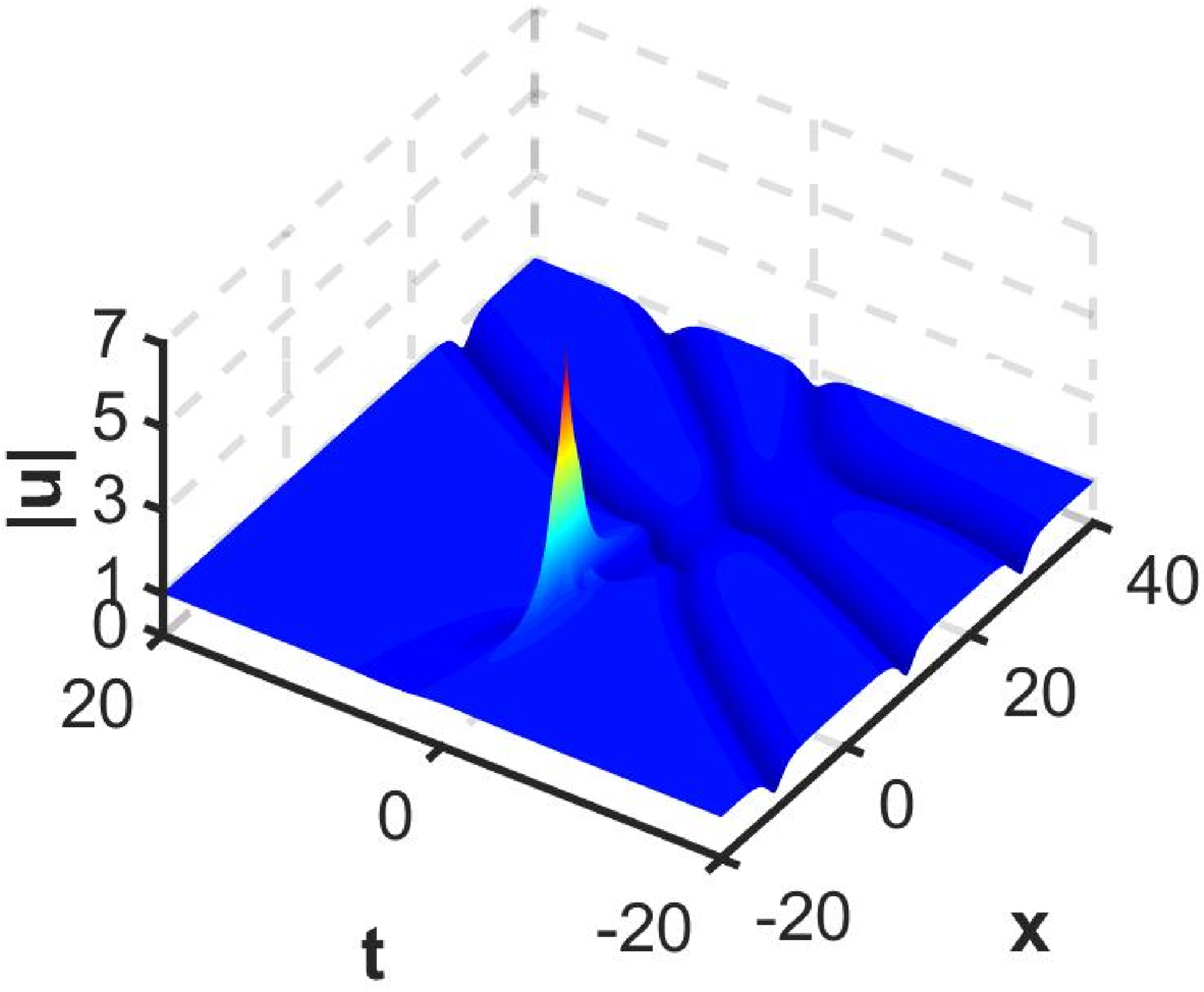}}\hspace{0.5cm}
\subfigure[]{\includegraphics[height=1.5in,width=1.9in]{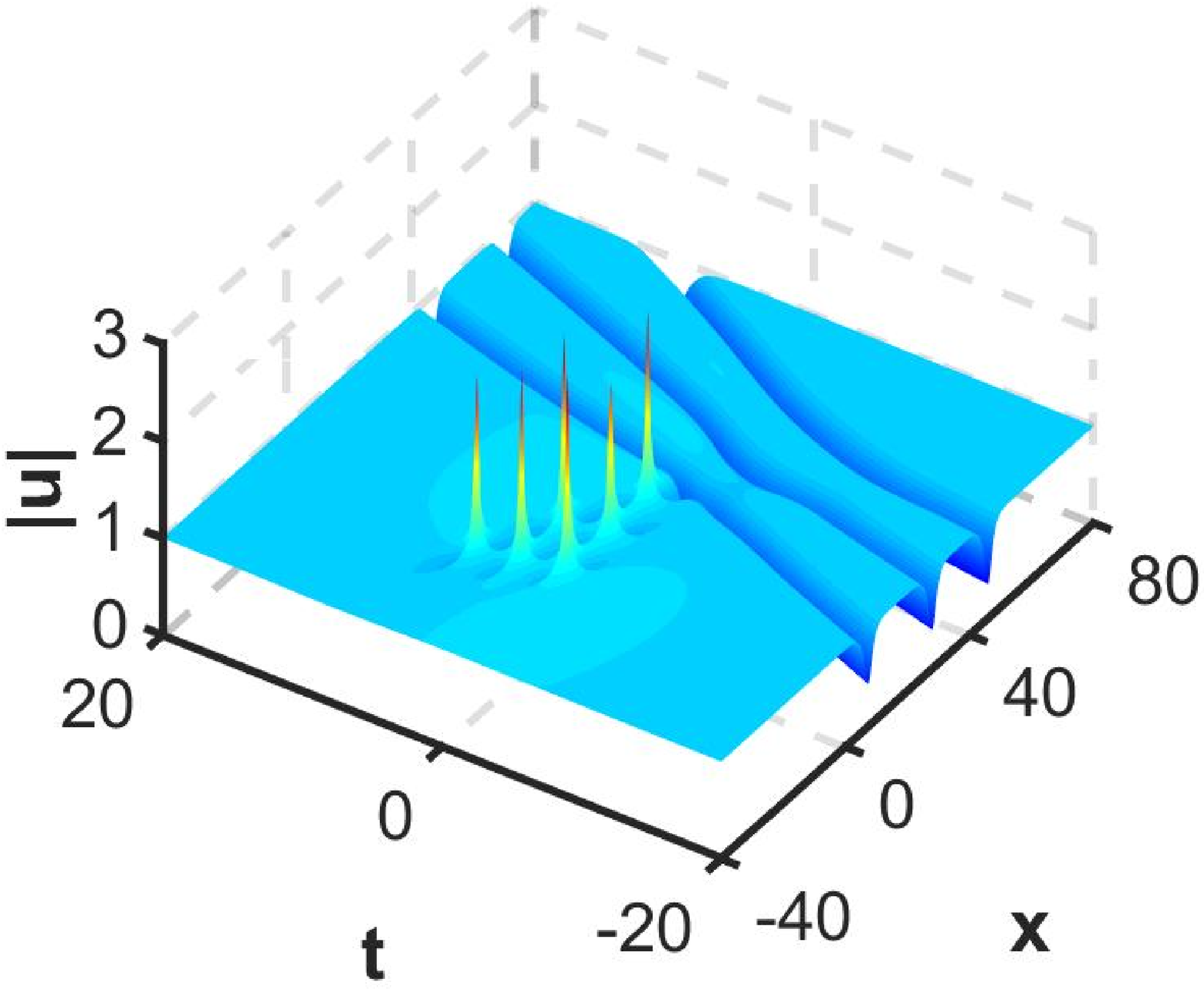}}\hspace{0.5cm}
\subfigure[]{\includegraphics[height=1.5in,width=1.9in]{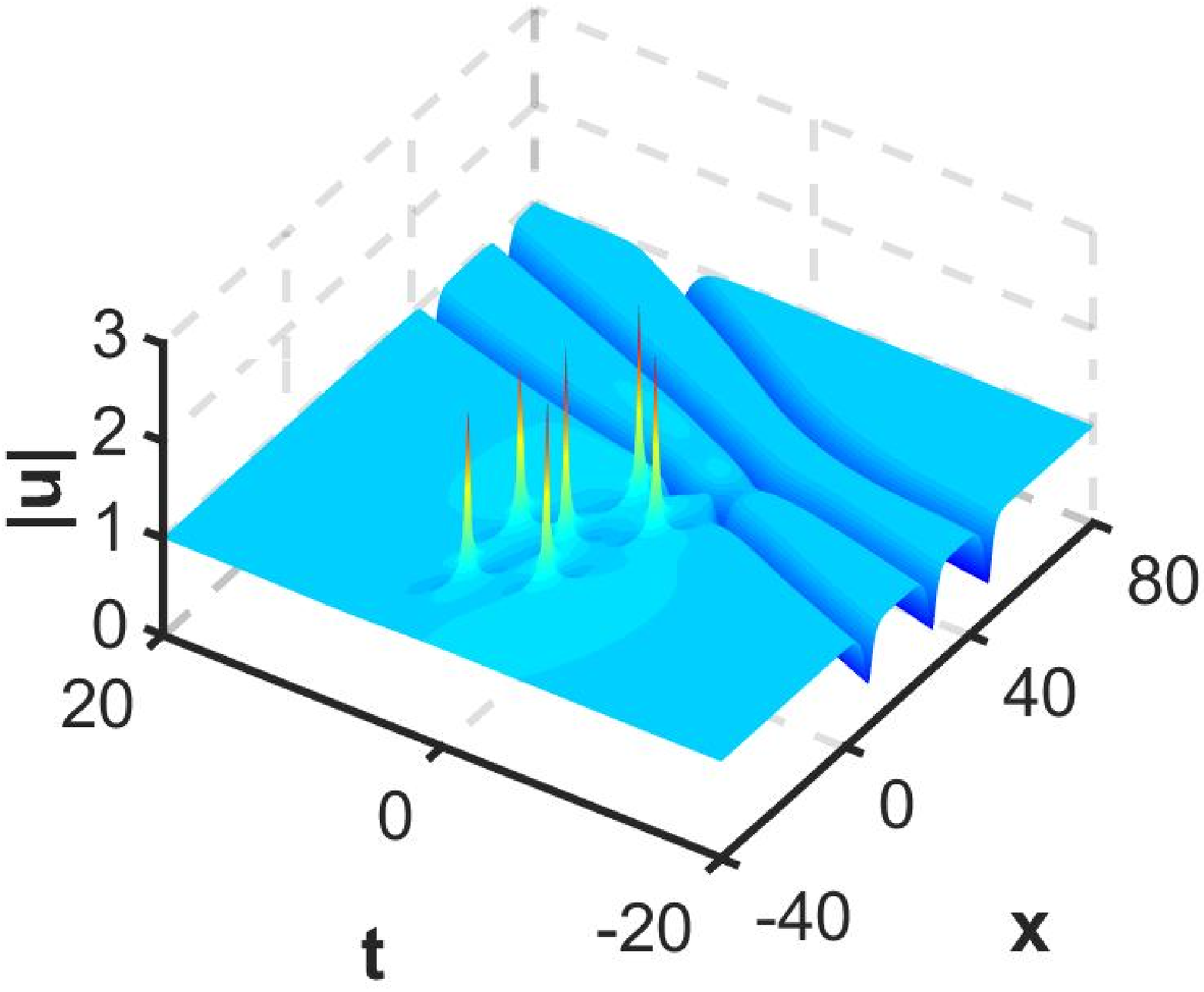}}\hspace{0.5cm}\\
\subfigure[]{\includegraphics[height=1.5in,width=1.9in]{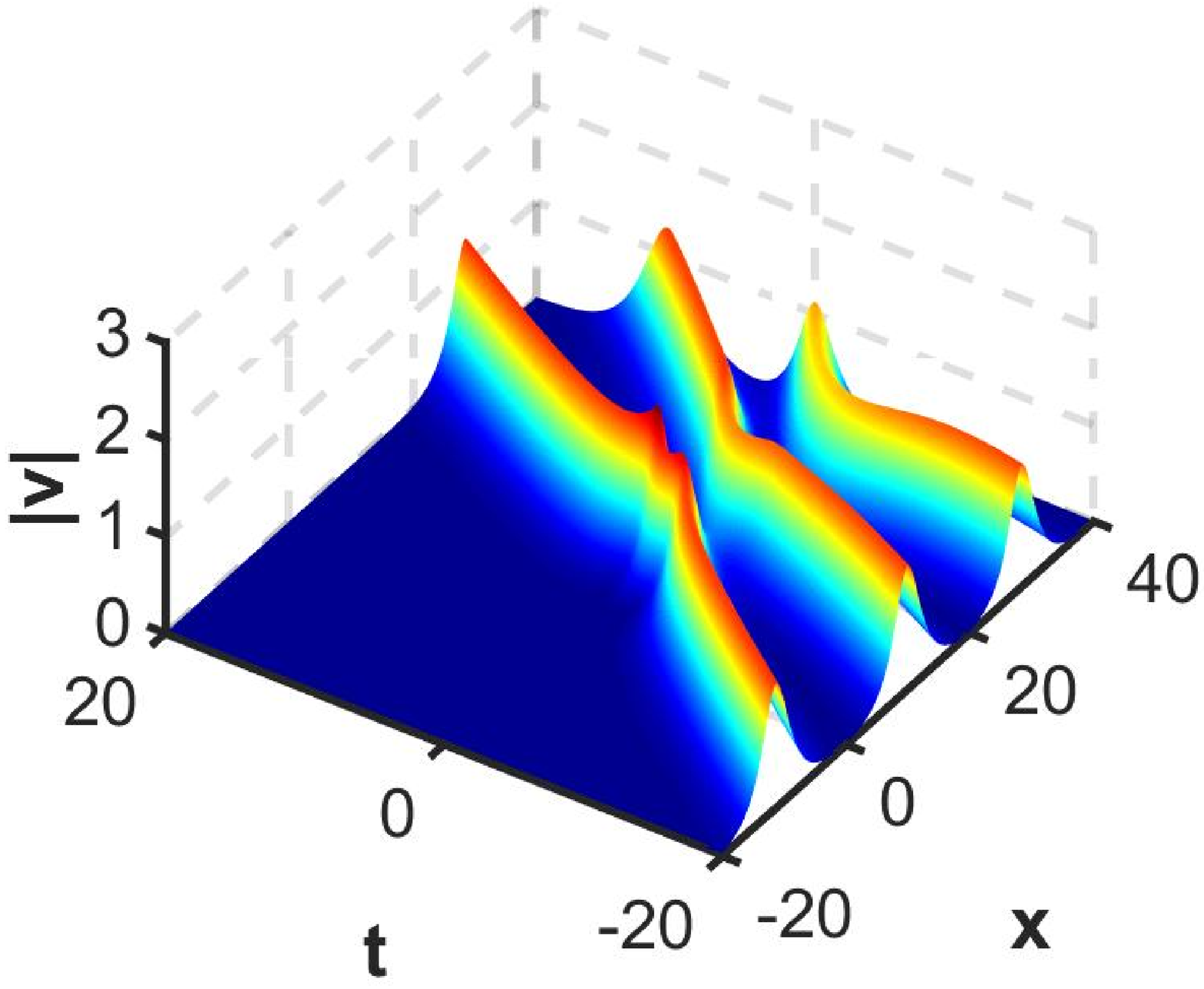}}\hspace{0.5cm}
\subfigure[]{\includegraphics[height=1.5in,width=1.9in]{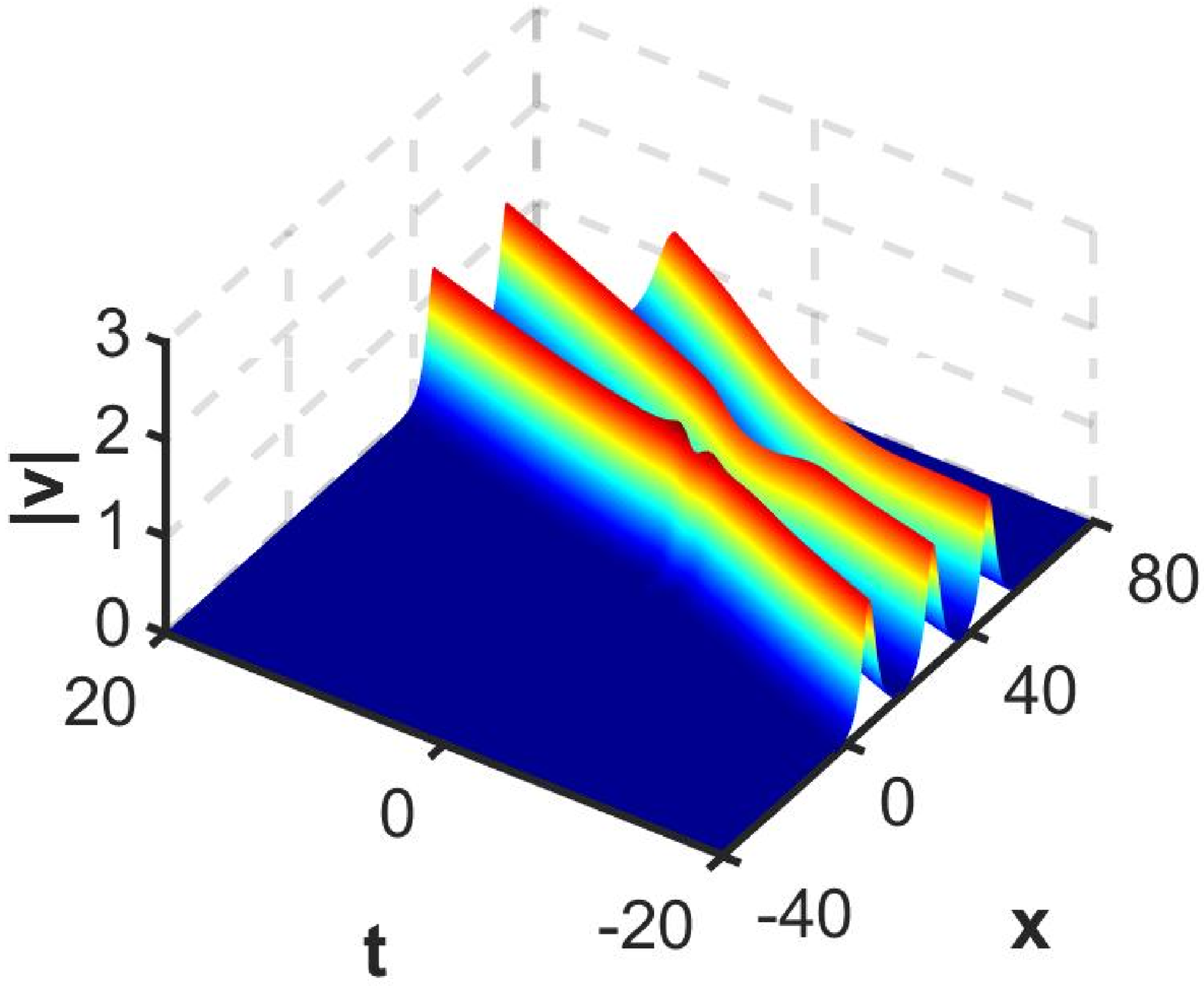}}\hspace{0.5cm}
\subfigure[]{\includegraphics[height=1.5in,width=1.9in]{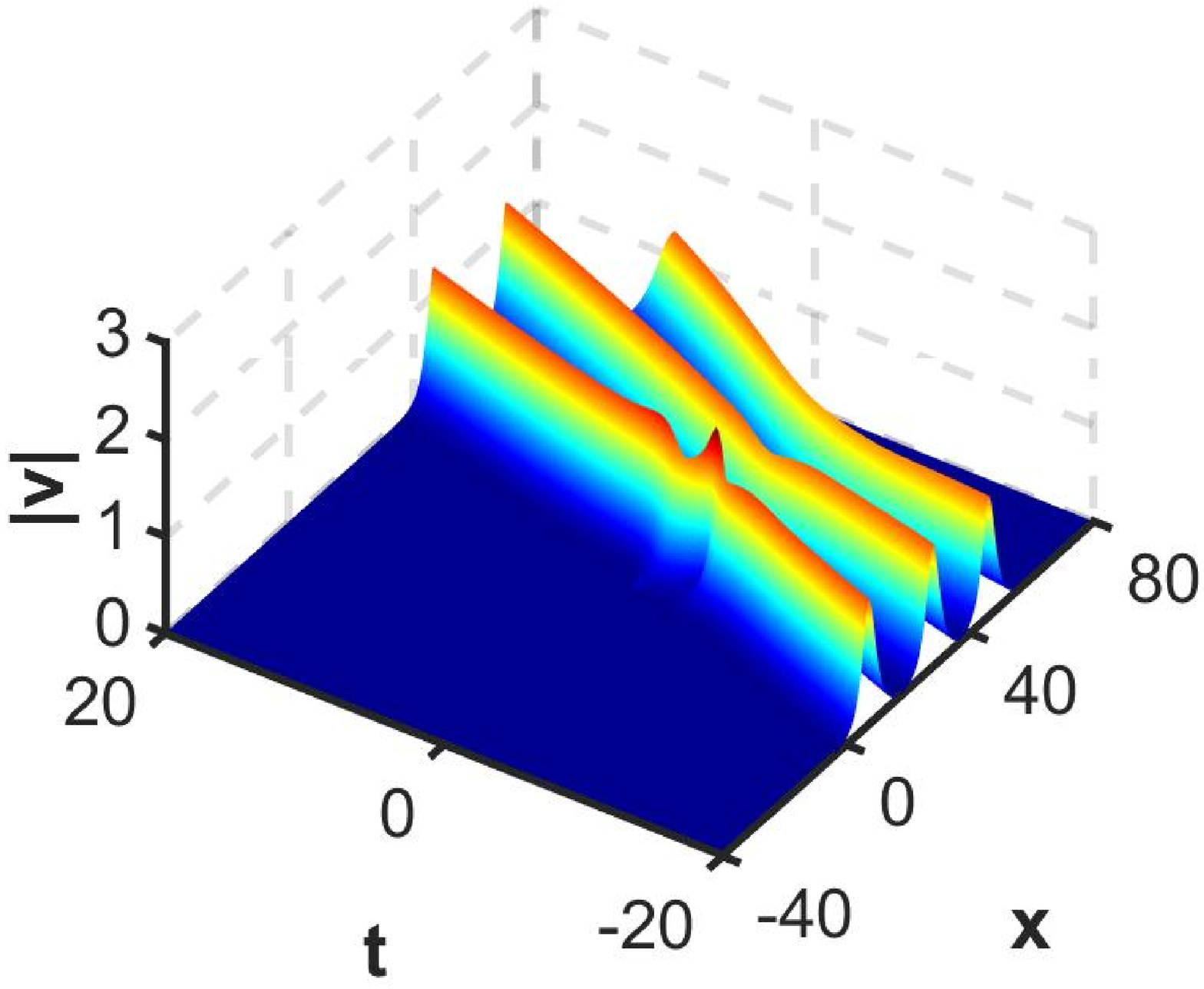}}
\caption{ The third-order interaction solution between three solitons and third-order rogue wave of the gc-FL equation \eqref{gcfl-eq} with the parameters $(\alpha,\beta,\gamma,c_1,c_2,d,m1,m2,n1,n2)$ from the left column to the right column are $(3,1,2,1,0,\frac{1}{1000},0,0,0,0)$, $(3,1,2,1,0,\frac{1}{10000000},100,0,100,0)$ and $(3,1,2,1,0,\frac{1}{10000000},10,10000,0,0)$, respectively. Top line: dark solitons interact with different structures of the third-order rogue waves; Bottom line: bright solitons interact with different structures of the third-order rogue waves.\label{Fig-gcfl-sr3}}
\end{figure*}

For $N=3$ in Eq. \eqref{gcfl-lws-8}, the third-order localized wave solution of the gc-FL equation \eqref{gcfl-eq} can be naturally deduced. In the case 1, we already know that the third-order rogue wave solution has three kinds of structures, just see bottom patterns in Fig. \ref{Fig-gcfl-rw}. Similarly, when parameters $(d,c_1)\neq(0,0)$ and $c_2=0$, three types of interaction solutions can be obtained, which are exhibited in Fig. \ref{Fig-gcfl-sr3}. Here, parameters $(m_i,n_i)$ control the merging and splitting of higher-order rogue waves. The parameter $d$ can regulate the distance between the soliton and the rogue wave to regulate the strength of the interaction. These above interaction structures can be degenerated into higher-order rogue wave solutions at $d=0$, namely, fundermental structure, triplet triangle structure and pentagon structure.


\textbf{Case 3. Breather + Rogue waves}

\begin{figure*}[!htbp]
\centering
\subfigure[]{\includegraphics[height=1.3in,width=1.6in]{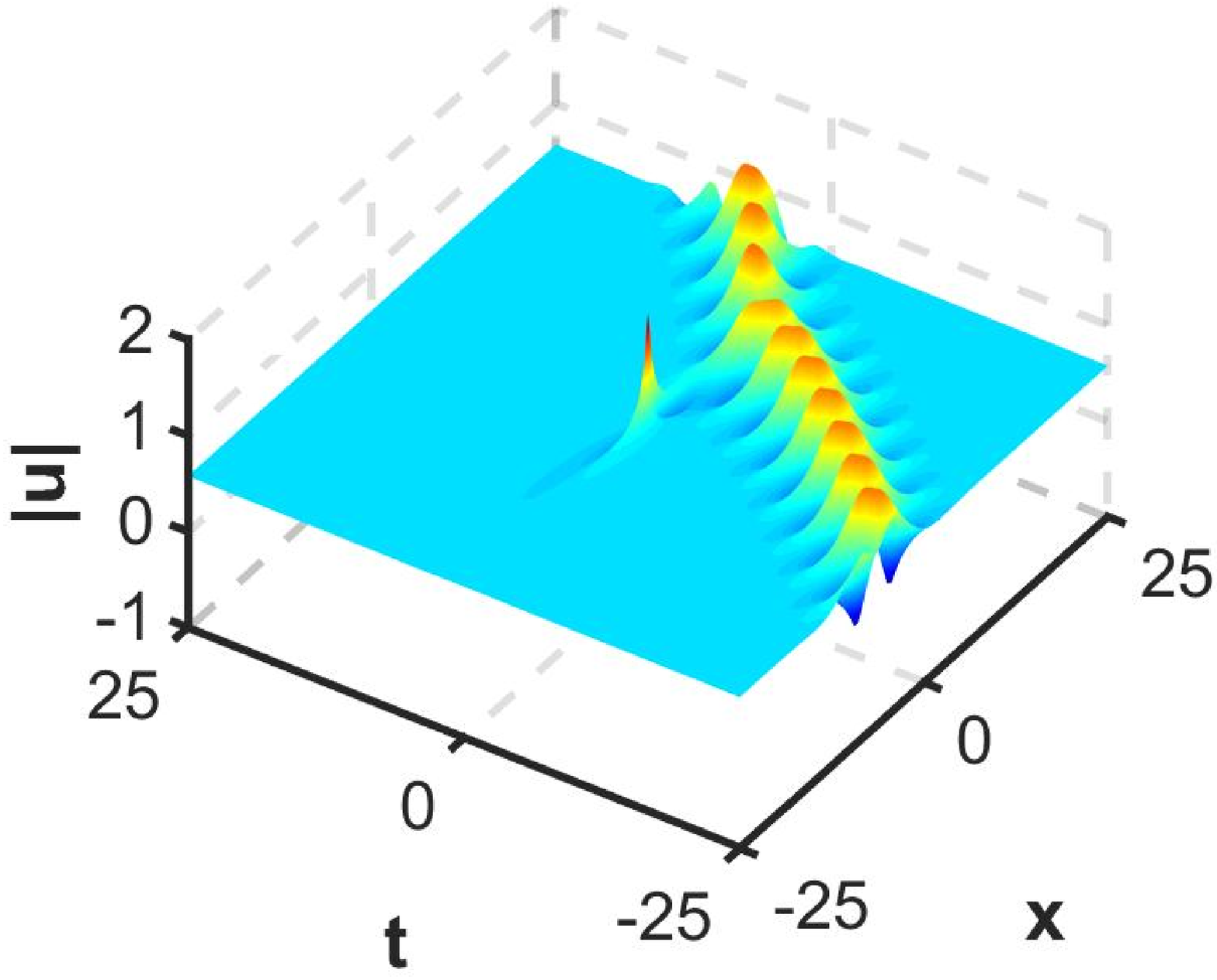}}\hspace{0.5cm}
\subfigure[]{\includegraphics[height=1.2in,width=1.3in]{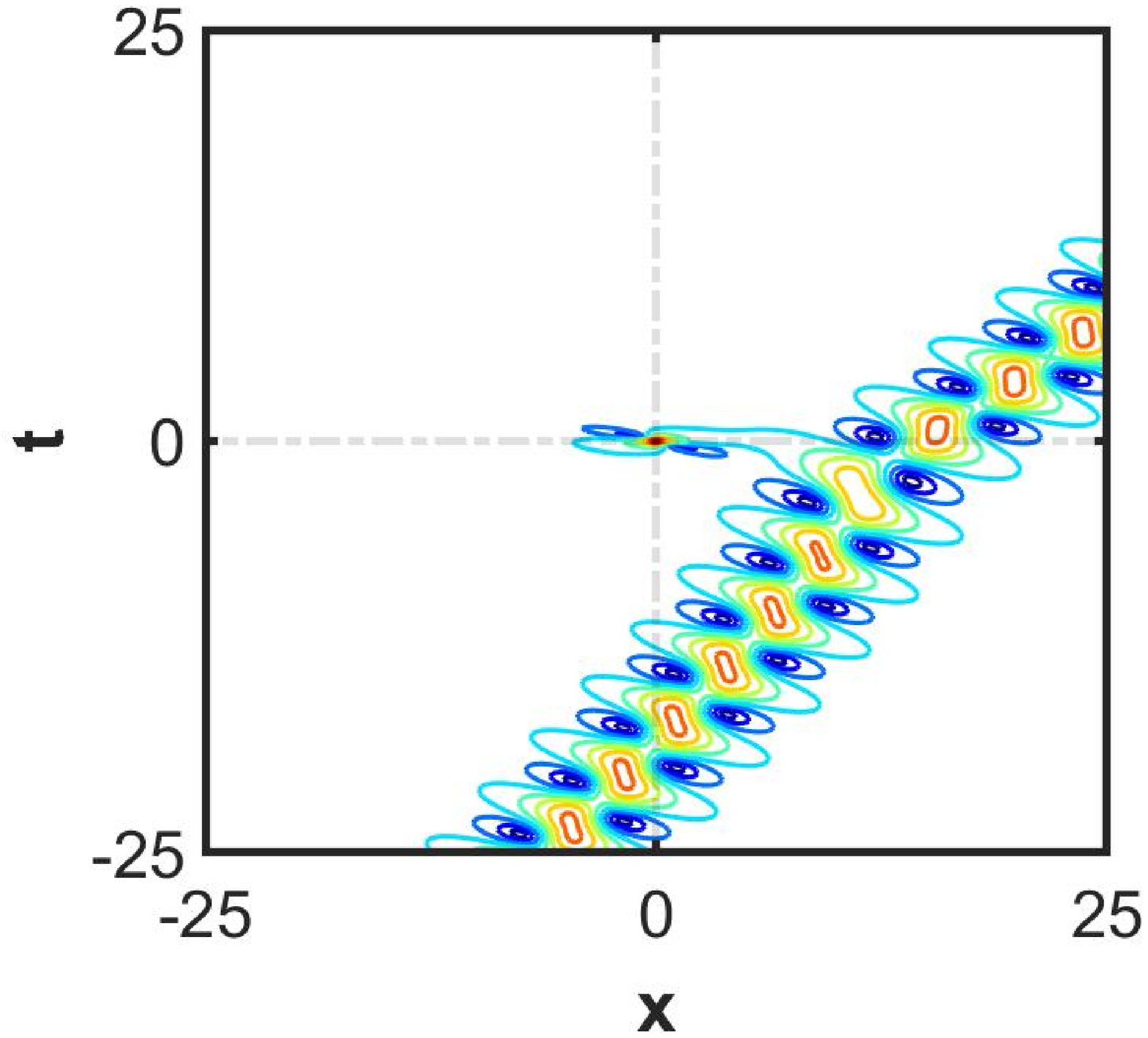}}\hspace{0.5cm}
\subfigure[]{\includegraphics[height=1.3in,width=1.6in]{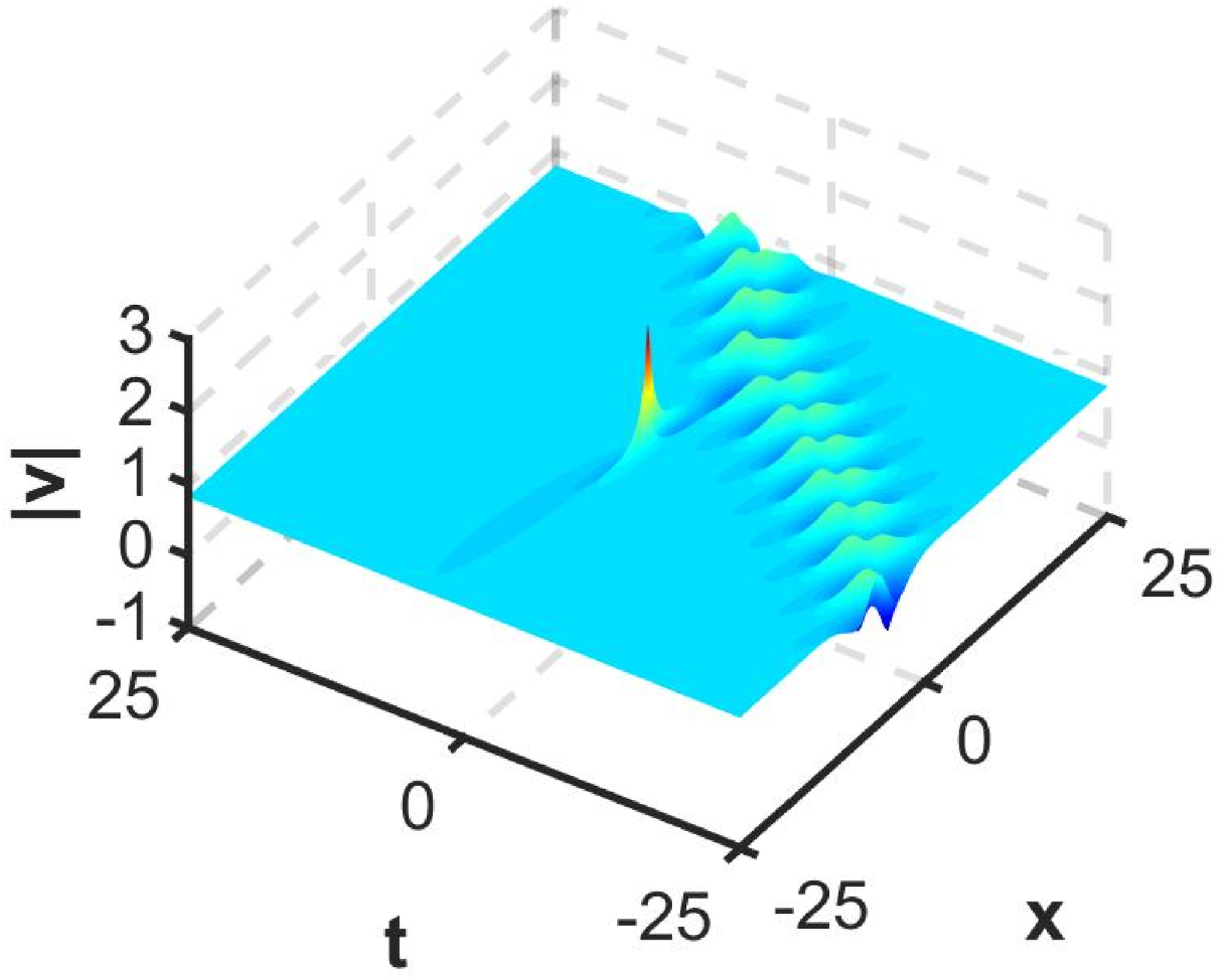}}\hspace{0.5cm}
\subfigure[]{\includegraphics[height=1.2in,width=1.3in]{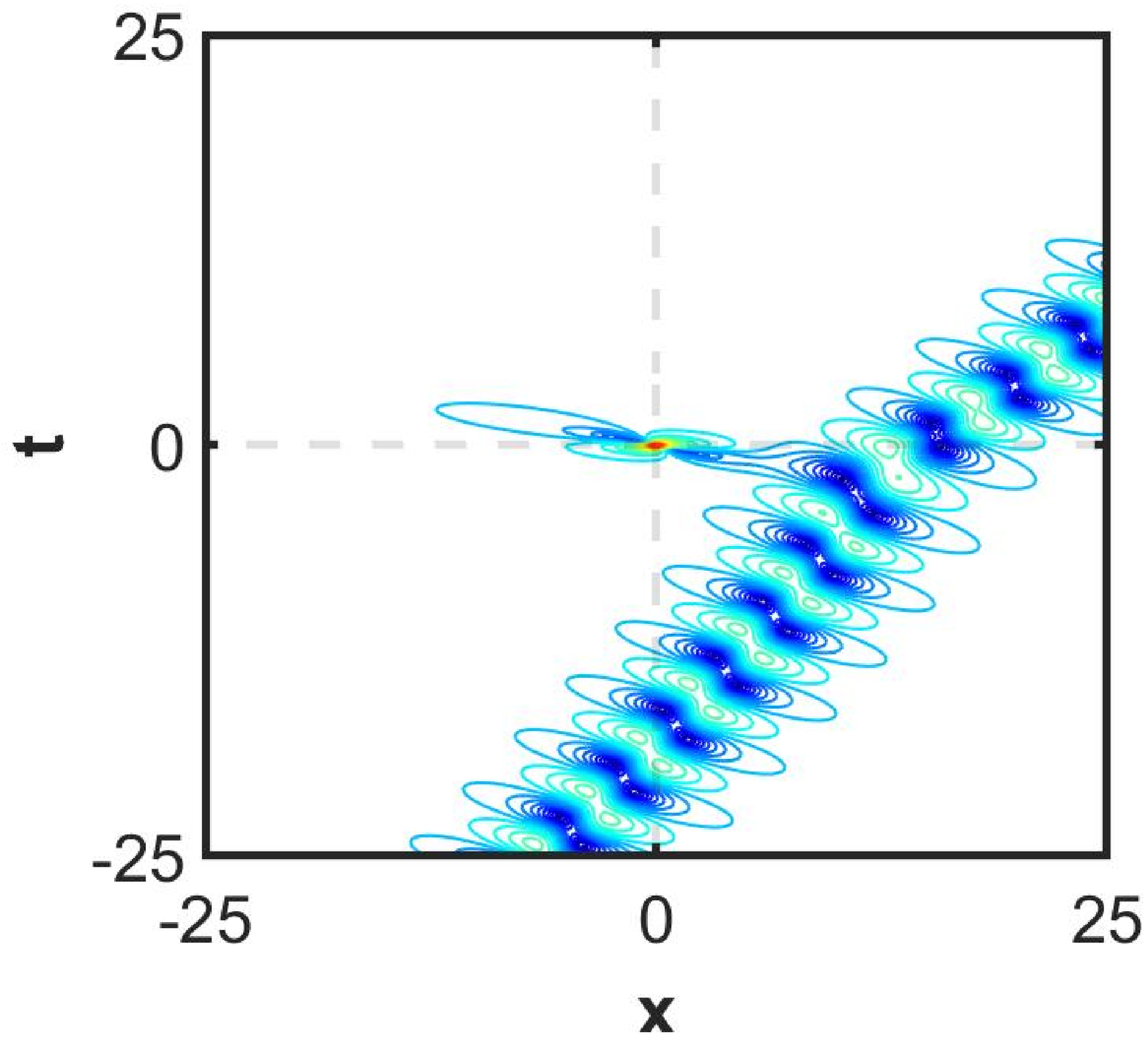}}\hspace{0.5cm}
\subfigure[]{\includegraphics[height=1.3in,width=1.6in]{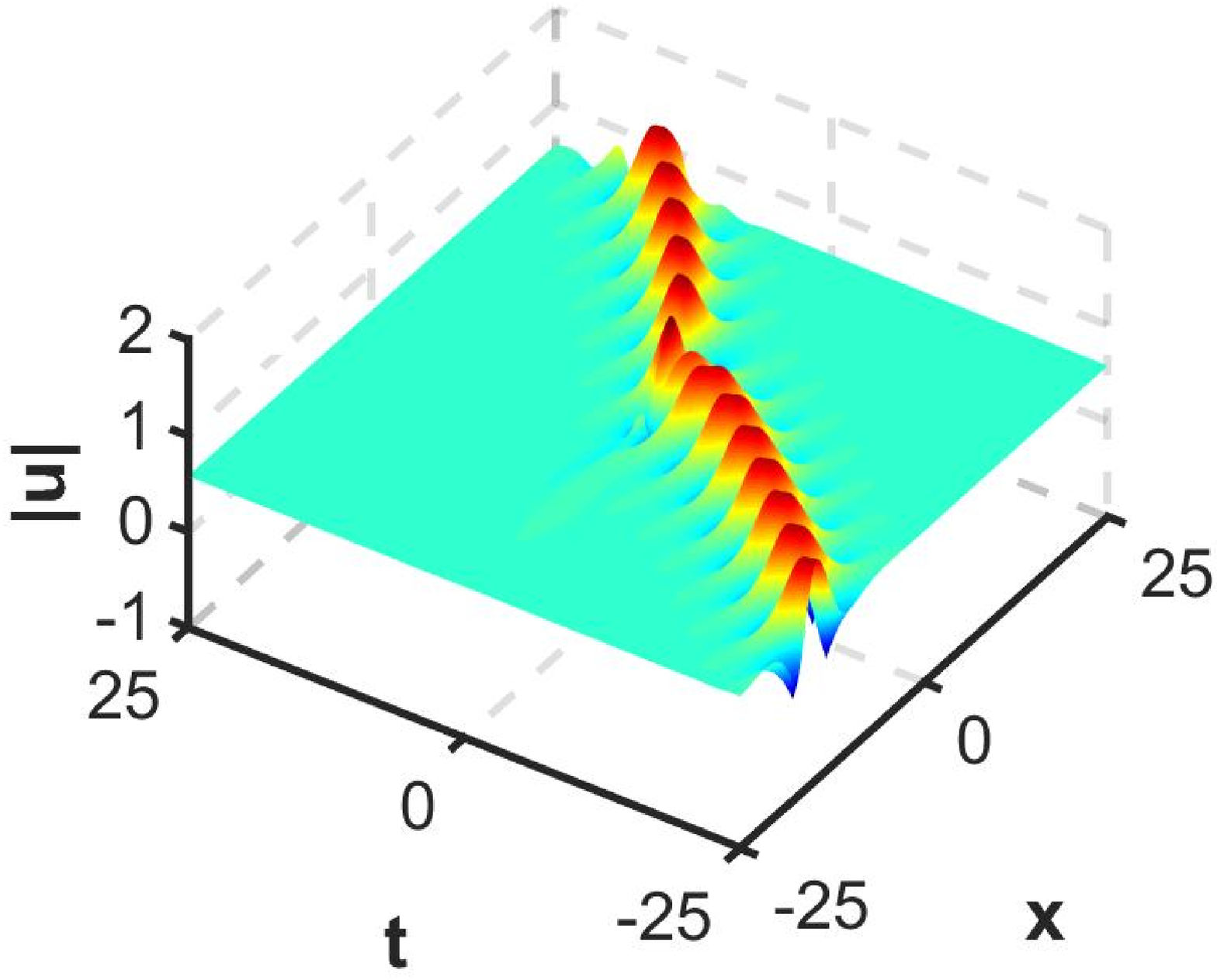}}\hspace{0.5cm}
\subfigure[]{\includegraphics[height=1.2in,width=1.3in]{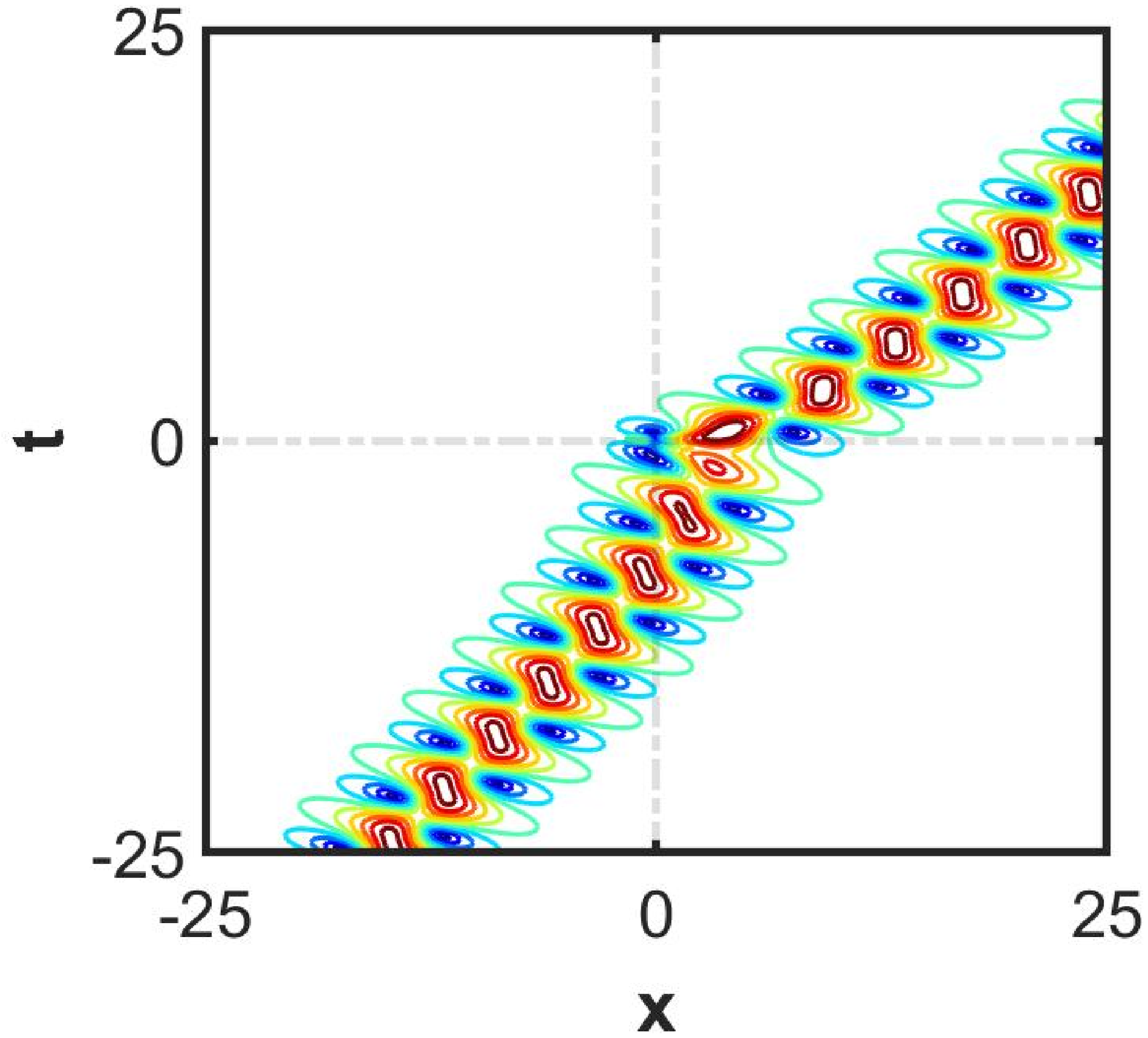}}\hspace{0.5cm}
\subfigure[]{\includegraphics[height=1.3in,width=1.6in]{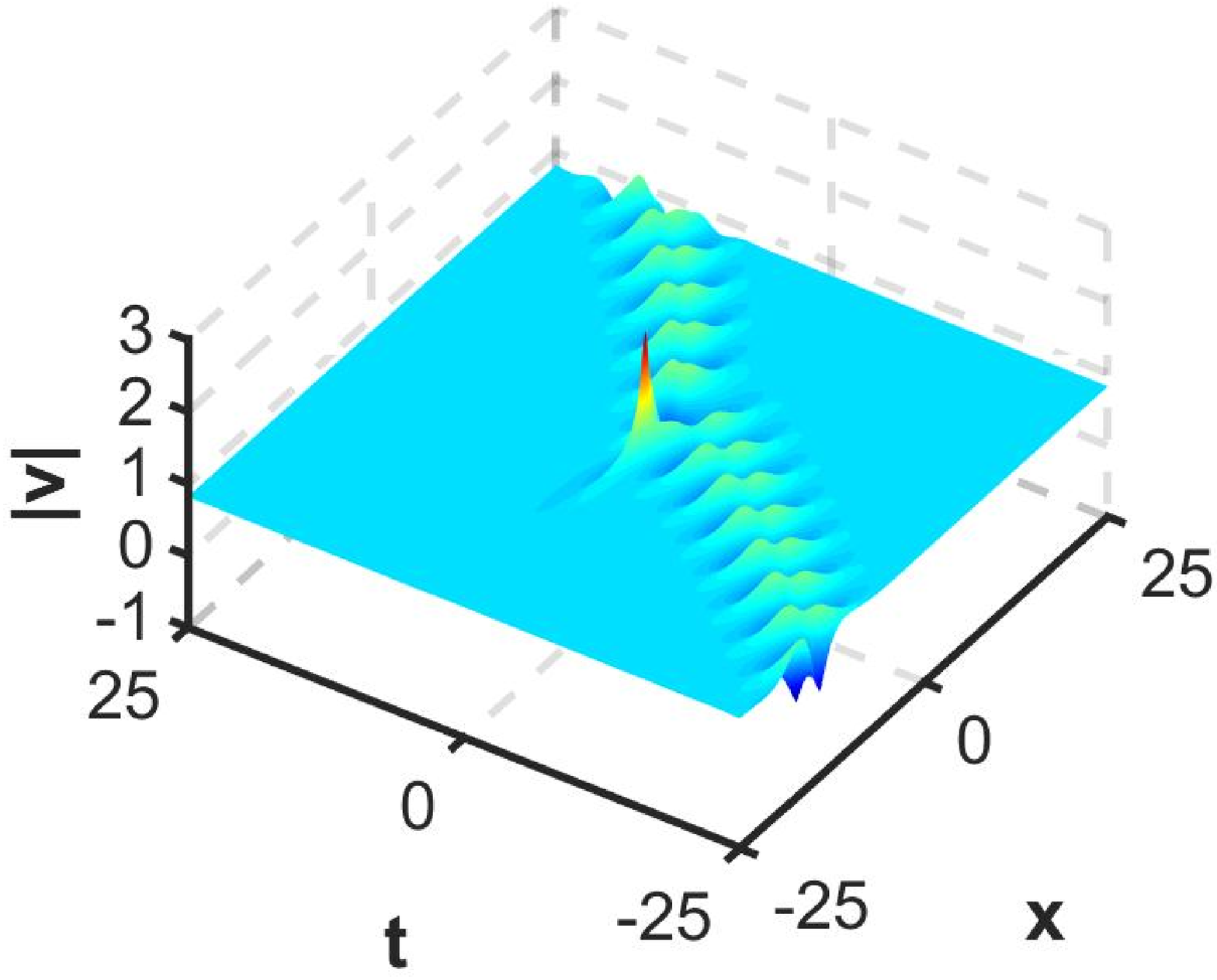}}\hspace{0.5cm}
\subfigure[]{\includegraphics[height=1.2in,width=1.3in]{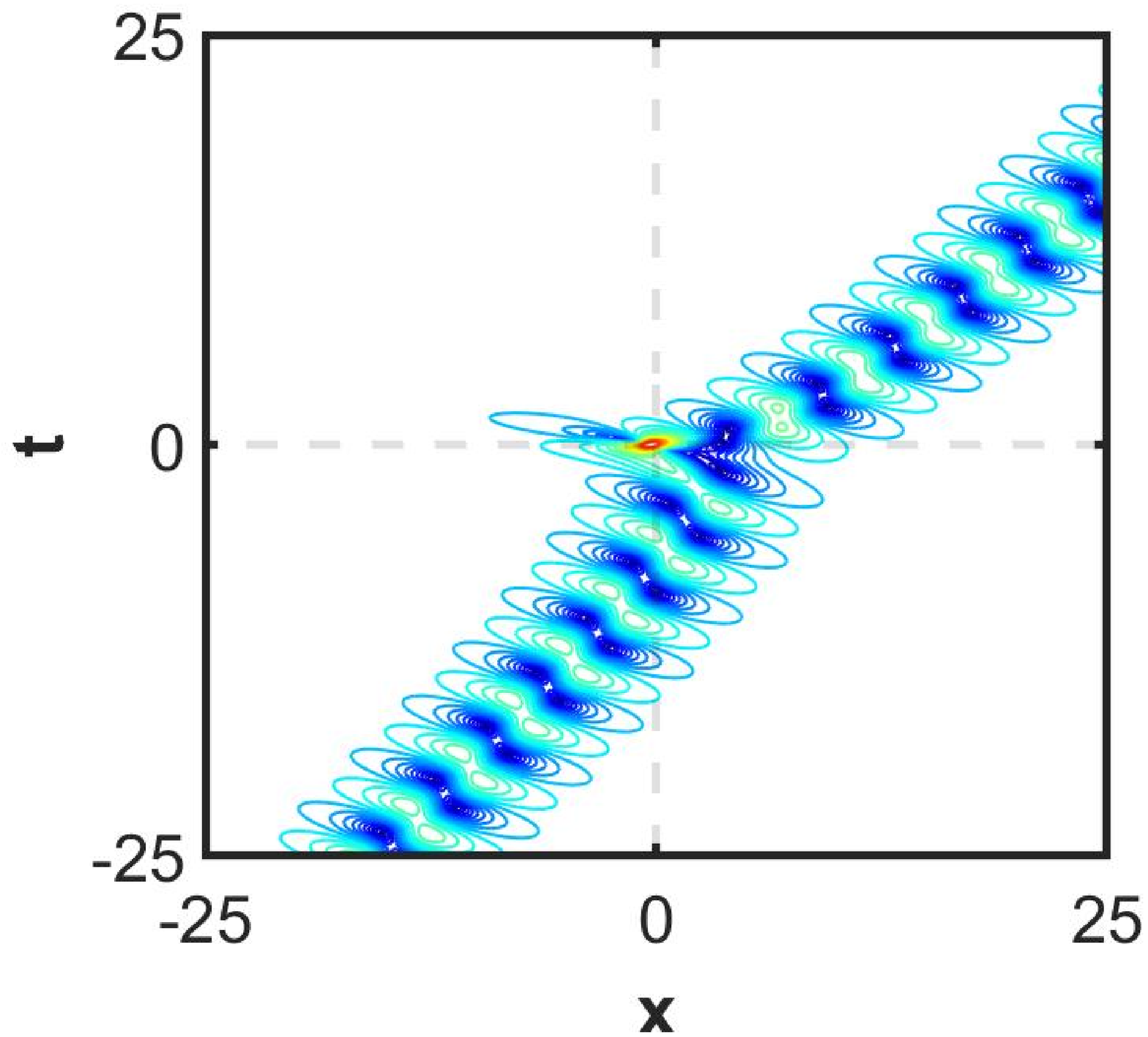}}
\caption{ The first-order interaction solution between a breather and a rogue wave of the gc-FL equation \eqref{gcfl-eq} with parameters: $(\alpha,\beta,\gamma,c_1,c_2)=(3,1,2,\frac{\sqrt{3}}{3},\frac{\sqrt{6}}{3})$.  Top line: 3D separate structure and corresponding contour pattern with $d=\frac{1}{100}$; Bottom line: 3D merge structure and corresponding contour pattern with $d=1$. \label{Fig-gcfl-br1}}
\end{figure*}

When parameters $(d,c_i)\neq(0,0)(i=1,2)$ in Eq. \eqref{gcfl-lws-10}, it gives rise to the semi-rational solution of the interaction between a breather and a rogue wave for the gc-FL equation. Here, the parameter $|d|$ plays the same role as in Case 2, namely, merge or separation between the breather and rogue wave during their interaction. The above dynamical behavior of the two components is shown in Fig. \ref{Fig-gcfl-br1}. Obviously, it can be seen from their contour patterns that the breather in the $u$ is different from one in $v$ component. The breather in the $u$ component has one peak and two valleys, while the $v$ component has two peaks and two valleys. When rogue wave and the breather are close to each other during the propagation process, it not only causes change in the amplitude, but also causes a certain phase deflection of the breather. When the interaction occurs, the energy of the rogue wave and breather solution will change. When variables $x$ and $t$ tend to infinity, it is not difficult to calculate the background amplitude ratio of the two components $u$ and $v$, which is $|\frac{c_1}{c_2}|$.


\begin{figure*}[!htbp]
\centering
\subfigure[]{\includegraphics[height=1.3in,width=1.6in]{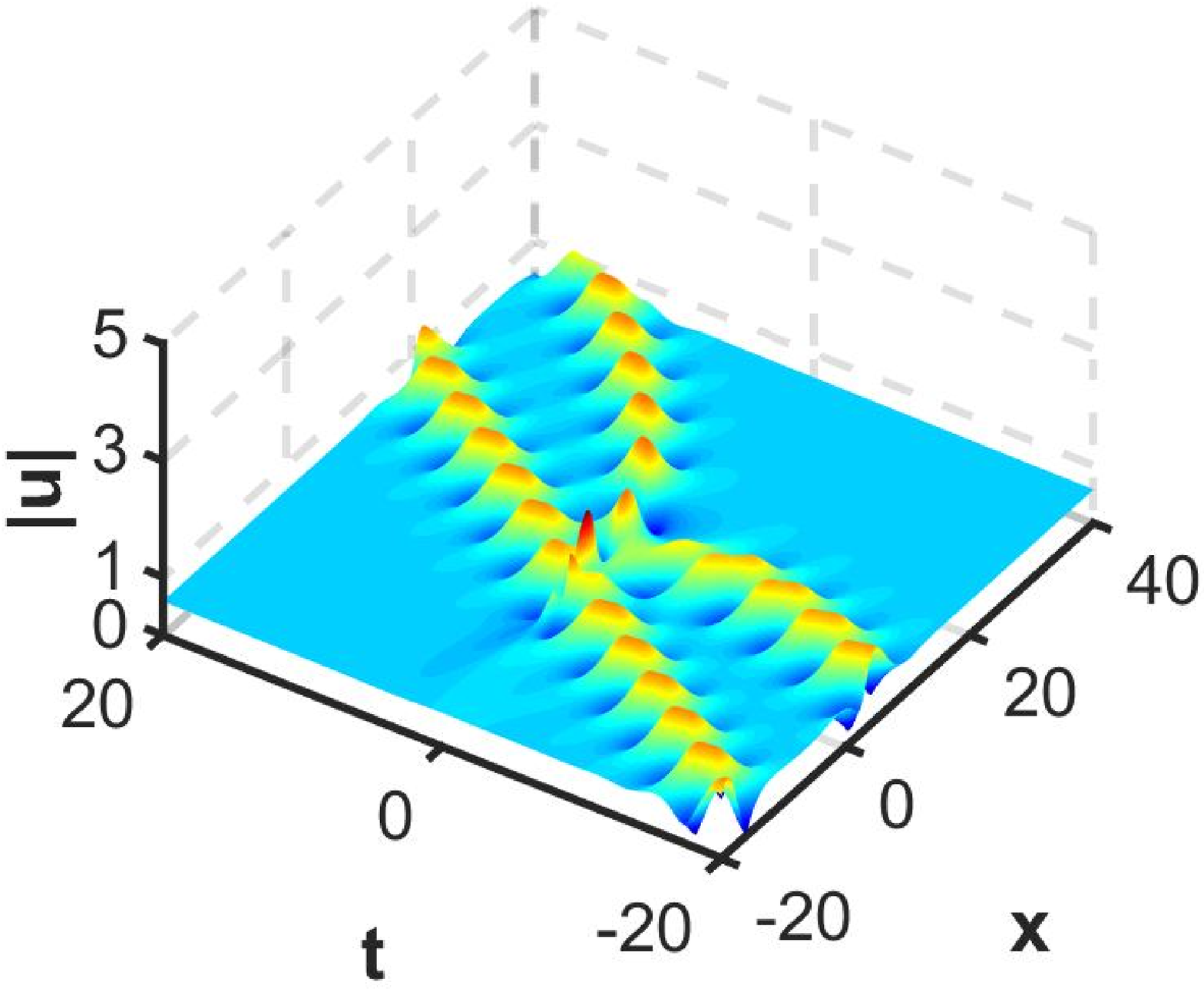}}\hspace{0.1cm}
\subfigure[]{\includegraphics[height=1.3in,width=1.6in]{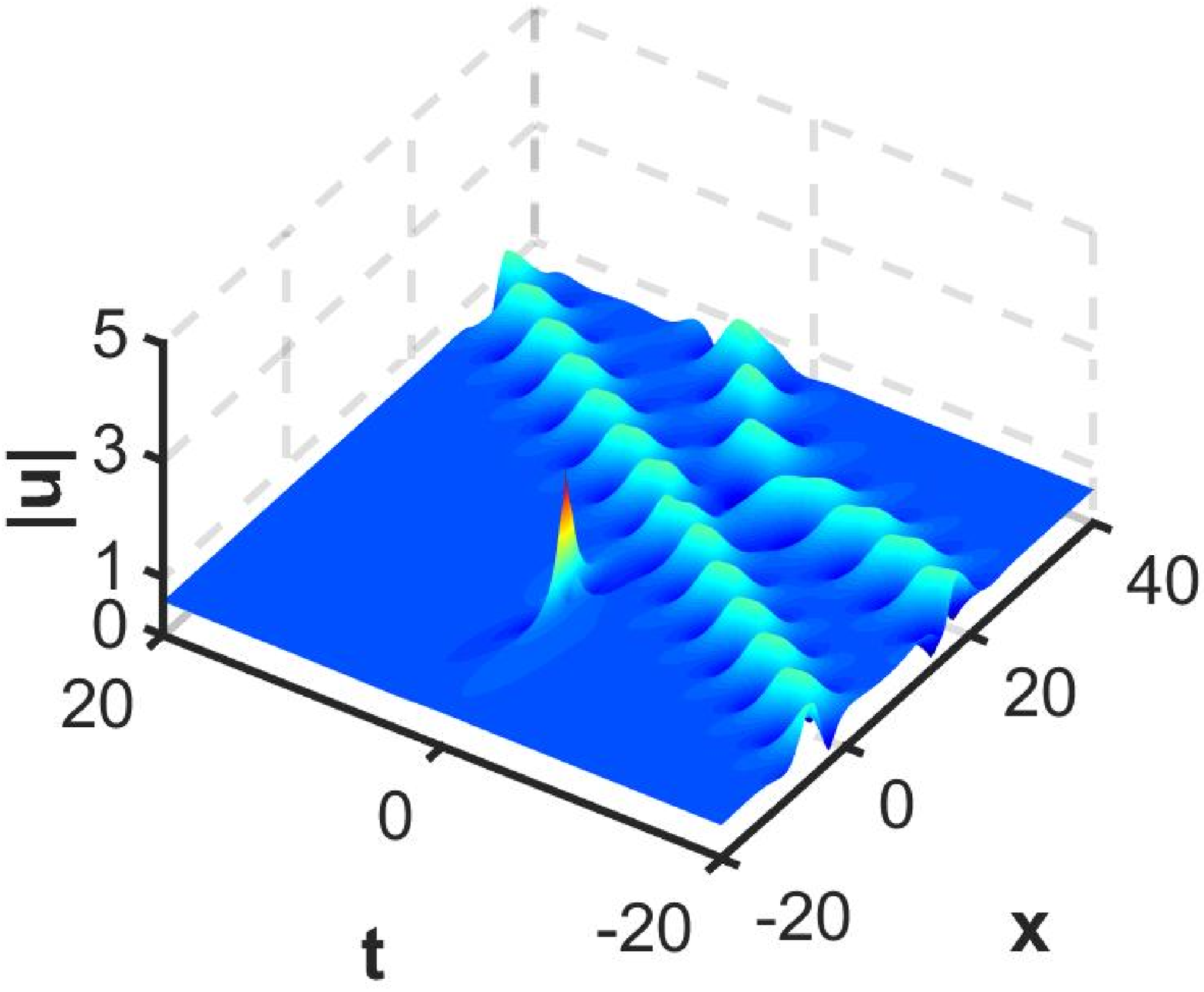}}\hspace{0.1cm}
\subfigure[]{\includegraphics[height=1.3in,width=1.6in]{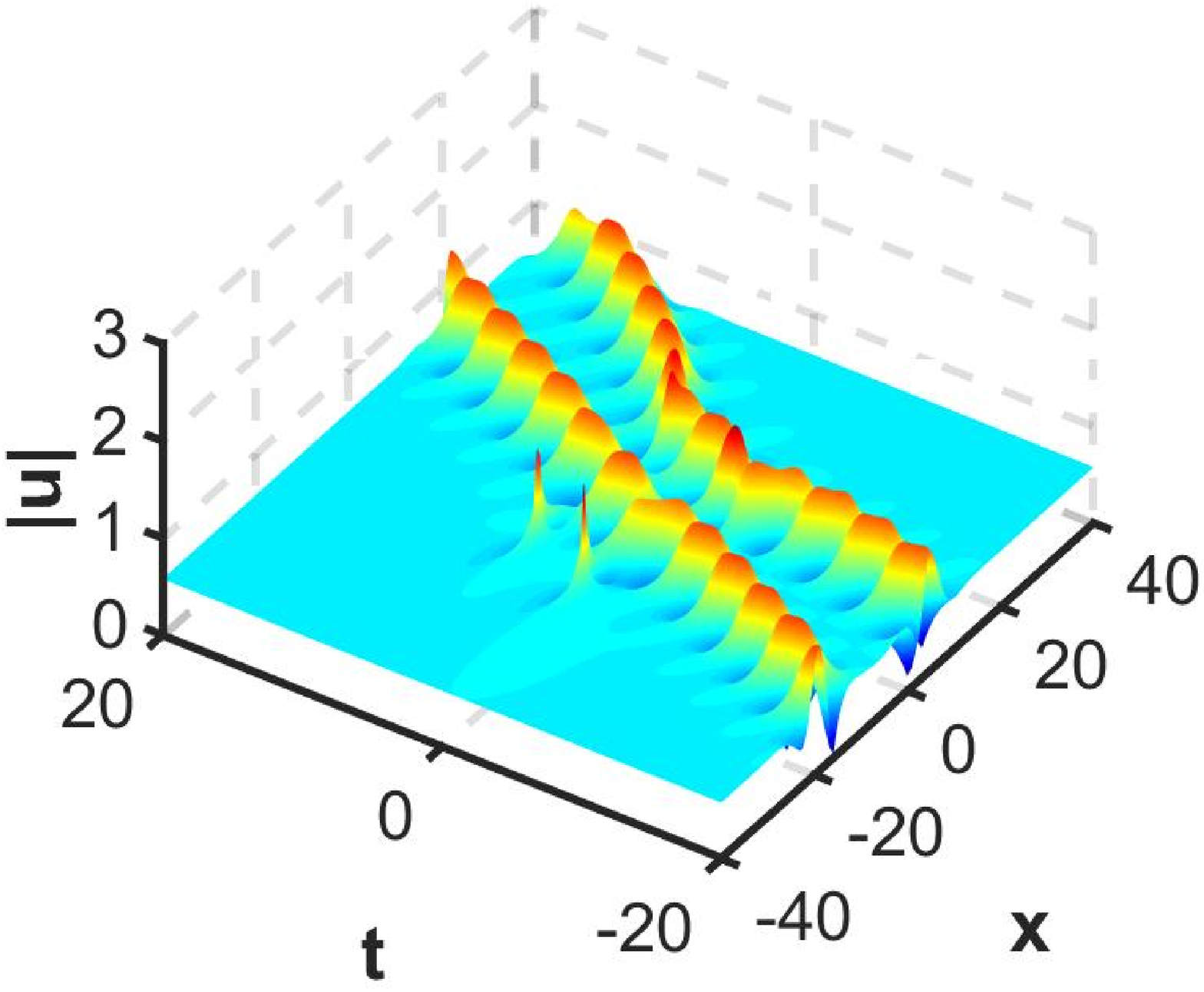}}\hspace{0.1cm}
\subfigure[]{\includegraphics[height=1.3in,width=1.6in]{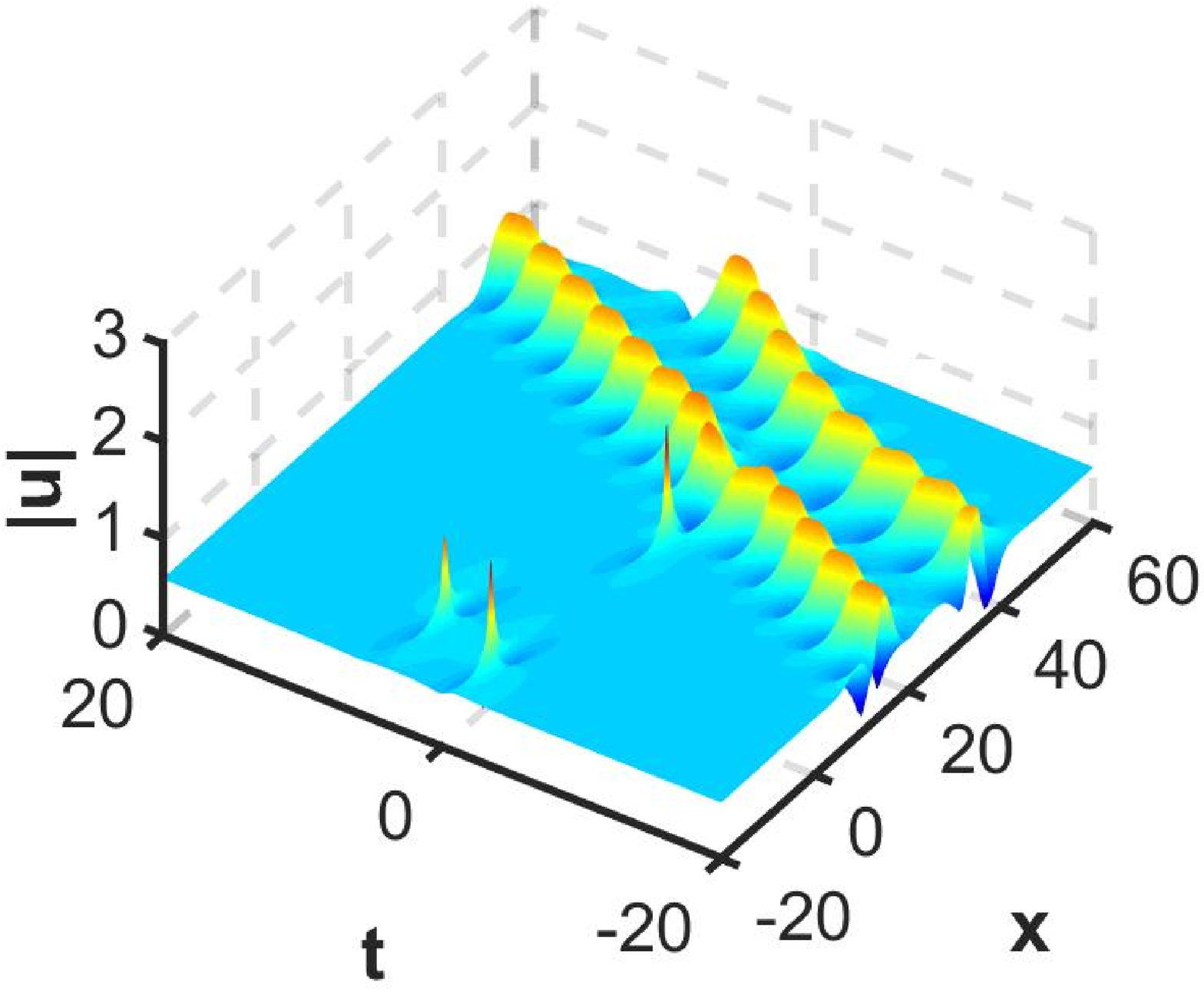}}\hspace{0.1cm}\\
\subfigure[]{\includegraphics[height=1.3in,width=1.6in]{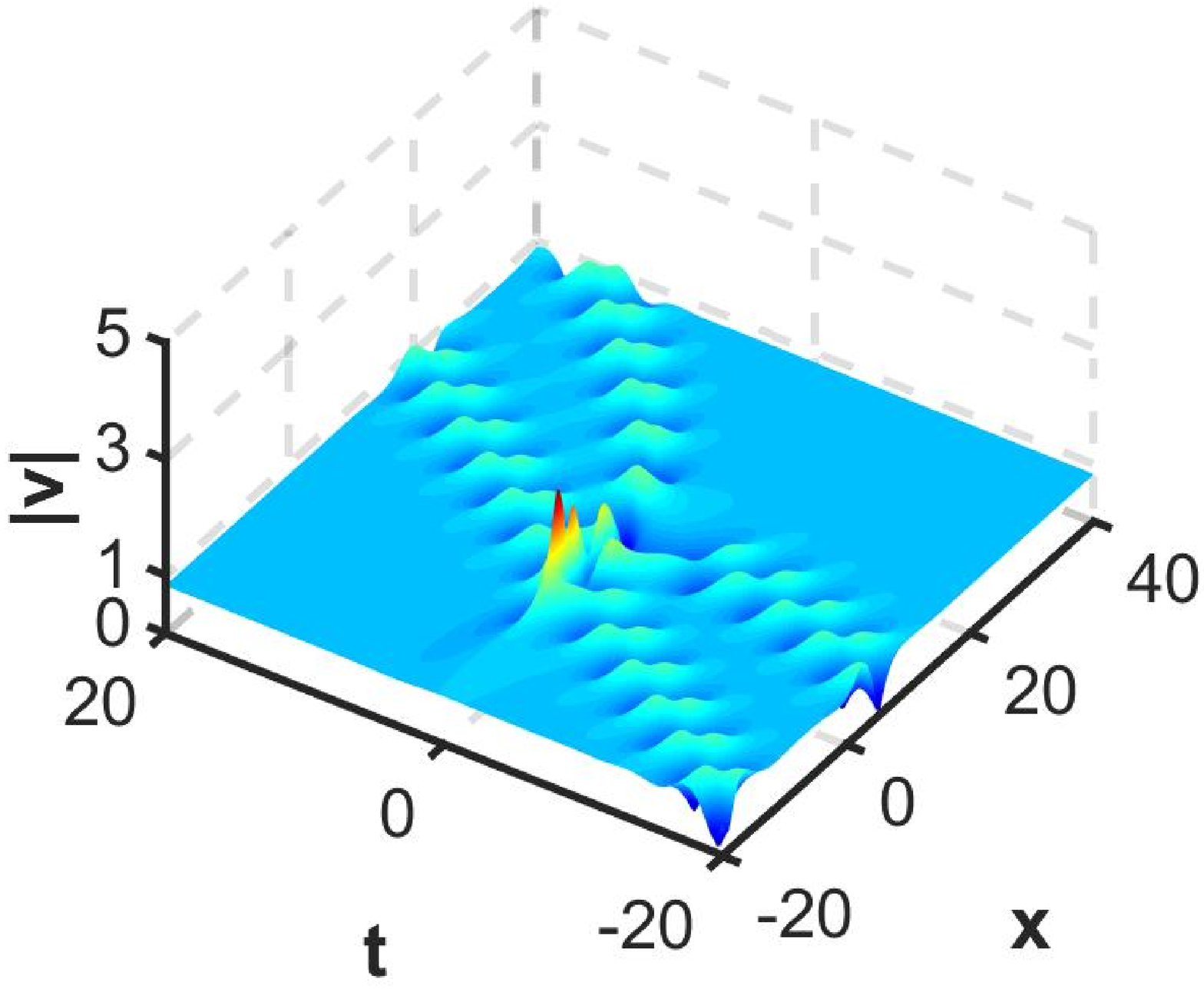}}\hspace{0.1cm}
\subfigure[]{\includegraphics[height=1.3in,width=1.6in]{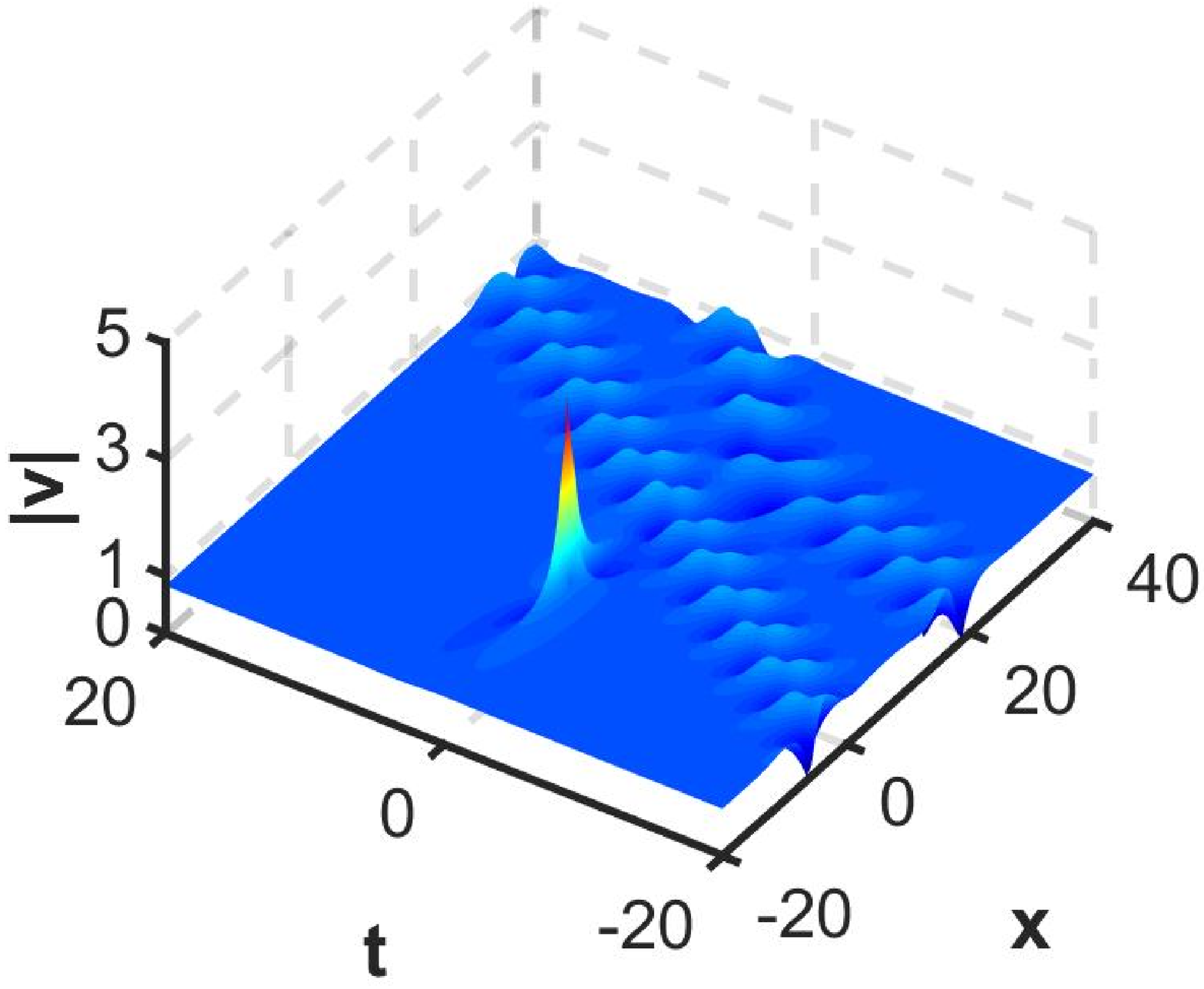}}\hspace{0.1cm}
\subfigure[]{\includegraphics[height=1.3in,width=1.6in]{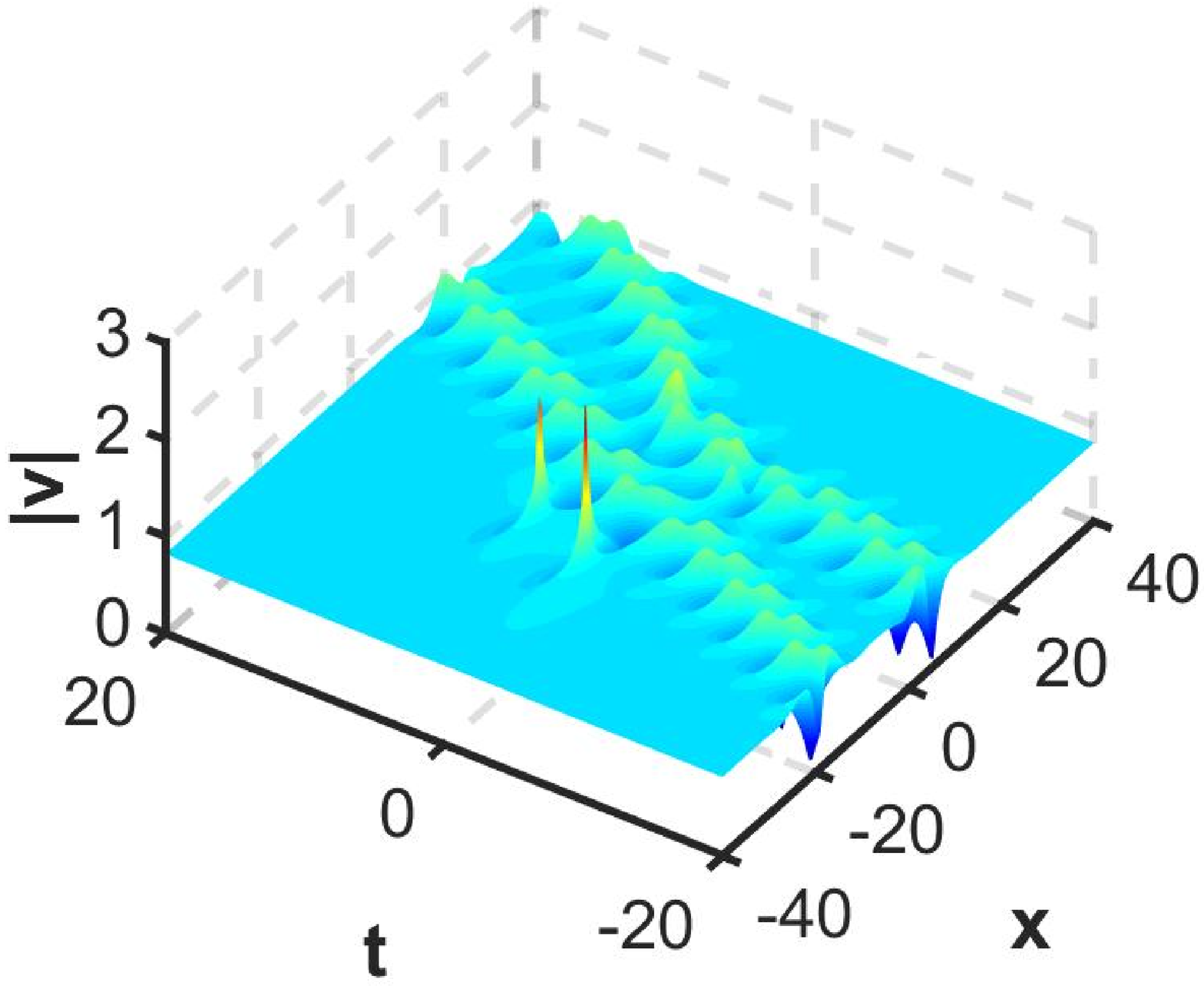}}\hspace{0.1cm}
\subfigure[]{\includegraphics[height=1.3in,width=1.6in]{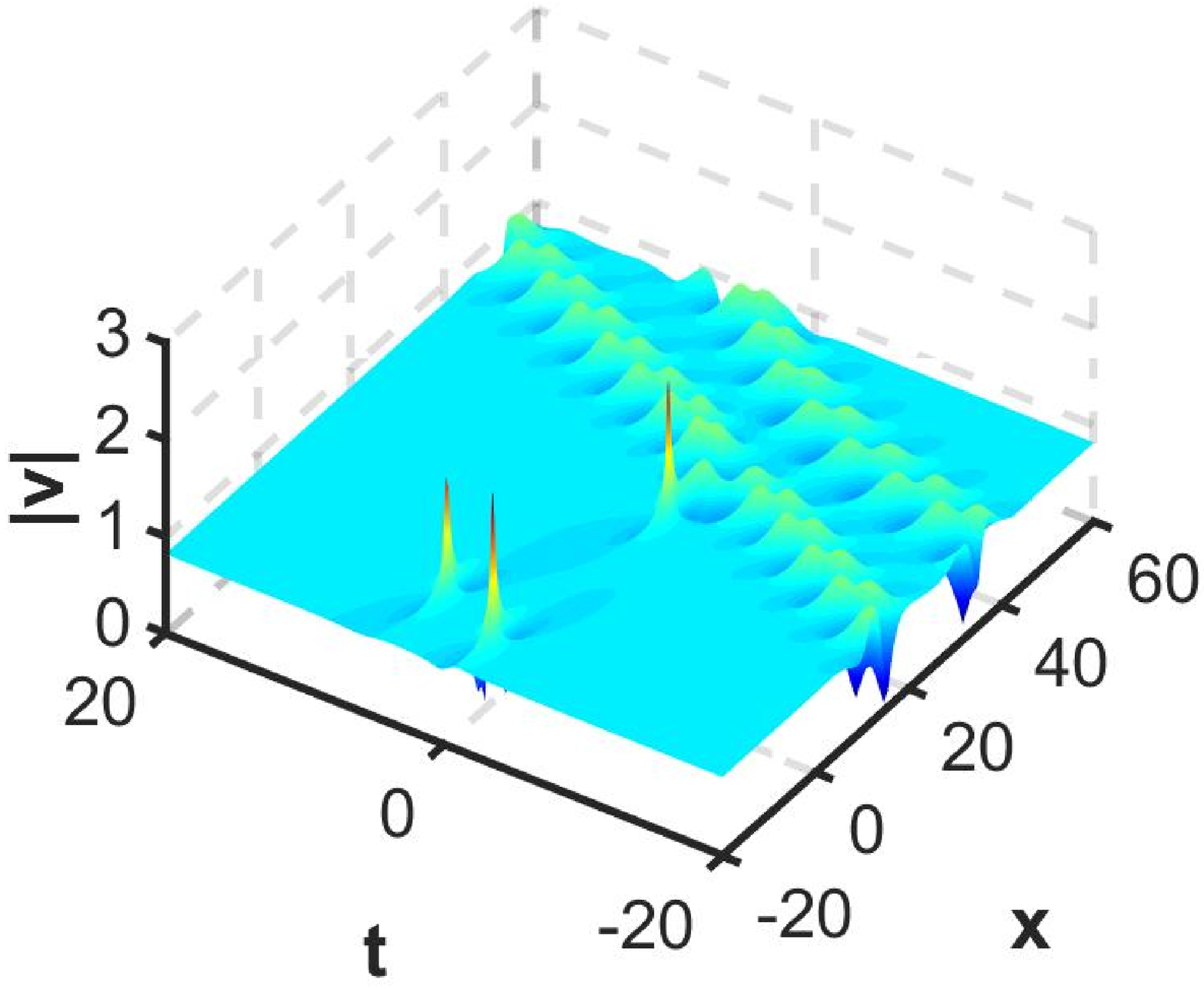}}
\caption{ The second-order interaction solution between two breathers and two-order rogue wave of the gc-FL equation \eqref{gcfl-eq} with parameters: $(\alpha,\beta,\gamma,c_1,c_2)=(3,1,2,\frac{\sqrt{3}}{3},\frac{\sqrt{6}}{3})$; From the left column to the right column the parameters $(d,m_1,n_1)$ are (1,0,0), $(\frac{1}{1000},0,0)$, (1,1000,0) and $(\frac{1}{10000000},1000,0)$, respectively. The odd columns show merge structure; the even columns show separate structure.\label{Fig-gcfl-br2}}
\end{figure*}
\begin{figure*}[!htbp]
\centering
\subfigure[]{\includegraphics[height=1.2in,width=1.5in]{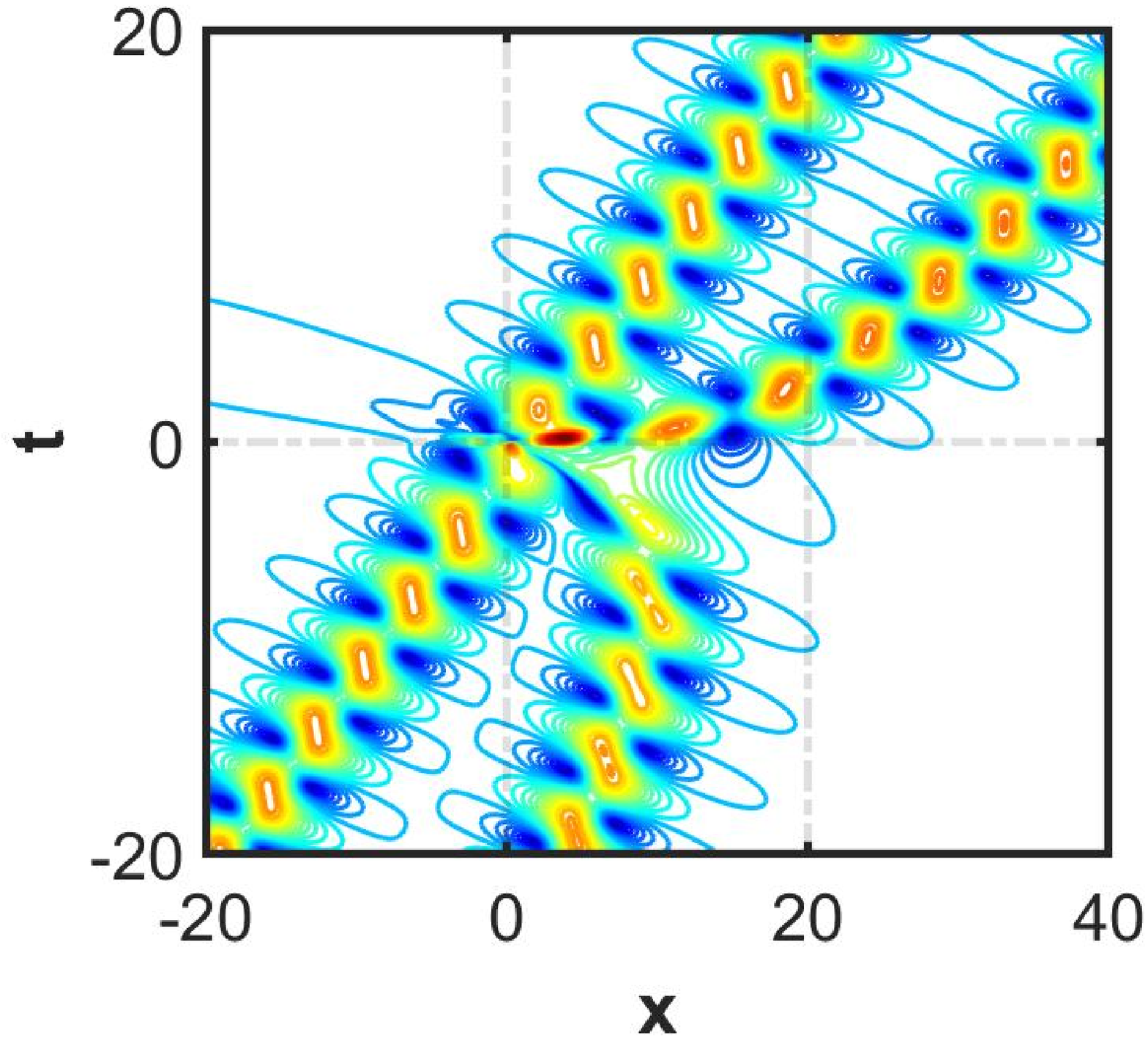}}\hspace{0.5cm}
\subfigure[]{\includegraphics[height=1.2in,width=1.5in]{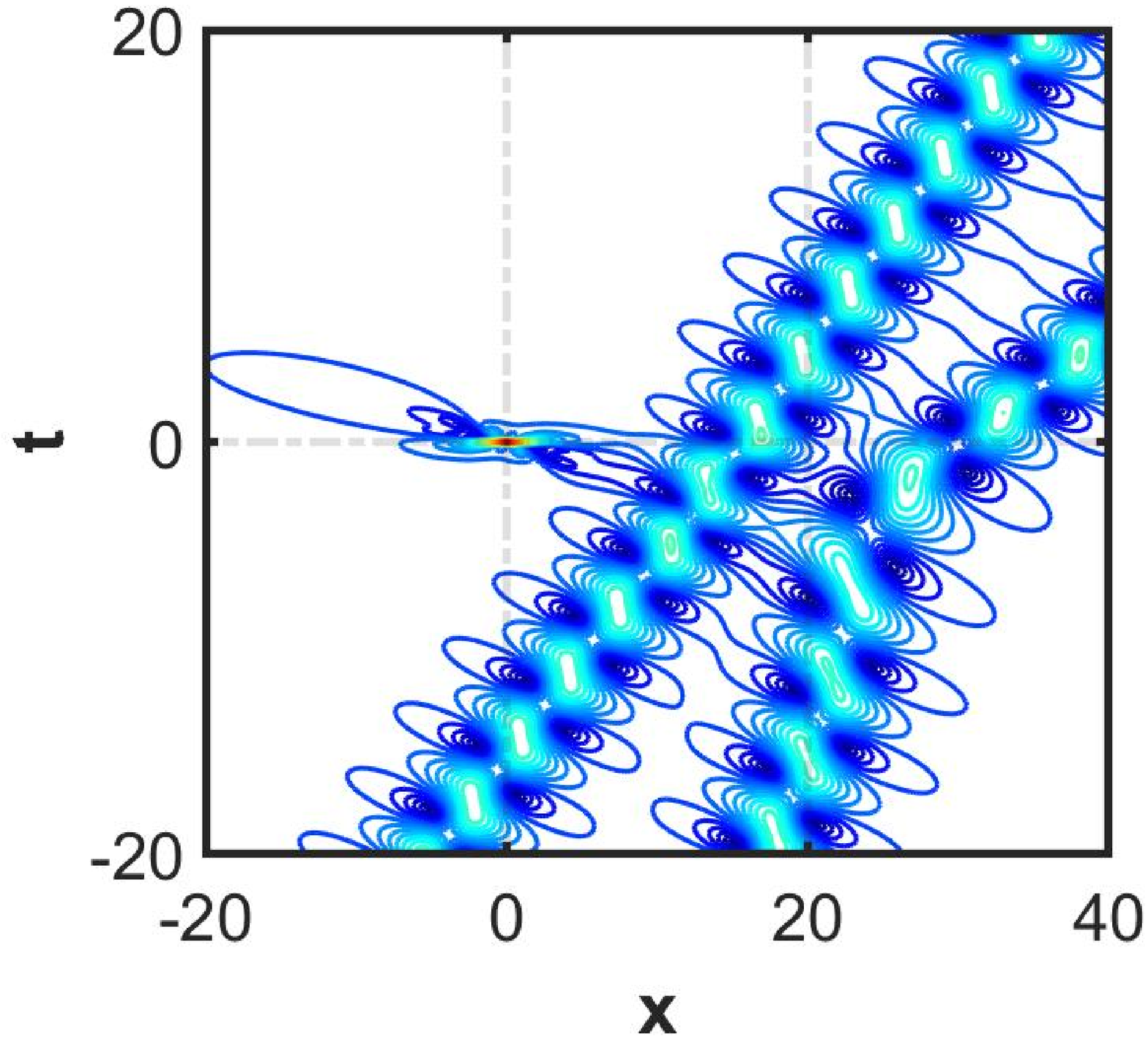}}\hspace{0.5cm}
\subfigure[]{\includegraphics[height=1.2in,width=1.5in]{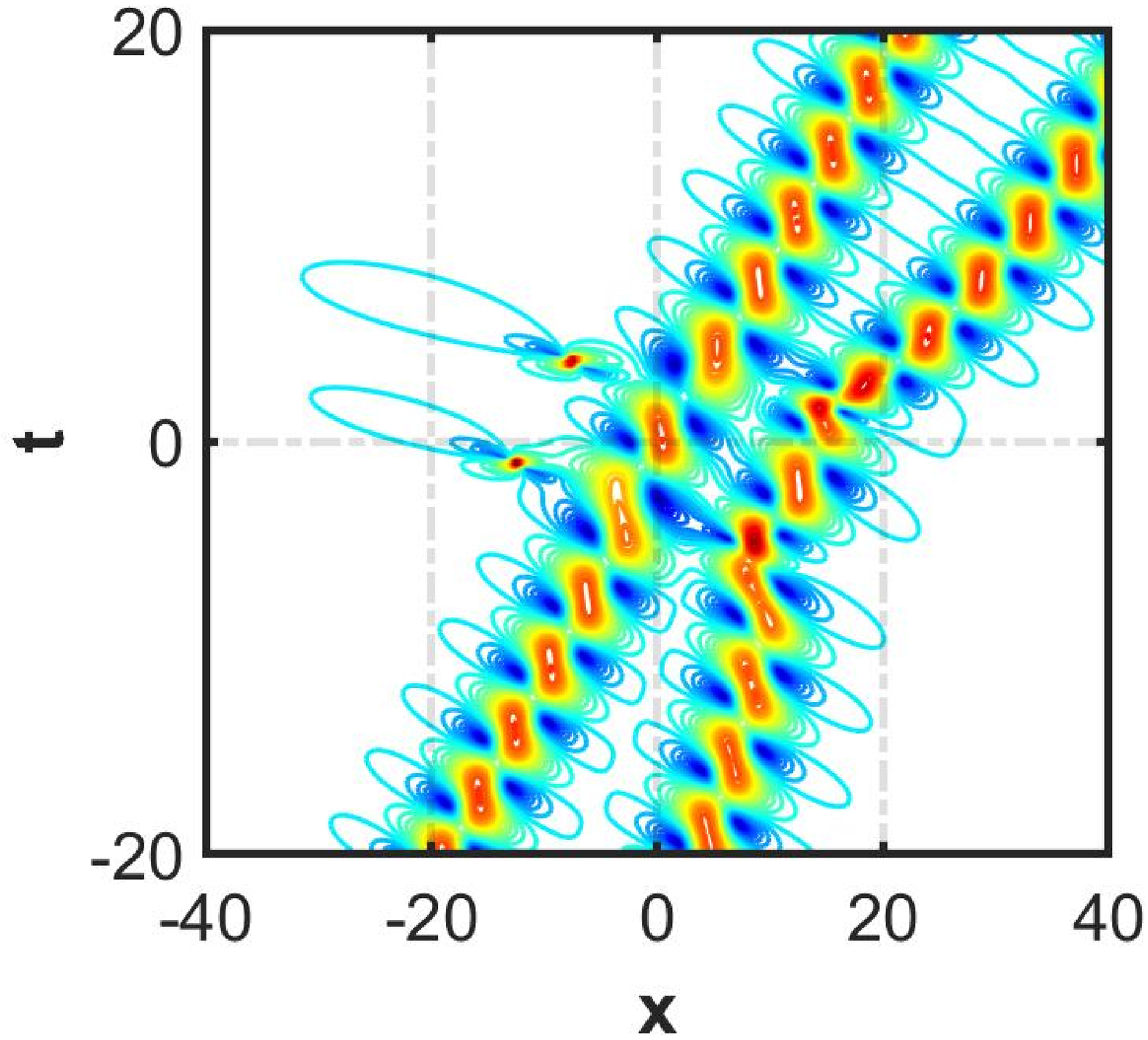}}\hspace{0.5cm}
\subfigure[]{\includegraphics[height=1.2in,width=1.5in]{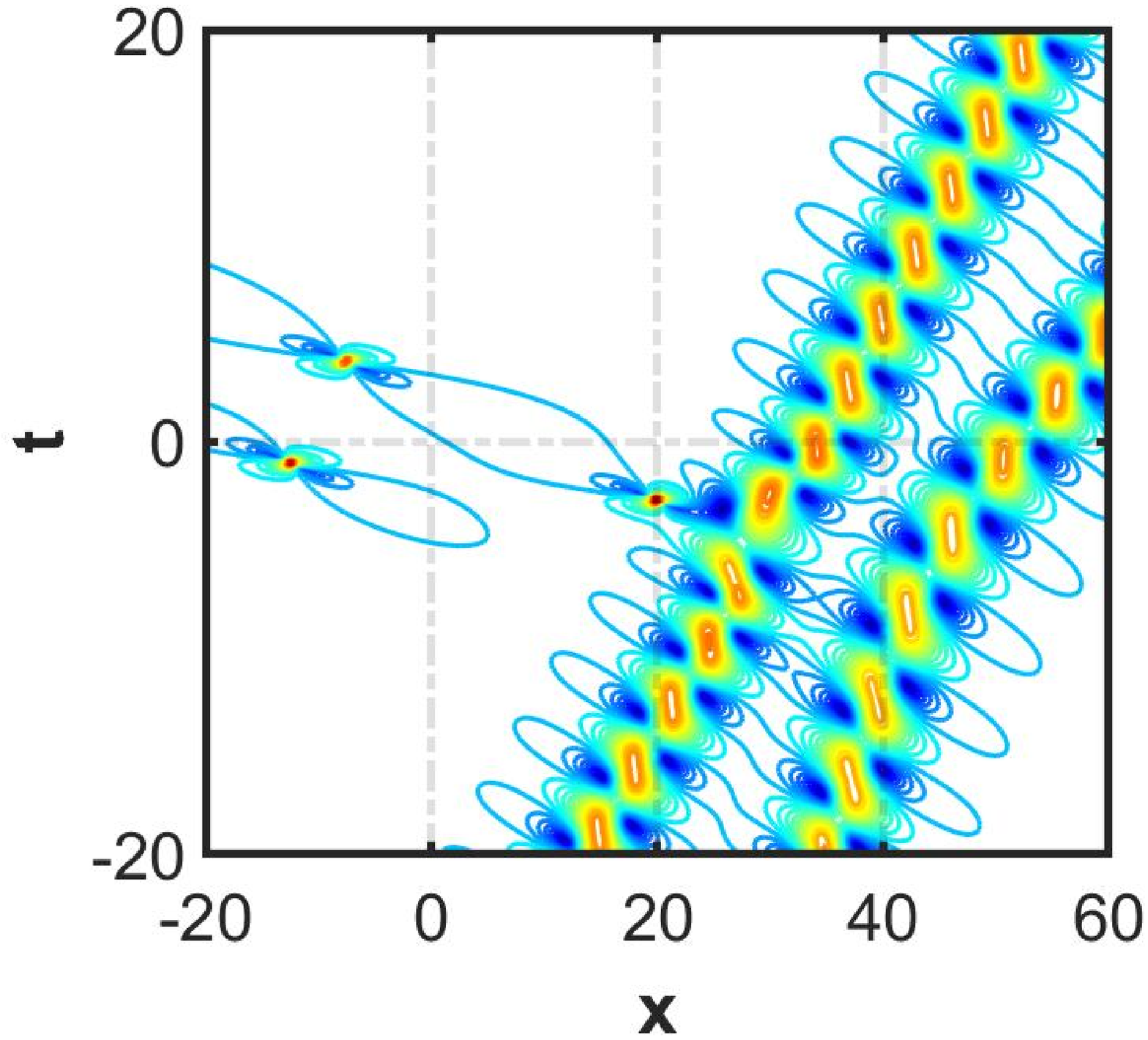}}\hspace{0.5cm}\\
\subfigure[]{\includegraphics[height=1.2in,width=1.5in]{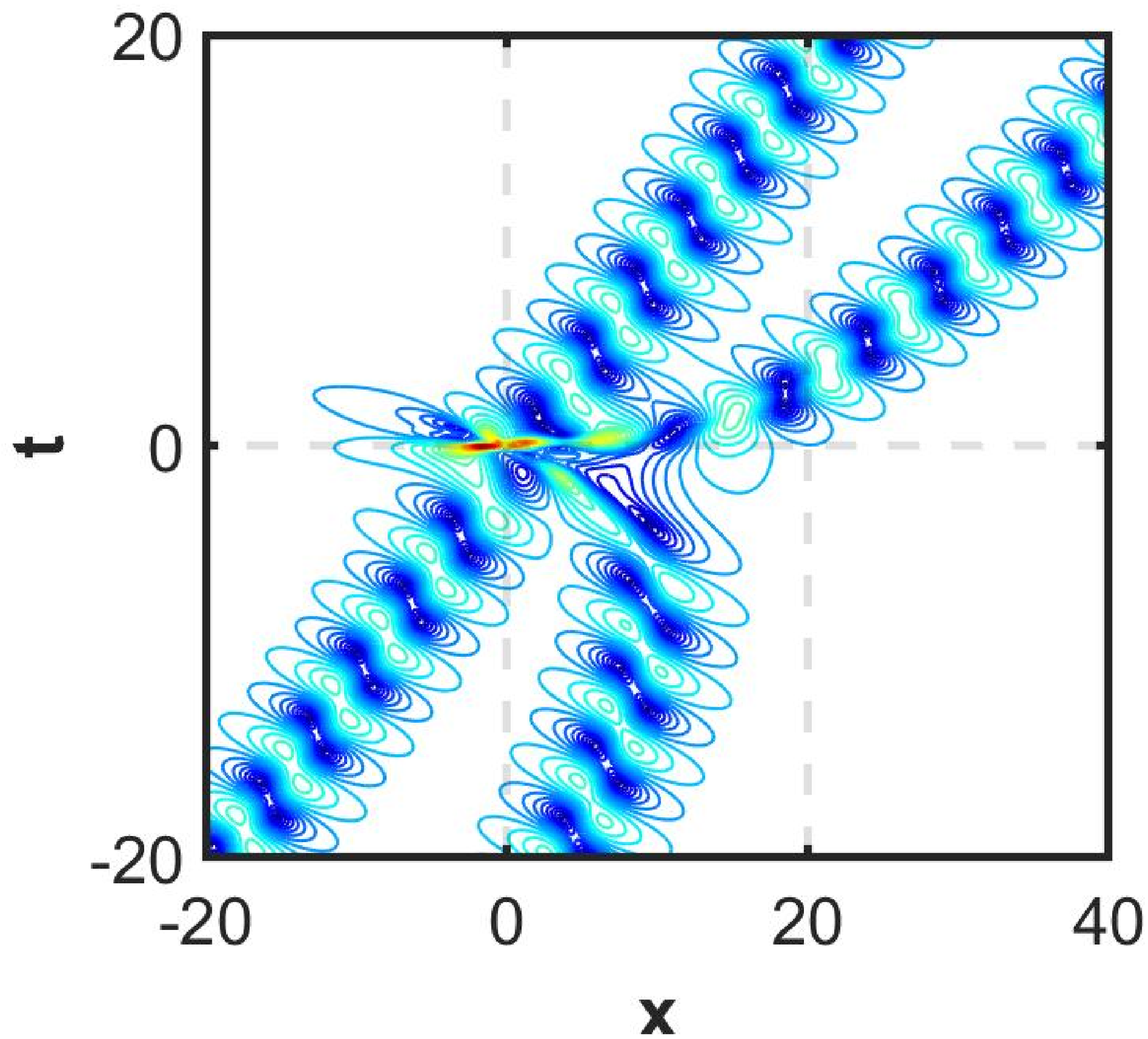}}\hspace{0.5cm}
\subfigure[]{\includegraphics[height=1.2in,width=1.5in]{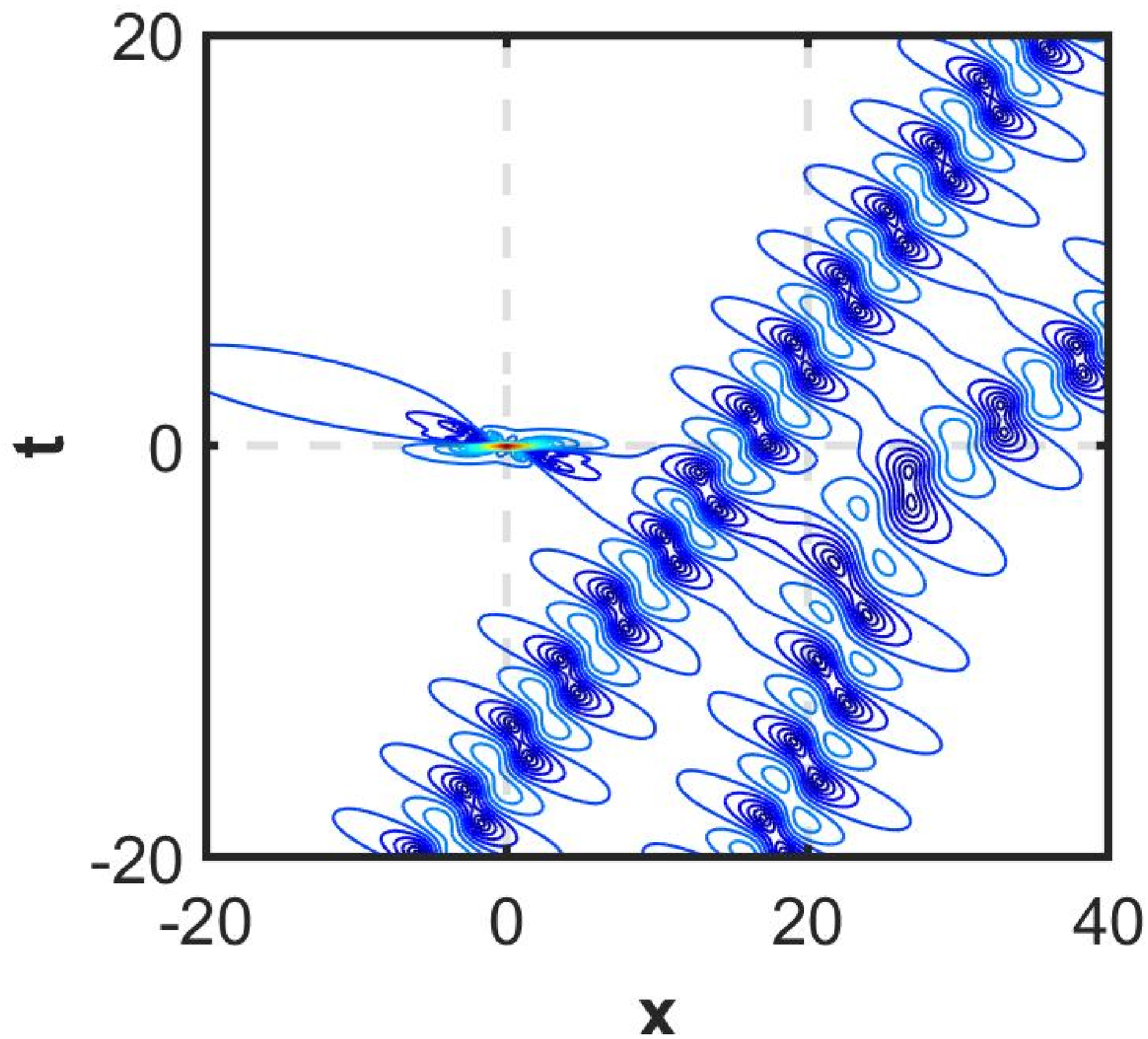}}\hspace{0.5cm}
\subfigure[]{\includegraphics[height=1.2in,width=1.5in]{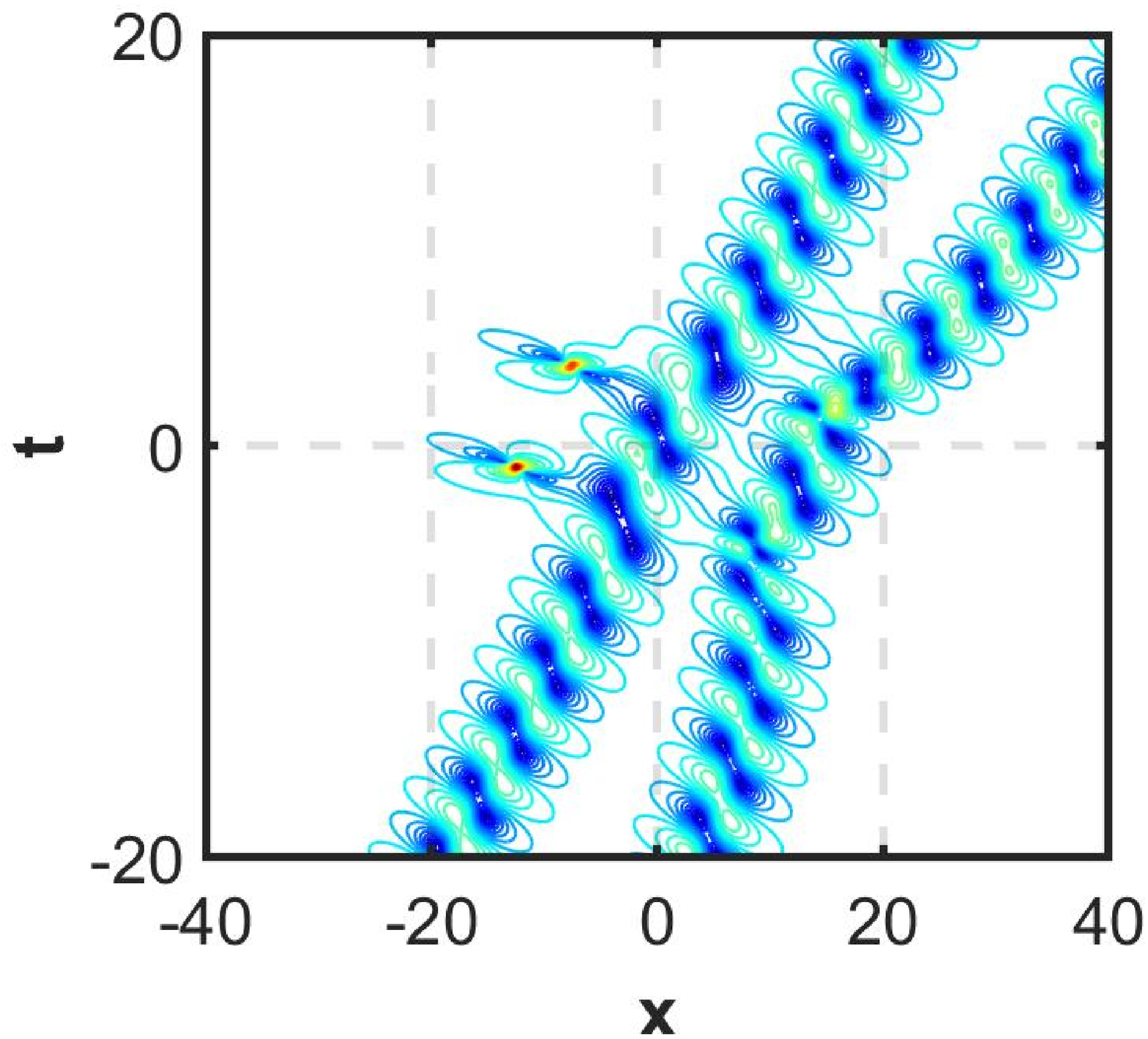}}\hspace{0.5cm}
\subfigure[]{\includegraphics[height=1.2in,width=1.5in]{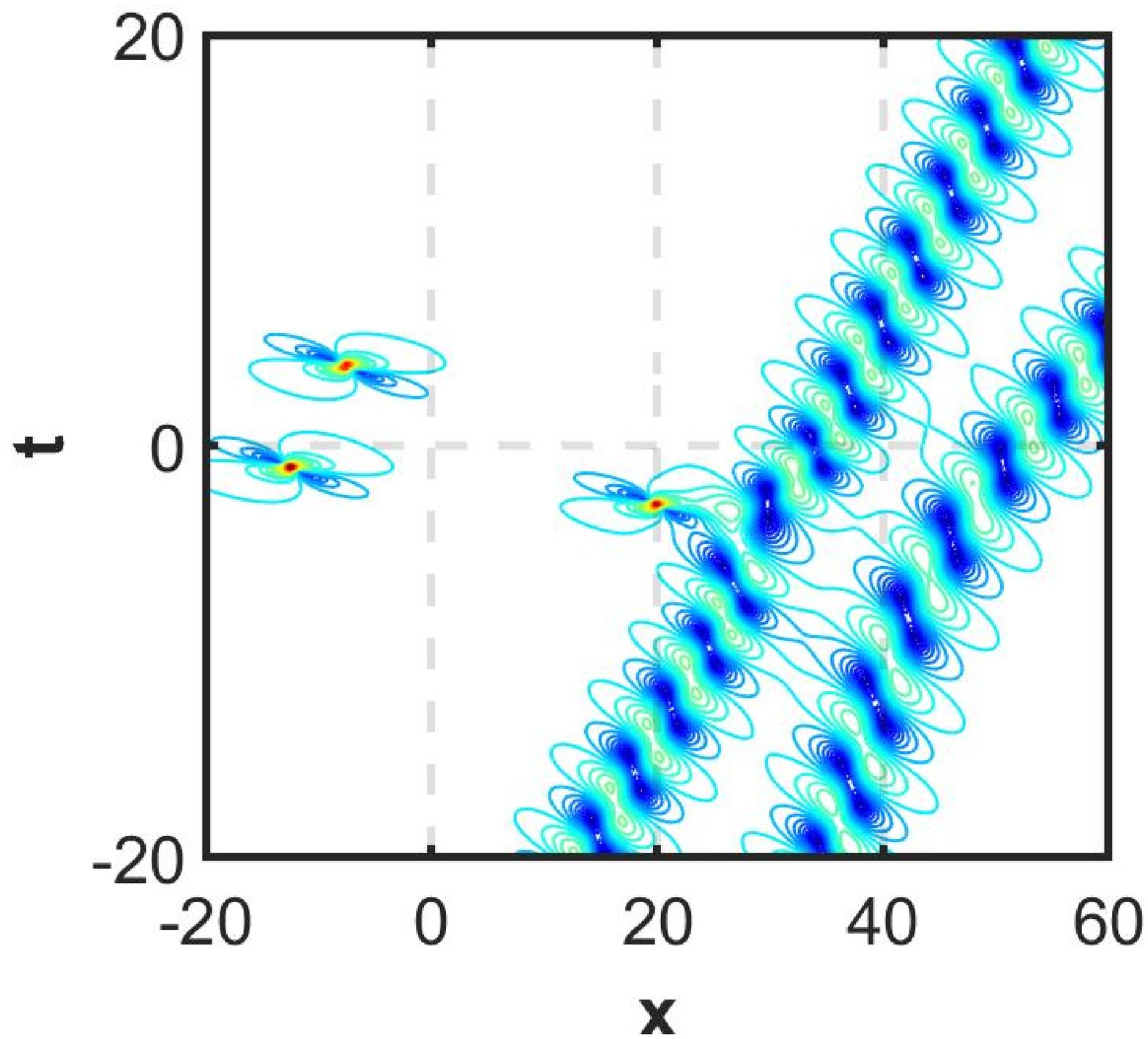}}
\caption{ The corresponding contour patterns of Fig. \ref{Fig-gcfl-br2}.\label{Fig-gcfl-br2c}}
\end{figure*}

\begin{figure*}[!htbp]
\centering
\subfigure[]{\includegraphics[height=1.5in,width=1.9in]{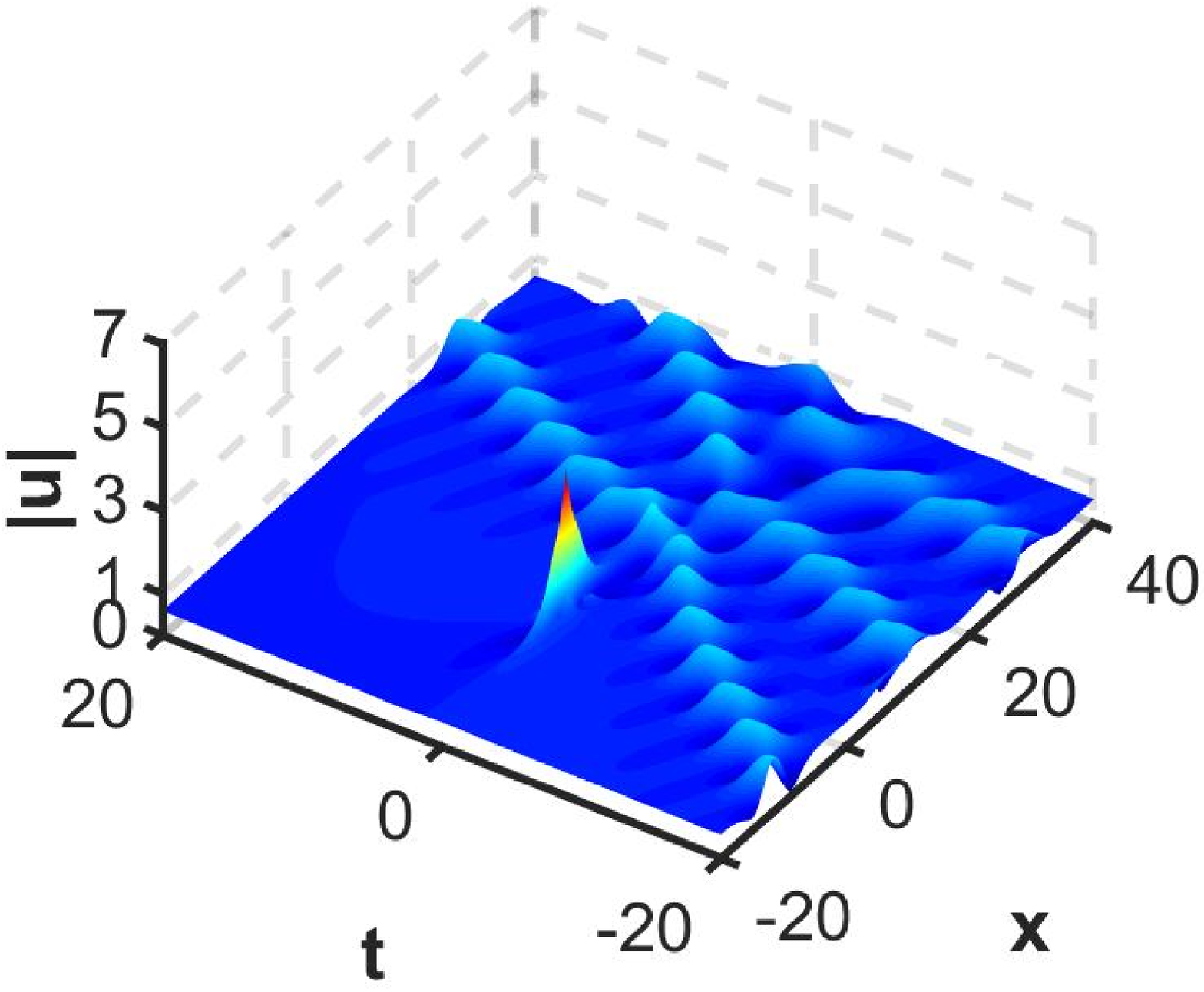}}\hspace{0.5cm}
\subfigure[]{\includegraphics[height=1.5in,width=1.9in]{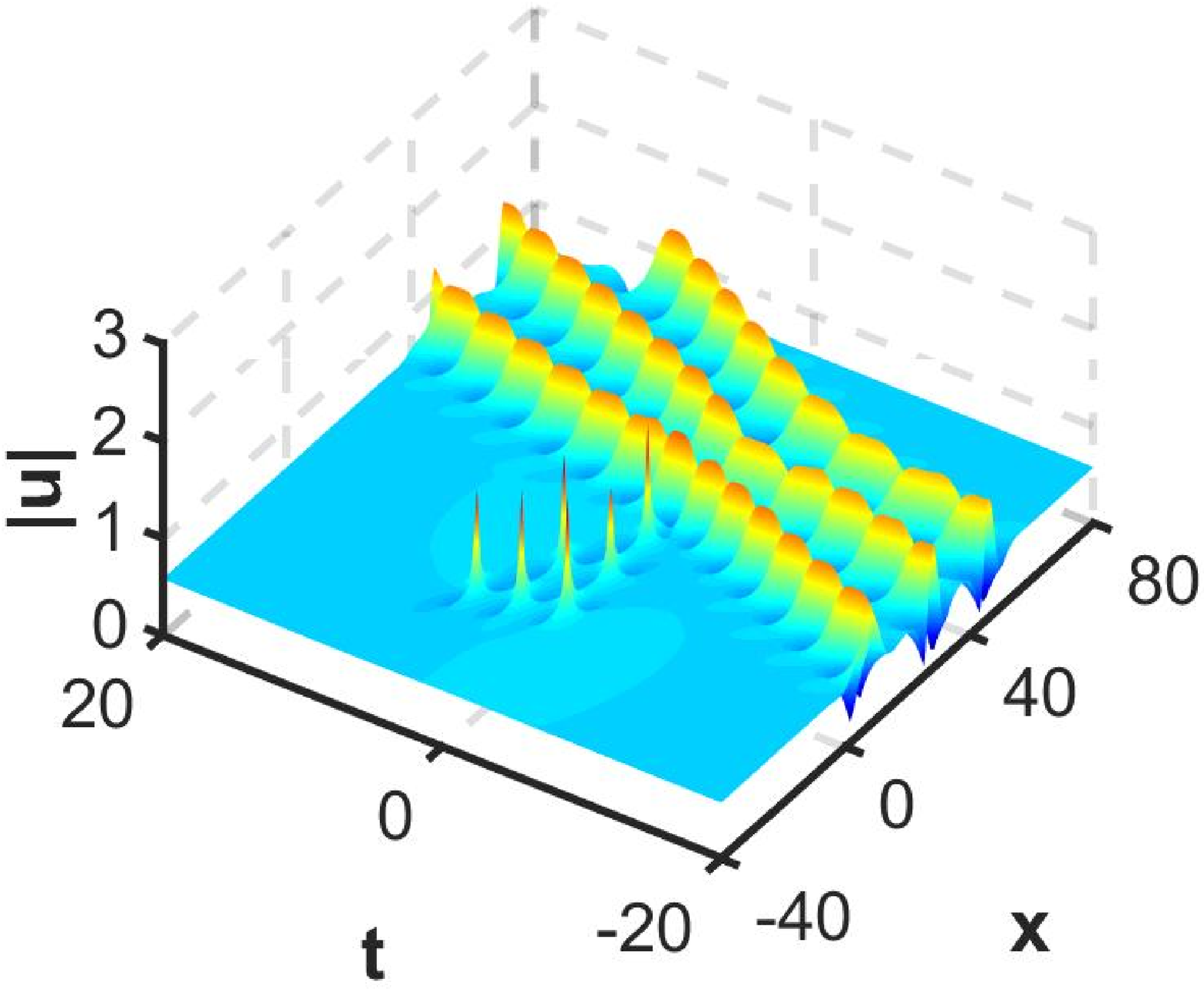}}\hspace{0.5cm}
\subfigure[]{\includegraphics[height=1.5in,width=1.9in]{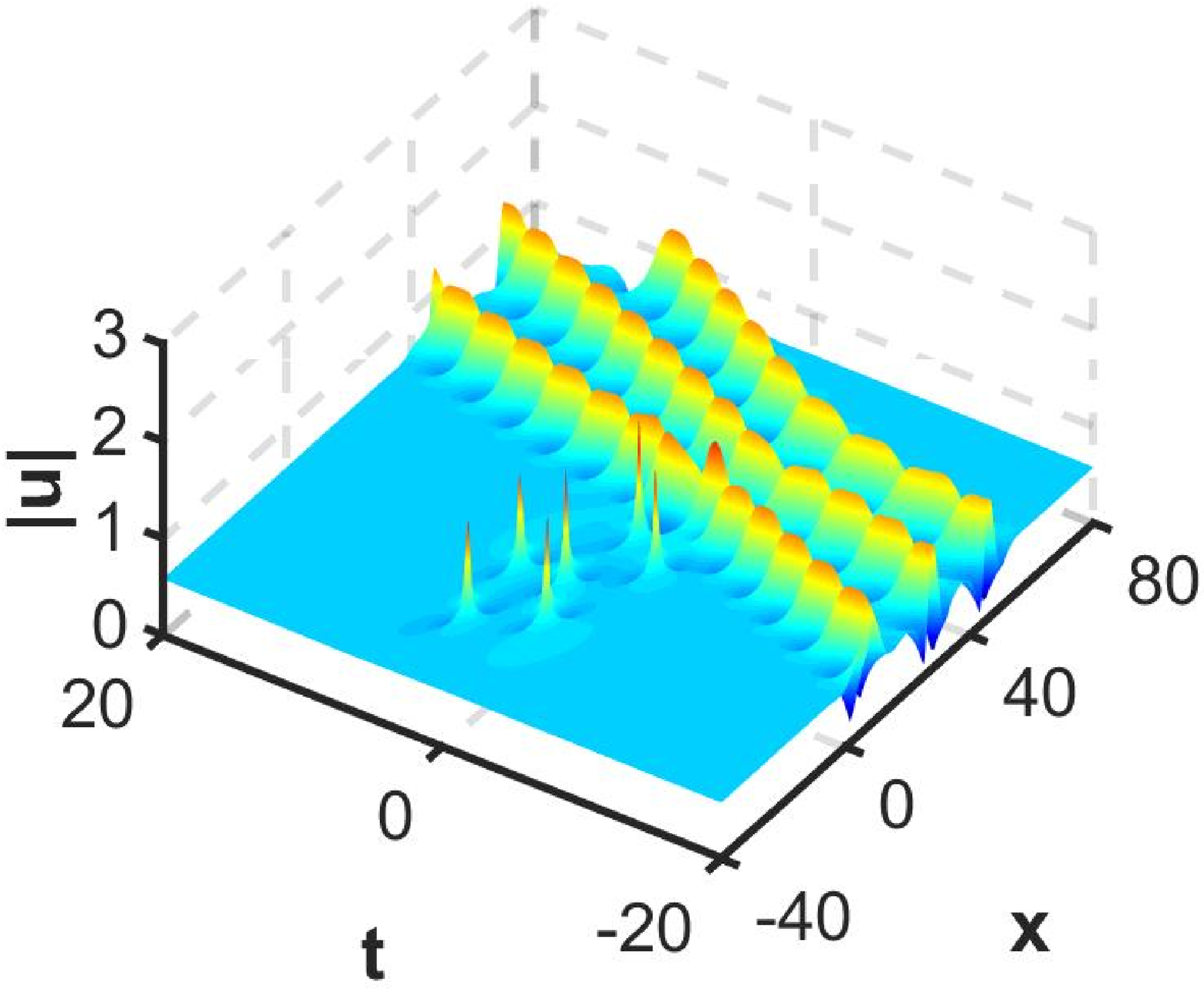}}\hspace{0.5cm}\\
\subfigure[]{\includegraphics[height=1.5in,width=1.9in]{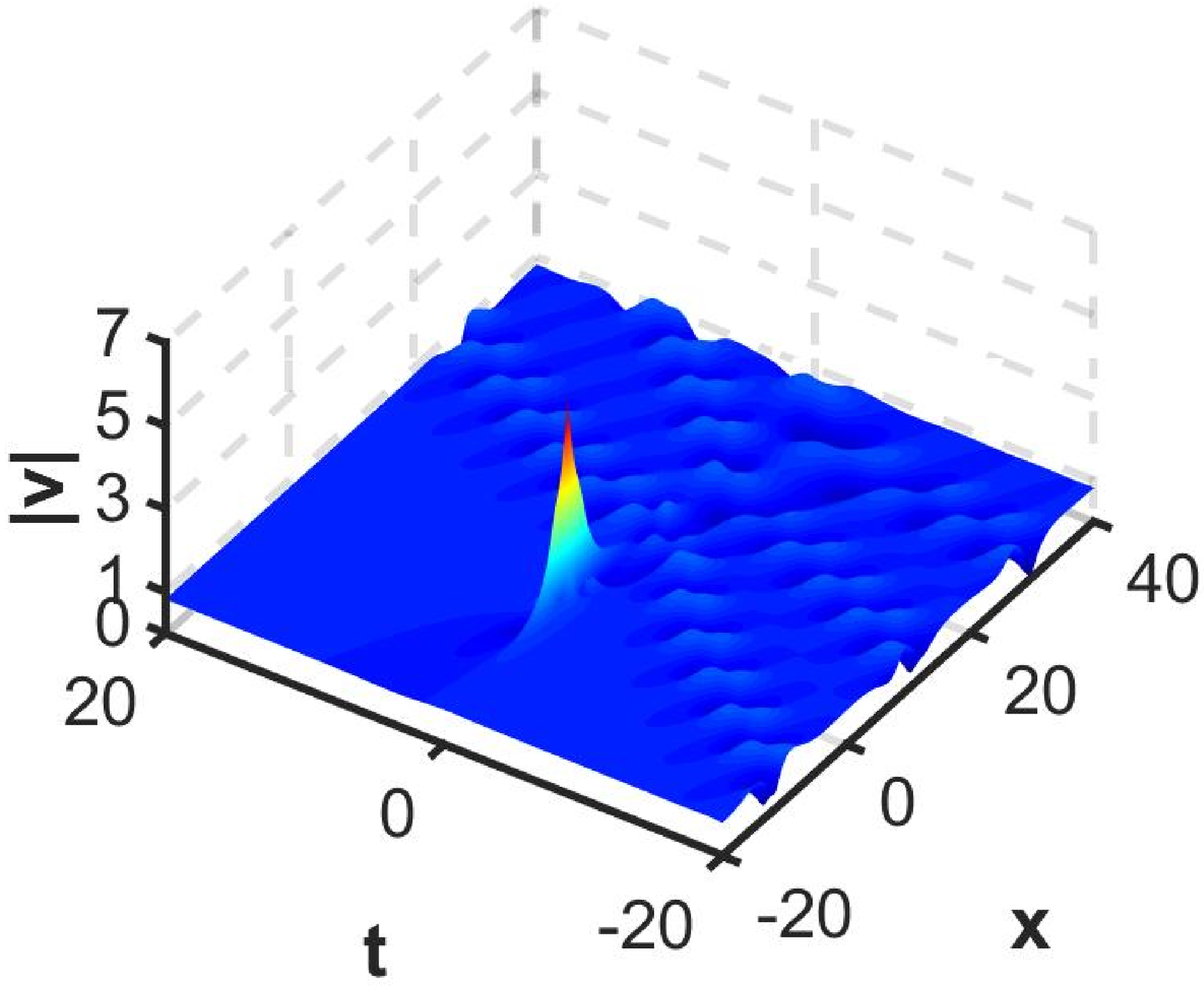}}\hspace{0.5cm}
\subfigure[]{\includegraphics[height=1.5in,width=1.9in]{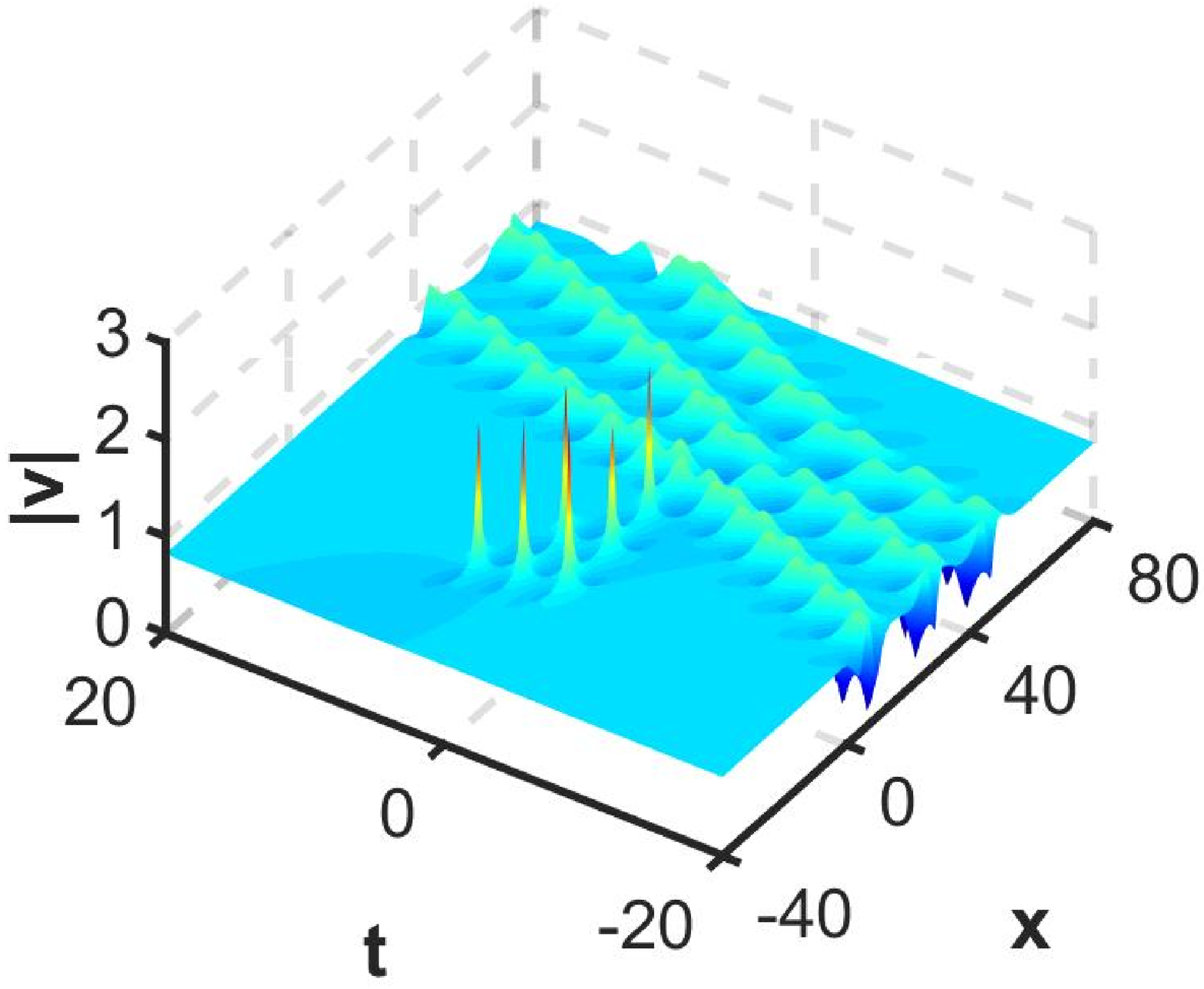}}\hspace{0.5cm}
\subfigure[]{\includegraphics[height=1.5in,width=1.9in]{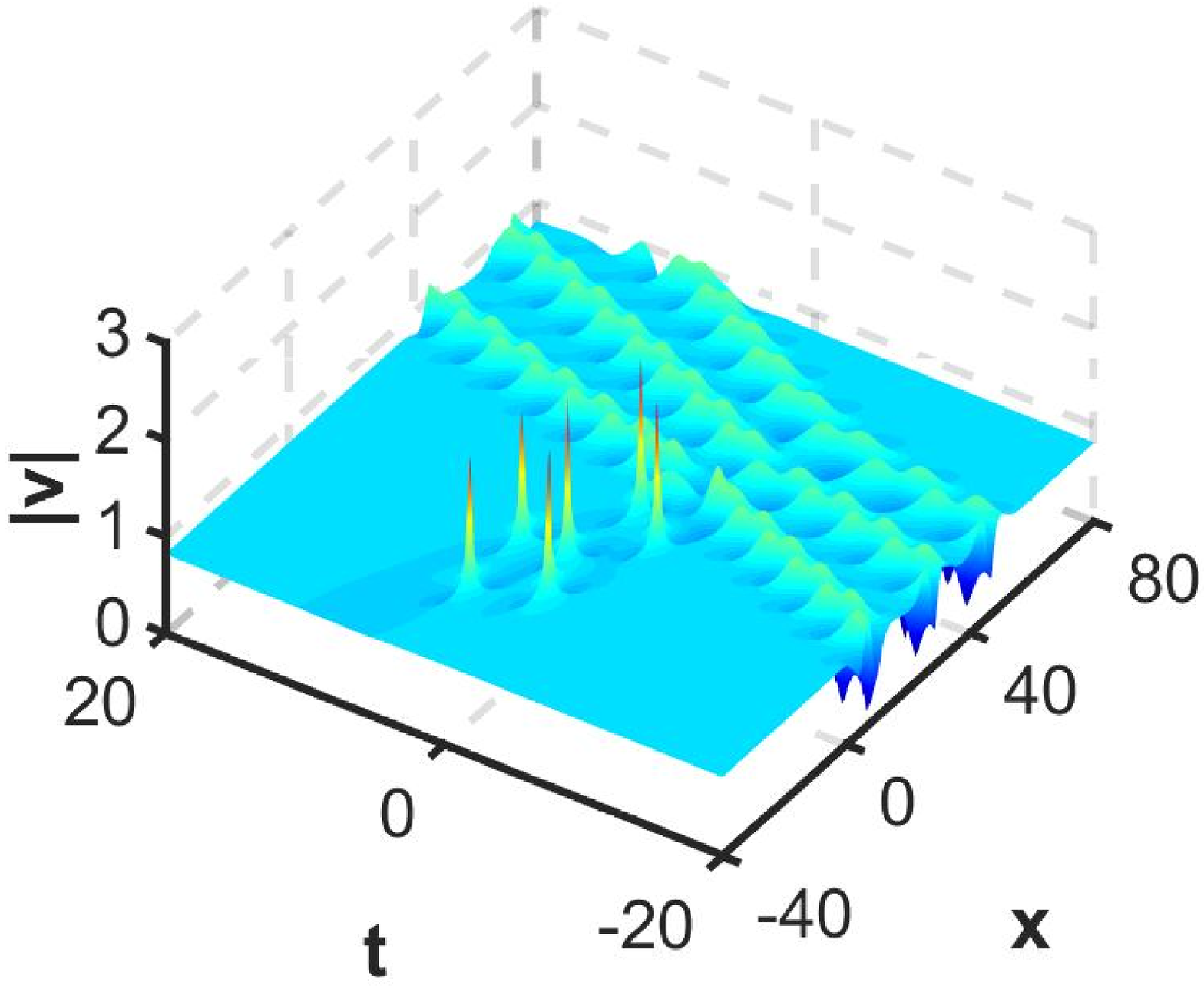}}
\caption{ The third-order interaction solution between three breathers and third-order rogue wave of the gc-FL equation \eqref{gcfl-eq} with parameters: $(\alpha,\beta,\gamma,c_1,c_2,d,m1,m2,n1,n2)$ from the left column to the right column are $(3,1,2,1,0,\frac{1}{1000},0,0,0,0)$, $(3,1,2,1,0,\frac{1}{10000000},100,0,100,0)$ and  $(3,1,2,1,0,\frac{1}{10000000},10,10000,0,0)$, respectively.\label{Fig-gcfl-br3}}
\end{figure*}
\begin{figure*}[!htbp]
\centering
\subfigure[]{\includegraphics[height=1.2in,width=1.5in]{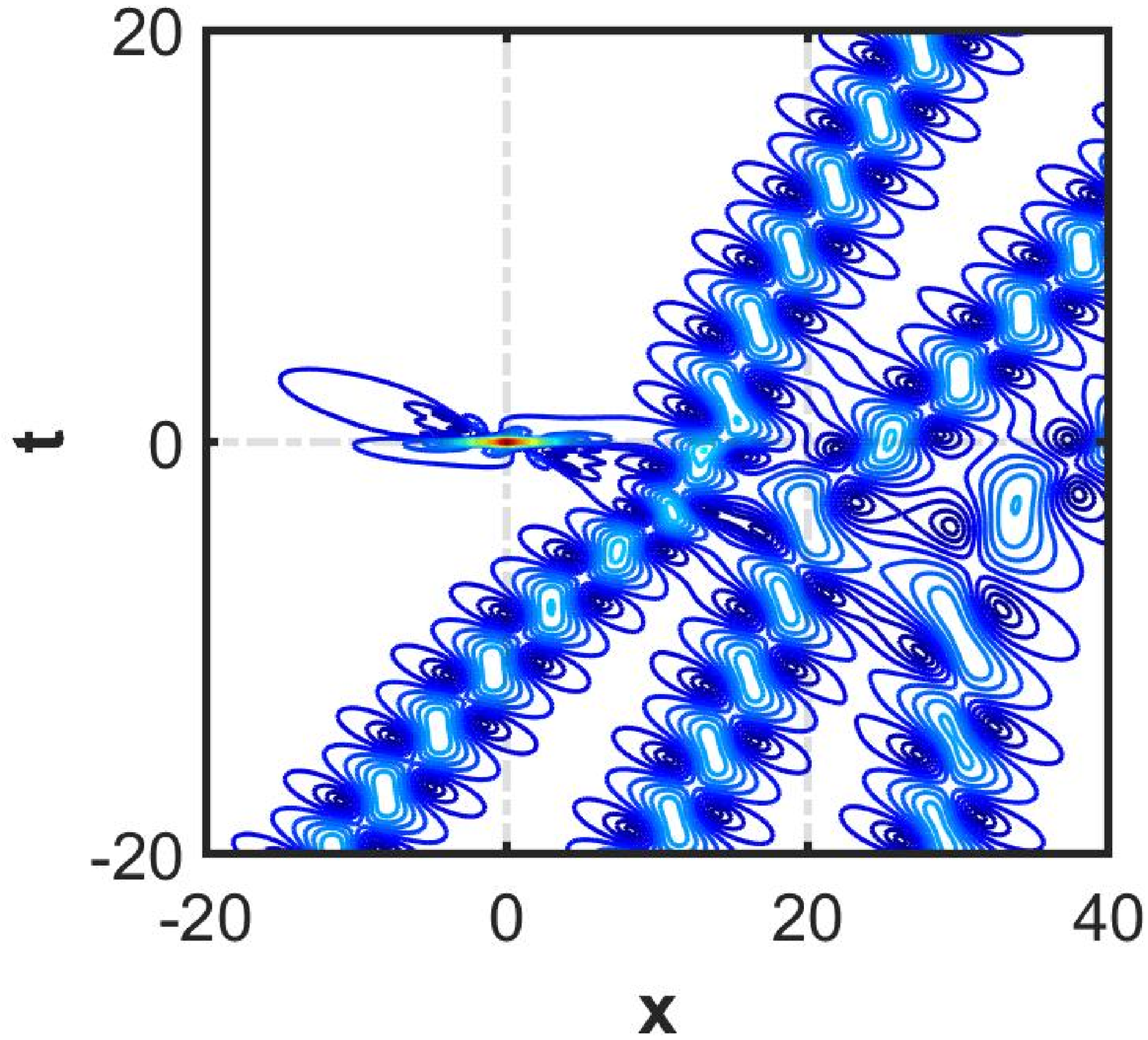}}\hspace{0.5cm}
\subfigure[]{\includegraphics[height=1.2in,width=1.5in]{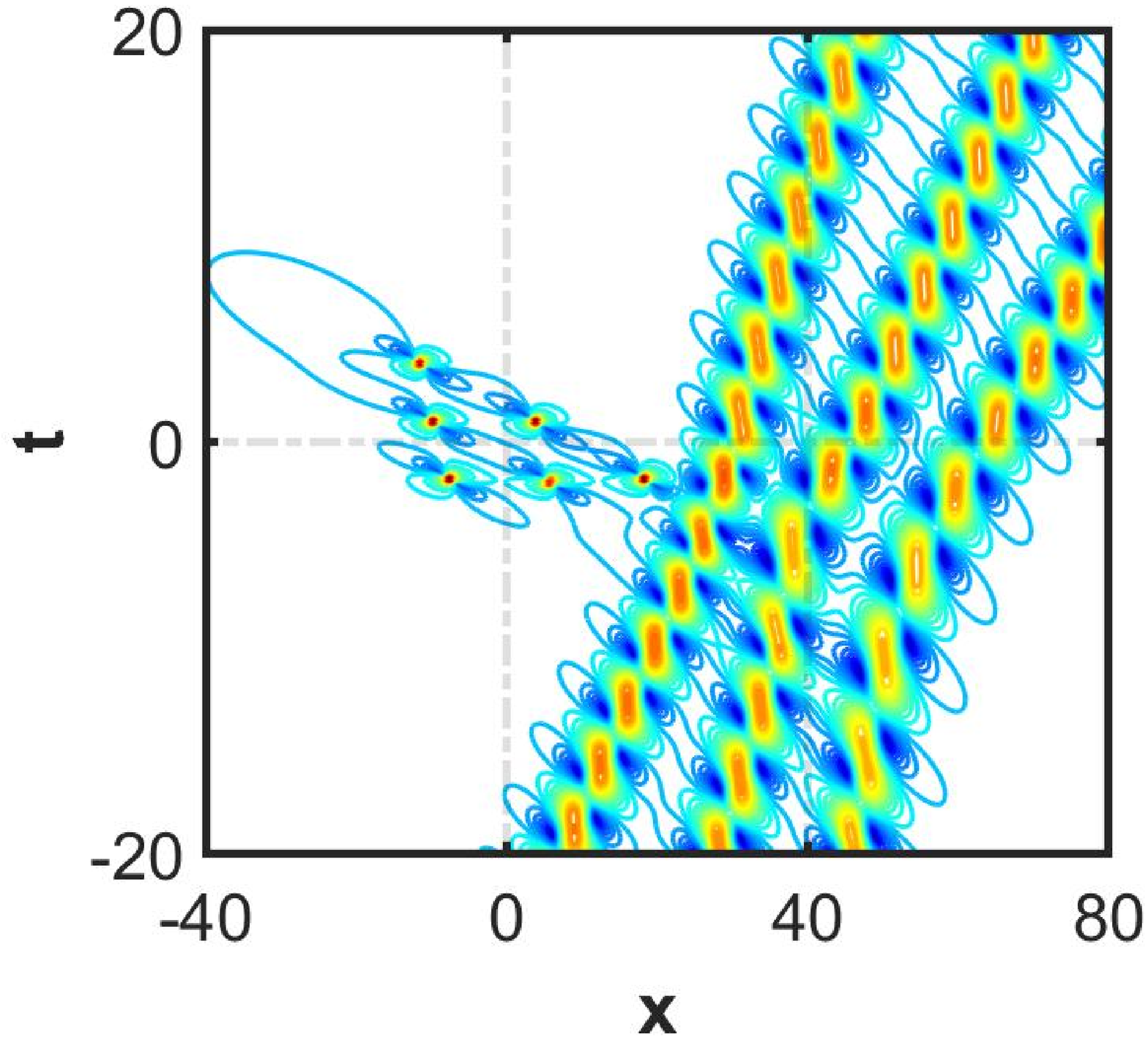}}\hspace{0.5cm}
\subfigure[]{\includegraphics[height=1.2in,width=1.5in]{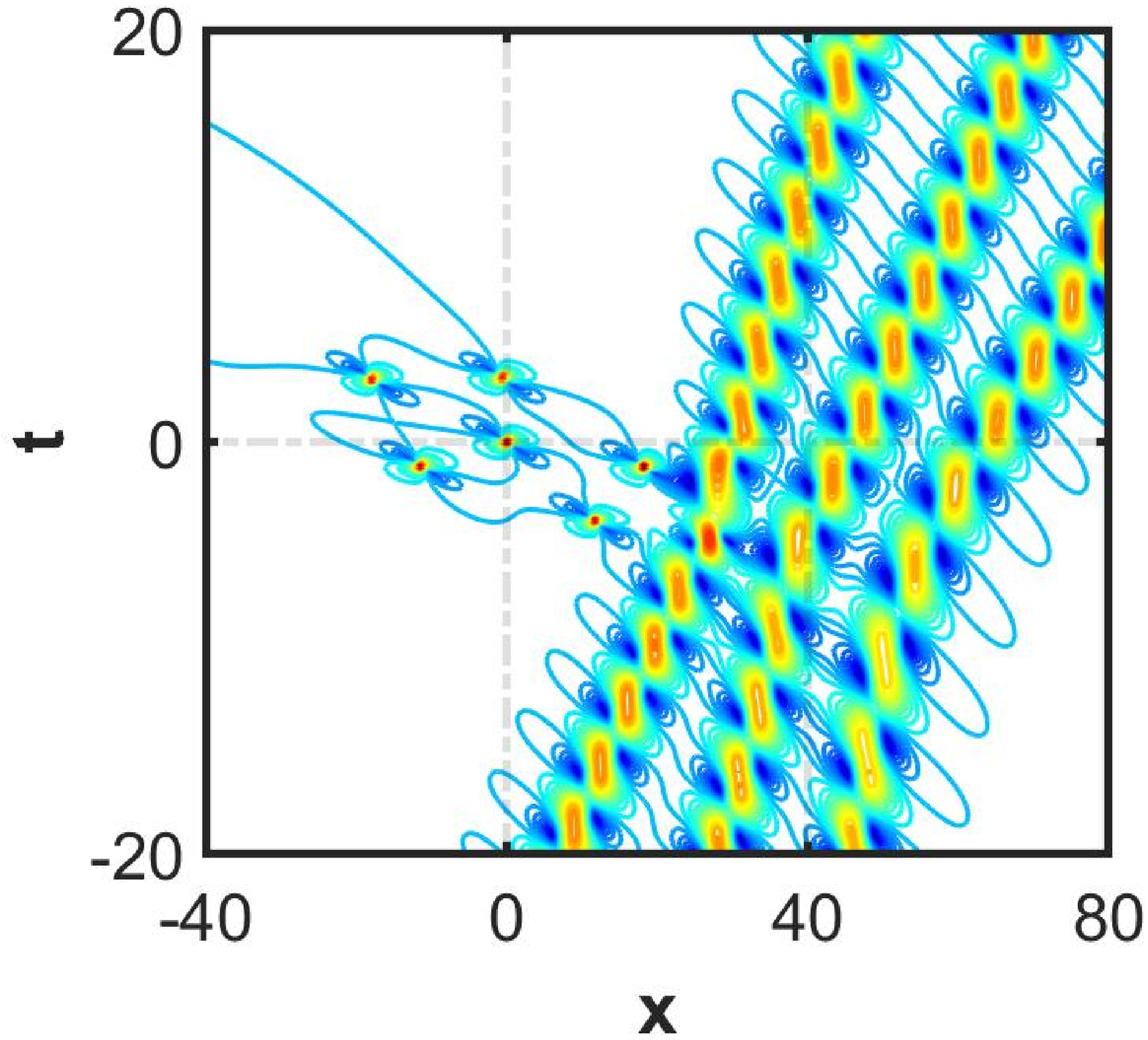}}\hspace{0.5cm}\\
\subfigure[]{\includegraphics[height=1.2in,width=1.5in]{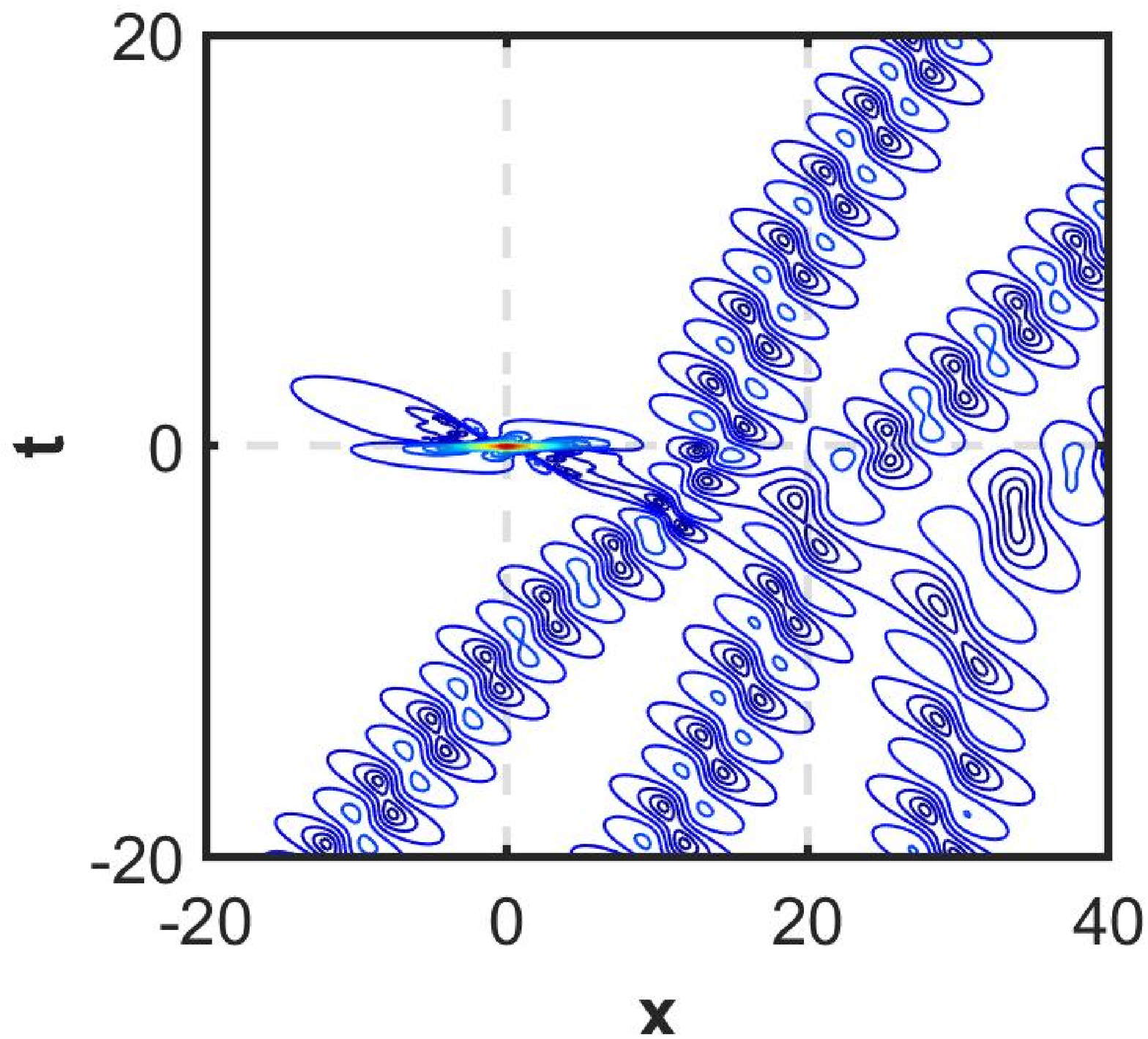}}\hspace{0.5cm}
\subfigure[]{\includegraphics[height=1.2in,width=1.5in]{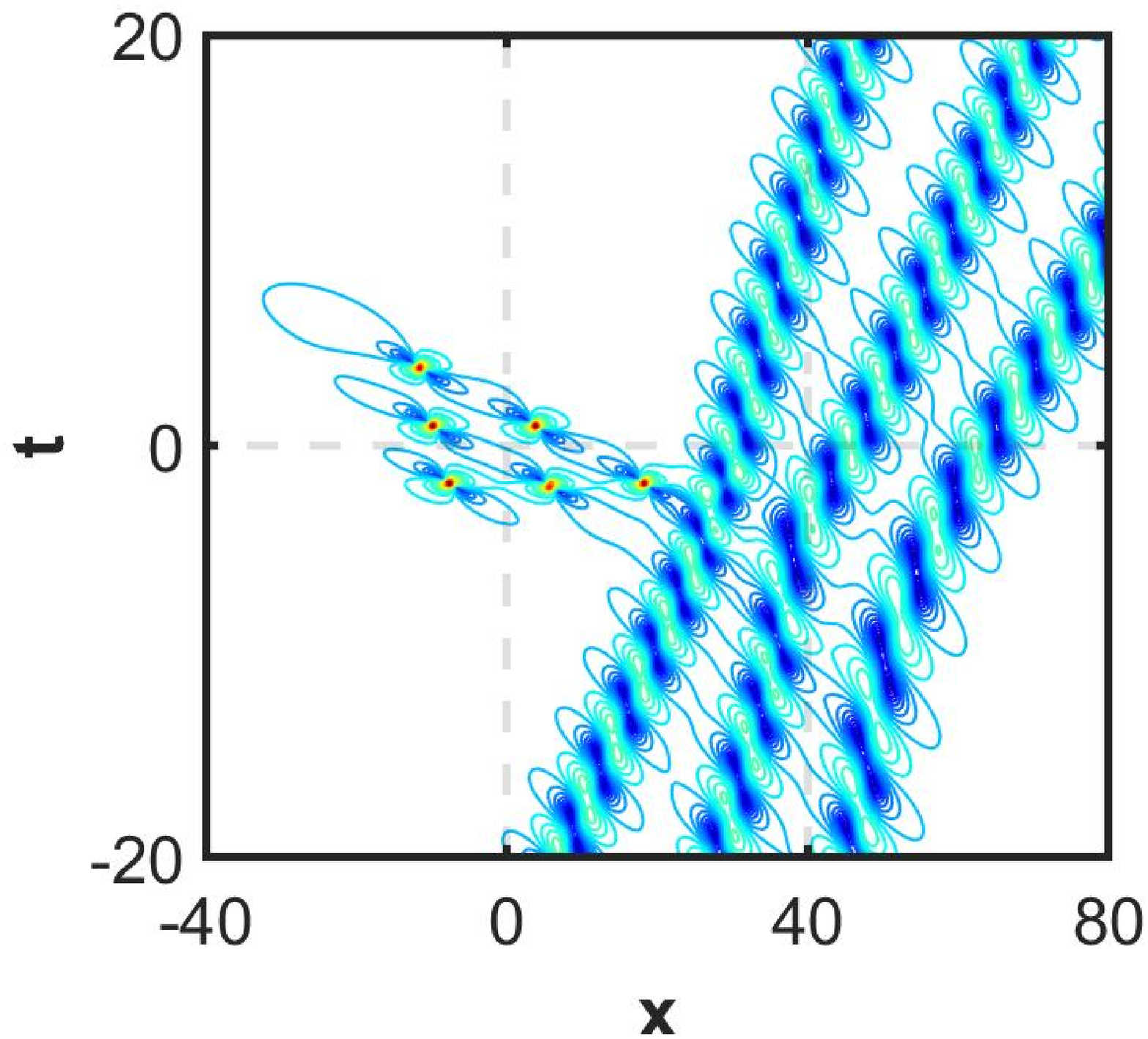}}\hspace{0.5cm}
\subfigure[]{\includegraphics[height=1.2in,width=1.5in]{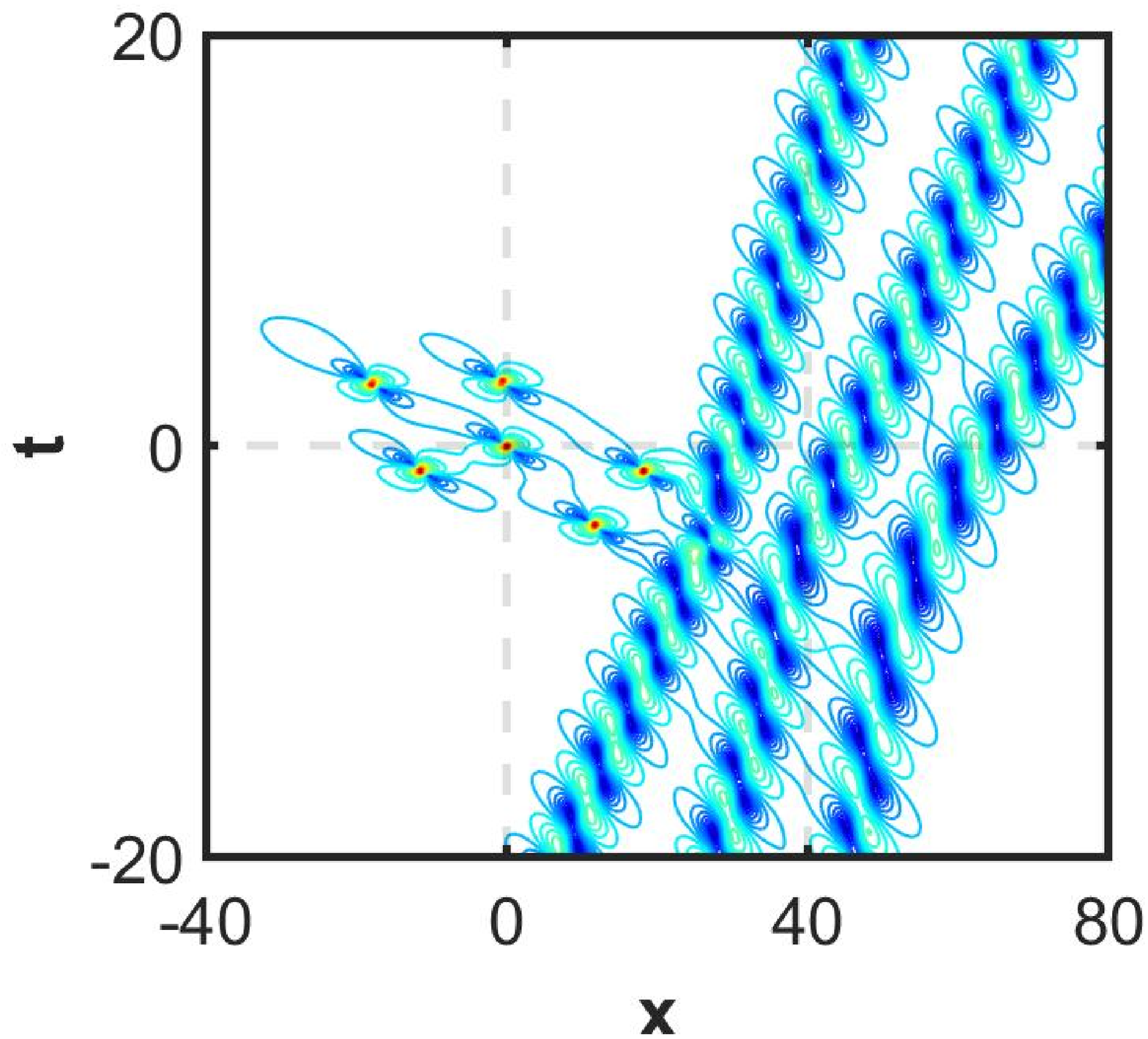}}
\caption{ The corresponding contour maps of Fig. \ref{Fig-gcfl-br3}.\label{Fig-gcfl-br3c}}
\end{figure*}

In what follows, we consider the higher-order case of Eq. \eqref{gcfl-lws-8} with $N=2,3$ and their separate structure forms. When $N=2$, the interaction between the second-order breather and the second-order fundamental rogue wave can be obtained, which are depicted in the first two columns of Fig. \ref{Fig-gcfl-br2} and Fig. \ref{Fig-gcfl-br2c}. These two structures are controllable by parameters $m_1$ and $n_1$. When $d=1$, they interact with each other strongly, i.e., the merge structure, as shown in the odd columns of Fig. \ref{Fig-gcfl-br2} and Fig. \ref{Fig-gcfl-br2c}. On the contrary, when $d=\frac{1}{10000000}$, they are in a separate state, as shown in the even columns of Fig. \ref{Fig-gcfl-br2} and Fig. \ref{Fig-gcfl-br2c}. Whether it is merge structure or separate structure, the phase transition of the breather farther away from the rogue wave changes seriously, while breather closer to the rogue wave does not undergo obvious phase transition. In addition, the phase change caused by the fundamental structure is greater than that caused by the triplet structure. When $N=3$, the interaction between the third-order breather and third-order rogue wave can be obtained. Only parameters $d,m_i,n_i (i=1,2)$ are adjusted here, and other parameters are consistent with the above case. Similar with case 1, there are also three different interaction solution, whose dynamical characteristics are demonstrated in Figs. \ref{Fig-gcfl-br3} and \ref{Fig-gcfl-br3c}. With the increase of $N$, the selection of parameters $m_i$ and $n_i$ has more combination forms, so more abundant higher-order rogue waves and semi-rational solutions can be obtained.


\section{Summary and discussions}

This paper mainly studies three aspects of the generalized coupled Fokas-Lenells equation: modulation instability analysis, infinitely many conservation laws and localized wave solutions.

Based on the linear analysis theory, we obtain the concrete expression of MI gain for the generalized coupled Fokas-Lenells equation. Through further analysis of the expression, the MI and MS distribution regions on frequency surface are depicted. Moreover, the constraint condition for the existence of rogue waves is obtained, namely $0<k<\frac{2}{C}$, which is consistent with the constraint condition obtained in the process of constructing the localized wave solution through the generalized Darboux transform.

According to the Riccati-type formulas of the spectral problem, infinitely many conservation laws are constructed to investigate the integrability of the generalized coupled Fokas-Lenells equation. In addition, we construct the generalized Darboux transform and give the compact determinant expression of the $N$-order localized wave solution. Under the MI analysis, there are mainly three structures:
\begin{itemize}
  \item $\Big\{d=0,~~c_i\neq0~~(i=1,2)\Big\}$ $\Longrightarrow$ rogue wave (fundamental, triangular, pentagon)
  \item $\Big\{d\neq0,~~c_i=0,~~c_j\neq0~~(i\neq j)\Big\}$ $\Longrightarrow$ rogue wave + bright-dark soliton
  \item $\Big\{d\neq0,~~c_i\neq0~~(i=1,2)\Big\}$ $\Longrightarrow$ rogue wave + breather.
\end{itemize}
Here, parameters $(m_i,n_i)$ control the structure of the rogue wave: separation or merge, which leads to the generation of the triangular and pentagon structure, while parameters $(\alpha,\beta)$ affect the phase deflection of the rogue wave. Owing to above diversity of rogue wave structures, the last two items can lead to the interaction between rogue wave with different structures and bright-dark solitons or breathers, where the value of $|d|$ controls the collision distance, which resulting in strong or weak interactions to realize their energy exchange.

\section*{Acknowledgment}
We would like to express our sincere thanks to members of our discussion group for their valuable comments. The project is supported by the National Natural Science Foundation of China (No. 11675054), Future Scientist/Outstanding Scholar training program of East China Normal University (No. WLKXJ2019-004) and Science and Technology Commission of Shanghai Municipality (No. 18dz2271000).







\end{document}